%% file: main.tex
\documentclass[11pt]{article}

\usepackage[T1]{fontenc}
\usepackage[margin=1in]{geometry}
\usepackage{amsmath,amsfonts,graphicx,amssymb,bm,amsthm,float,subfigure}
\usepackage{xcolor}
\usepackage{appendix}
\usepackage[hidelinks]{hyperref}
\usepackage{booktabs}
\usepackage{multirow}
\usepackage{makecell}
\usepackage[ruled, vlined]{algorithm2e}
\usepackage{natbib}
\usepackage{cleveref}
\usepackage{thmtools,thm-restate}
\usepackage{setspace}
\usepackage{tablefootnote}
\usepackage{crossreftools}
\newtheorem{theorem}{Theorem}
\newtheorem{proposition}{Proposition}
\newtheorem{definition}{Definition}
\newtheorem{lemma}{Lemma}
\newtheorem{corollary}{Corollary}
\newtheorem{example}{Example}

\crefname{assumption}{Assumption}{Assumption}
\crefname{algorithm}{Algorithm}{Algorithm}

\newcommand{\ql}{{\mathrm{QL}}}

\begin{document}

\title{Online Fair Allocation with Best-of-Many-Worlds Guarantees}

\author{
Zongjun Yang \quad Luofeng Liao \quad Yuan Gao \quad Christian Kroer \\[4pt]
Columbia University \\[2pt]
\texttt{\{zy2684, ll3530, yg2541, ck2945\}@columbia.edu}
}

\date{}
\maketitle
\onehalfspacing

\begin{center}
\textit{Major Revision at Management Science}
\end{center}

\begin{abstract}
We investigate the online fair allocation problem with sequentially arriving items under various input models, with the goal of balancing fairness and efficiency. We propose the unconstrained PACE (Pacing According to Current Estimated utility) algorithm, a parameter-free allocation dynamic that requires no prior knowledge about the input, while using only integral allocations. PACE attains convergence or approximation guarantees under stationary, stochastic-but-nonstationary, and adversarial input models, thereby achieving the first best-of-many-worlds guarantee in online fair allocation. Beyond theoretical bounds, PACE is simple, efficient, and decentralized, and is thus likely to perform well on a broad range of real-world inputs.
Numerical results support the conclusion that PACE works well under a variety of input models. We find that PACE performs very well on two real-world datasets, even under the true temporal arrivals in the data, which are highly nonstationary.
\end{abstract}

\input{best-of-many-worlds/introduction}

\input{best-of-many-worlds/related-work}

\input{best-of-many-worlds/preliminaries}

\input{best-of-many-worlds/algorithm}

\input{best-of-many-worlds/stationary}

\input{best-of-many-worlds/nonstationary}

\input{best-of-many-worlds/adversarial}

\input{best-of-many-worlds/applications}

\input{best-of-many-worlds/experiments}

\section*{Acknowledgements}
This work was supported by the Office of Naval Research awards N00014-22-1-2530 and N0014-23-1-2374, and the National Science Foundation awards IIS-2147361 and IIS-2238960.

\bibliographystyle{plainnat}
\bibliography{ref}

\newpage
\appendix
\input{best-of-many-worlds/appendix-related-work}
\input{best-of-many-worlds/review}
\input{best-of-many-worlds/proof-iid}
\input{best-of-many-worlds/proof-non}

\input{best-of-many-worlds/proof-adv}

\input{best-of-many-worlds/proof-applications}
\input{best-of-many-worlds/appendix-experiments}

\end{document}

%% file: best-of-many-worlds/introduction.tex
\section{Introduction}
\label{section: introduction}
We study the problem of \emph{online fair allocation}, where items arrive sequentially over $t$ time steps and must be allocated among $n$ agents with heterogeneous preferences, with the goal of balancing fairness and efficiency. At each time step, the decision maker observes the agents' values for the arriving item and makes an irrevocable allocation decision. We assume throughout that agents have linear utility functions.

Many marketplaces and Internet-based systems face the challenge of allocating a large volume of incoming items fairly and efficiently, often at high frequency and in real time. In the most direct interpretation, the items are time-sensitive resources that become available sequentially and must be allocated promptly to satisfy heterogeneous needs. Examples include allocating perishable food donations to food banks~\citep{gorokh2021remarkable,sinclair2022sequential} and computational resources to distributed clients in cloud platforms~\citep{wei2010game,wang2014multi}. More broadly, the framework also applies to online platforms where the arriving items are not goods or resources per se, but platform users whose engagement generates utility for content providers through downstream outcomes such as clicks or other conversions. For example, in notification optimization, opportunities to engage users arise as users check their notification updates, and the platform must allocate user attention across content generated by different stakeholders with competing objectives~\citep{kroer2023fair}. In nonprofit operations, similar allocation scenarios arise in platforms that match volunteers to neighborhoods~\citep{manshadi2023redesigning} or blood donors to donation centers~\citep{mcelfresh2020matching}, where volunteers or donors are scarce resources that must be allocated equitably to different recipients through recommendations or notifications. In these applications, balancing fairness and efficiency is both an operational priority for the platform and a central question raised by the literature~\citep{mcelfresh2020matching,manshadi2021fair,kroer2023fair}. Given the large-scale nature of these systems, it is important to develop online fair allocation algorithms that are simple, efficient, and robust.

To formalize the balance between fairness and efficiency, we focus on Nash welfare (NW)~\citep{nash1950bargaining}, defined as the geometric mean of agents' utilities. This objective is especially attractive for fair allocation because, under linear utilities, the offline NW-maximizing allocation satisfies a collection of fairness and efficiency properties. On the fairness side, it satisfies canonical properties such as envy-freeness and proportionality. On the efficiency side, it is strongly Pareto optimal and core stable; moreover, the Nash welfare objective itself is an economically meaningful and widely studied measure of collective welfare~\citep{varian1974equity,moulin2004fair}. In the offline setting, the NW-maximizing allocation can be computed efficiently through the classical Eisenberg--Gale (EG) convex program~\citep{eisenberg1959consensus}. Our goal in the online setting is then to design a simple and computationally efficient algorithm that, despite making irrevocable allocation decisions without knowing future arrivals, achieves tight approximations to this collection of fairness and efficiency guarantees: envy-freeness, proportionality, strong Pareto optimality, core stability, and Nash welfare maximization.

In real-world allocation problems, the arrival process of items may not follow a stationary stochastic pattern, but can instead be nonstationary or even adversarial\footnote{Such arrivals can also model settings where agents' preferences or availabilities change over time. For example, an agent's temporary absence can be captured by zeroing out her valuations over a period of time.}. Since the input model is often unknown in advance, online algorithms tailored to a specific arrival model may be undesirable. For example, assuming that items are sampled from a fixed distribution at each time step may enable the strongest convergence guarantees, but this assumption may not be suitable in all settings. Conversely, one may seek worst-case guarantees against adversarial inputs, yet such guarantees can be overly conservative when arrivals are generated by a more benign process. Existing online fair allocation algorithms for Nash welfare are designed specifically for either stochastic or worst-case inputs, and may fail to achieve desirable guarantees when the input does not follow the assumed model. For stochastic inputs, \citet{gao2021online} provide an algorithm that converges asymptotically to the NW-maximizing utilities, but even moderately adversarial inputs may lead to highly unfair allocations. For adversarial inputs under suitable normalization conditions, \citet{banerjee2022online,huang2022online} develop online approximation algorithms with Nash welfare competitive-ratio guarantees, but these algorithms can be inefficient on typical inputs because they make highly conservative decisions, such as allocating a fixed fraction of every item uniformly across agents. Prior to our work, it was not known whether a simple and robust algorithm could achieve strong fairness and efficiency guarantees simultaneously across multiple input models, from the ideal \textit{i.i.d.} case to adversarial arrivals.

\subsection{Our Contributions}

\paragraph{The unconstrained PACE algorithm.}
We propose PACE (Pacing According to Current Estimated utility)\footnote{Our algorithm is an unconstrained version of a previous version of PACE by \citet{gao2021online}, which is prior-dependent and sensitive to adversarial input; see detailed discussions in \Cref{section: algorithm}. We will refer to their algorithm as ``constrained PACE''.}, a simple online allocation algorithm based on proportional utility increments. In the standard unweighted setting, PACE follows a highly intuitive decision rule: allocate each arriving item to the agent whose utility would increase by the largest proportion after receiving it:
\begin{equation*}
    i^\tau \in \arg\max_{i\in[n]} \frac{v_i^\tau}{U_i^{\tau-1}} .
\end{equation*}
This rule directly encodes the balance between fairness and efficiency. The numerator $v_i^\tau$ captures the immediate efficiency gain from assigning item $\tau$ to agent $i$, while the denominator $U_i^{\tau-1}$ captures how well agent $i$ has already been served in previous allocations. By maximizing the ratio between current value and accumulated utility, PACE can be viewed as interpolating between the efficiency extreme of allocating to the agent with the largest instantaneous value and the conservative fairness extreme of serving only the worst-off agent so far. This simple structure gives PACE several desirable features for large-scale online applications: it uses only integral allocations, requires no prior knowledge of the arrival process, value distribution, horizon length, or benchmark solution, and can be implemented in a decentralized manner using only agent-level utility estimates.

Besides its intuitive interpretation in terms of fairness and efficiency, PACE also closely resembles the widely-used pacing strategies found in the practice of large-scale online advertising, where budget-constrained agents participate in repeated auctions by shading their values by a series of pacing multipliers. PACE can be interpreted exactly as agents bidding for the $t$ items with the multiplier inversely proportional to their cumulative utility $U_i^{\tau-1}$. In \Cref{section: fppe}, we concretely formulate PACE as multi-agent bidding dynamics under budget constraints, and provide a set of theoretical guarantees \textit{w.r.t.} the benchmark of \textit{first-price pacing equilibrium} (FPPE, \citet{conitzer2022pacing}) under various arrival models. 

\paragraph{Best-of-many-worlds results.}
We consider the fairness notions of \textit{envy-freeness} and \textit{proportionality}, and the efficiency notions of \textit{Nash welfare approximation} and \textit{core stability}\footnote{Core stability implies strong Pareto optimality by considering the grand coalition, so we do not state separate Pareto optimality results.} in the online setting, in the asymptotic sense of the arrival length $t$. Our guarantees show that PACE performs well under a broad range of inputs, and is thus a natural and robust dynamic for online fair allocation. The main input models and results are summarized in \Cref{table: overview-linear}.

For stationary data (\Cref{section: stationary}), where item values are independently and identically distributed (\textit{i.i.d.})\ according to an unknown distribution, PACE achieves vanishing $\widetilde{O}(1/\sqrt{t})$ time-averaged additive errors for envy-freeness, proportionality, Nash welfare optimality, and core stability. These guarantees follow from a stronger underlying convergence result: the average utility vector achieved by PACE converges to the NW-maximizing utilities in a mean-square sense at an $\widetilde{O}(1/t)$ rate. This utility convergence is important beyond its role in the analysis, because it shows that the algorithm does not merely perform well according to aggregate fairness and efficiency metrics; rather, each agent's realized long-run utility becomes stable, predictable, and calibrated to the offline fair allocation benchmark.

For nonstationary-yet-stochastic data (\Cref{section: nonstationary}), we study two input models: ergodic data and blockwise independent data. The ergodic model captures settings where temporal correlations may be present locally but decay in the long run, with the mixing parameter $\iota$ measuring how quickly the arrival process returns close to its long-run average behavior and $\delta$ measuring the remaining distributional deviation. The blockwise independent model captures structured nonstationarity such as periodicity or mild corruption, where the maximum block length $|\mathcal P|_\infty$ measures the duration of a correlated or seasonal regime and the variation parameter $\delta$ measures how far the block distributions deviate from the overall average. Under both models, we show that the guarantees for the \textit{i.i.d.} case are preserved up to additive errors that depend smoothly on the corresponding nonstationarity parameters, highlighting the robustness of PACE beyond stationary environments.

For adversarial data (\Cref{section: adversarial}), unrestricted worst-case arrivals preclude meaningful guarantees for any online algorithm~\citep{banerjee2022online}, so we study two natural and practically motivated relaxations. First, under a non-extreme input condition, characterized by a bounded ratio between the largest and smallest non-zero value observed by each fixed agent, PACE achieves multiplicative guarantees for all fairness and efficiency metrics, with up-to-constant dependence on $t$ and logarithmic dependence on the extremity parameter $\varepsilon$. Practically, a larger $\varepsilon$ means that each stakeholder's relevant items are of similar scale in value, while a very small $\varepsilon$ corresponds to settings with rare, exceptionally valuable opportunities mixed with much lower-value ones, where worst-case fairness guarantees are inherently harder. Second, under a seed utility assumption, which captures systems where agents have accumulated initial stakes of value at least $\xi$ before the online procedure begins, PACE obtains multiplicative guarantees for the fairness and efficiency metrics with logarithmic dependence on $t/\xi$.

\begin{table}[!htbp]
\renewcommand{\arraystretch}{1.5}
\centering
\small
\begin{tabular}{c|c|c}
\Xhline{1.2pt}
\multicolumn{2}{c|}{Input Model}
& \makecell{Envy, Proportionality, NW, and Core Stability} 
\\ 
\Xhline{1.2pt}

\multicolumn{2}{c|}{\makecell{Stationary\\(\Cref{section: stationary})}}
& \makecell{ $\widetilde{O}(\sqrt{t})$-additive\\
(with $\widetilde{O}(1/t)$ utility convergence)}
\\ 
\hline

\multirow{2}{*}{\makecell{Nonstationary\\(\Cref{section: nonstationary})}}
& \makecell{Ergodic input\\$\mathsf{C}^{\mathrm{E}}(\delta,\iota)$}
& \makecell{$\widetilde{O}\!\left(\sqrt{\iota t}+\sqrt{\delta} t\right)$-additive\\
(with $\widetilde{O}\!\left(\iota/t+\delta\right)$ utility convergence)}
\\ 
\cline{2-3}

& \makecell{Block input\\$\mathsf{C}^{\mathrm{B}}(\mathcal{P},\delta)$}
& \makecell{$\widetilde{O}\!\left(|\mathcal{P}|_\infty\sqrt{t}+\sqrt{\delta}t\right)$-additive\\
(with $\widetilde{O}\!\left(|\mathcal{P}|_\infty^2/t+\delta\right)$ utility convergence)}
\\ 
\hline

\multirow{2}{*}{\makecell{Adversarial\\(\Cref{section: adversarial})}}
& \makecell{Non-extreme input\\parameter $\varepsilon$}
& \makecell{$O\!\left(\log(1/\varepsilon)\right)$-multiplicative}
\\ 
\cline{2-3}

& \makecell{Seed utility $\xi$}
& \makecell{$O((1/\xi)\cdot \log (t/\xi))$-multiplicative}
\\ 
\Xhline{1.2pt}
\end{tabular}

\caption{\small Summary of theoretical bounds for linear utilities.}
\label{table: overview-linear}
\vspace{-0.5em}
\begin{center}

\begin{minipage}{1.0\textwidth}
\centering
\footnotesize
\par
$\mathsf{C}^{\mathrm{E}}(\delta,\iota)$: ergodic arrivals whose $\iota$-step conditional distribution is within total variation distance $\delta$ from average.
\par
$\mathsf{C}^{\mathrm{B}}(\mathcal{P},\delta)$: block-wise independent arrivals with partition $\mathcal{P}$ and average block-distribution deviation at most $\delta$.
\par
$\varepsilon$: lower bound on the ratio between an agent's smallest and largest positive valuation.
\par
$\xi$: initial seed utility for each agent.
\end{minipage}
\end{center}
\end{table}

\paragraph{{\textbf{Application to fair matching in blood donation platforms.}} } 
{Our framework also applies to fair matching problems on platforms where users arrive over time and must be matched or routed fairly across recipients. A representative example is a blood donation platform, such as Facebook Blood Donation~\citep{american2019new}, where potential donors are matched to compatible recipients such as hospitals, blood banks, or donation sites. In this setting, fairness is an operational concern: a platform that only maximizes the immediate likelihood of successful donations may repeatedly favor some recipients while leaving others with legitimate demand underserved. We build on the matching setting of \citet{mcelfresh2020matching} where each day a dynamic donor-recipient compatibility graph appears, with edge weights encoding expected donation probabilities. The daily graph may be sparse and may exhibit strong intra-day correlation due to common operational factors such as location, donor eligibility, blood-type demand, and site capacity. We show that a donor-by-donor application of PACE achieves $\widetilde{O}(1/D)$ mean-square convergence in recipient matched weights over $D$ days, and consequently obtains the corresponding fairness and efficiency guarantees for the induced fair allocation problem (\Cref{section: app}). The resulting policy is operationally simple: rather than repeatedly solving a full daily matching problem or estimating the full arrival distribution, the platform checks each active donor once and immediately recommends the donor to the compatible recipient whose accumulated service level would increase by the largest priority-adjusted proportion.}

\paragraph{Technical Highlights }   
For the adversarial analysis, we use the primal interpretation of PACE as an integral first-order greedy algorithm for Nash welfare maximization. Unlike several previous algorithms for online NW maximization~\citep{banerjee2022online,huang2022online} or envy minimization~\citep{benade2018make}, PACE does not rely on conservative uniform or set-aside allocations or their randomized versions. This requires more delicate worst-case analysis, since a complete set of fairness and efficiency metrics needs to be investigated without explicitly reserving resources for agents. Our key structural observation is a \textit{consistency} property of PACE: if one restricts attention to any subset of agents and to the items allocated to them by PACE, then re-running PACE on this restricted instance produces the same allocation. This property allows us to reduce general adversarial instances to two-agent instances, and then characterize worst-case arrival sequences through a family of linear programs. 
This helps enable multiplicative fairness and efficiency guarantees without relying on conservative decisions.

For our analysis under stochastic input, we show that PACE's allocation rule can be interpreted as a type of \textit{dual-averaging} update~\citep{xiao2009dual} on the dual Eisenberg-Gale program, yet with a composite regularizer and without domain boundedness and strong convexity. Due to these technical distinctions, applying the analysis of \citet{xiao2009dual} for non-strongly convex functions only gives an $\widetilde{O}(1/\sqrt{t})$ rate (corresponding to $\widetilde{O}(t^{3/4})$ in the fairness and efficiency metrics), and even this still requires boundedness.\footnote{For the unboundedness issue, \citet{gao2021online} provide a direct fix of projecting the dual multipliers to a fixed bounded set. However, doing this safely requires prior knowledge, and thus adds vulnerability to adversarial input, which our paper aims to avoid in the first place.} To this end, 
we develop an \textit{implicit bounding} technique, which uses the primal interpretation of PACE to show that, with high probability, every agent's time-averaged utility becomes bounded away from zero after a short initial period. Equivalently, the induced dual variables remain in a bounded region where the composite term behaves as a strongly convex regularizer. This links the fairness nature in the allocation rule to the stability required by the dual learning analysis.

\nocite{anari2016nash,bach2019universal,nash1950bargaining, kaneko1979nash,eisenberg1959consensus,cole2017convex,cole2018approximating,barman2018finding,garg2020approximating,garg2018approximating,li2022constant,barman2020tight,chaudhury2021fair}

%% file: best-of-many-worlds/related-work.tex
\subsection{Relationship to the Literature}
\label{section: related-work}

Here we discuss the most closely related work; see \Cref{appendix: related-work} for a more comprehensive review.

\paragraph{Constrained PACE.}
For \textit{i.i.d.}\ arrivals, \citet{gao2021online} develop the constrained PACE algorithm and show that it asymptotically attains the NW-maximizing solution~\citep{gao2023infinite} with an $O(\log t/t)$ mean-square convergence rate. However, they assume a uniform pre-normalization of all agent valuations, as well as prior knowledge of the range of each underlying optimal utility $u_i^*$. This information is used to enforce projection onto a compact domain. We show that this projection is unnecessary for stochastic inputs and can be problematic in the adversarial setting. Our algorithm removes this prior-knowledge requirement while retaining the same convergence rate under \textit{i.i.d.} input and yielding new robust worst-case bounds; see \Cref{section: algorithm} for a detailed comparison. 

\paragraph{Online Nash welfare maximization.}
For general adversarial arrivals, no online algorithm can achieve meaningful Nash welfare guarantees in the worst case~\citep{banerjee2022online}. This motivates using assumptions to enable meaningful guarantees. A line of prior work focused on adversarial robustness, often by using additional predictions, regularity assumptions, or conservative allocation components. Accordingly, these algorithms are not designed to provide convergence guarantees under stochastic inputs; our focus is instead on a prior-free non-conservative dynamic that preserves stochastic-input convergence. \citet{azar2010allocate} adopt an assumption similar to the first part of our adversarial analysis: the minimum nonzero valuation of each agent is at least an $\varepsilon$ fraction of that agent's maximum valuation. They develop guarantees on utility ratios for a greedy-based resolving algorithm, which imply comparable guarantees for Nash welfare. However, their $O(\log(nt/\varepsilon))$ upper bound retains a dependence on the horizon length $t$, so the ratio grows to infinity as the horizon length scales. \citet{banerjee2022online} assume access to a prediction of each agent's total sum of utilities and propose the set-aside greedy (SAG) algorithm, which achieves $O(\log n)$ and $O(\log t)$ competitive-ratio guarantees. However, SAG uses a conservative design: a fixed fraction of each item is allocated uniformly across agents, which can be inefficient for non-adversarial inputs. Our analysis for the seed-utility setting recovers a set-aside-type guarantee with the same dependence on $t$. \citet{huang2022online} assume that the input is $\lambda$-balanced or $\mu$-impartial, where $\lambda$ and $\mu$ characterize desired regularity properties of the input; their competitive-ratio upper bounds are logarithmic in these parameters and in $n$; their guarantees implicitly depend on the horizon length $t$, are not scale invariant, and rely on input parameters that are NP-hard to compute in advance. Notably, non-conservative greedy-style algorithms have appeared previously in this literature, but only with $t$-dependent guarantees~\citep{azar2010allocate} or only for binary valuations~\citep{huang2022online}.

\begin{table}[t]
\renewcommand{\arraystretch}{1.5}
\centering
\caption{\small Comparison with existing online NW maximization algorithms; \textit{i.i.d.}\ and adversarial input models.}
\label{tabel: compare-with-existing}
\small 
\begin{tabular}{c|cc!{\vrule width 1.2pt}cc}
\Xhline{1.2pt}
\multirow{2}{*}{}           & \multicolumn{2}{c!{\vrule width 1.2pt}}{\textit{i.i.d.}}                                                                    & \multicolumn{2}{c}{Adversarial}                                   \\ \cline{2-5} 
                            & \multicolumn{1}{c|}{Assumption}                                  & \makecell{Additive\\ sub-optimality}        & \multicolumn{1}{c|}{Requirement}                     & \makecell{Competitive \\ratio}          \\ \Xhline{1.2pt}
        \citet{gao2021online}                    & \multicolumn{1}{c|}{\makecell{Known \\bounds of $u_i^*$}} & $\widetilde{O}(\sqrt{t})$                  & \multicolumn{2}{c}{$\times$}                                      \\ \hline
                            \citet{azar2010allocate}& \multicolumn{2}{c!{\vrule width 1.2pt}}{$\times$}                                                                      & \multicolumn{1}{c|}{Non-extreme input}               & $O(\log t)$ \\ \hline
                            \citet{banerjee2022online}& \multicolumn{2}{c!{\vrule width 1.2pt}}{$\times$}                                                                      & \multicolumn{1}{c|}{Known monopolist utility}      & $O(1)$      \\ \hline
                            \citet{huang2022online}& \multicolumn{2}{c!{\vrule width 1.2pt}}{$\times$}                                                                      & \multicolumn{1}{c|}{\makecell{Known bounds of\\ impartial/balanced parameters }} & $O(1)$      \\ \hline
\multirow{2}{*}{This paper} & \multicolumn{1}{c|}{\multirow{2}{*}{None}}                         & \multirow{2}{*}{$\widetilde{O}(\sqrt{t})$\tablefootnote{Also extends to non-stationary-but-stochastic inputs; depends on nonstationary modulus.}} & \multicolumn{1}{c|}{Non-extreme input}               & $O(1)$      \\ \cline{4-5} 
                            & \multicolumn{1}{c|}{}                                             &                                & \multicolumn{1}{c|}{Seed utility}                    & $O(\log t)$ \\ \hline
\end{tabular}
\end{table}

\paragraph{Online fair allocation vs. online resource allocation.}
We remark that there are fundamental differences between our online fair allocation framework and the well-studied \textit{online resource allocation} problem, where best-of-many-worlds guarantees are also known~\citep{balseiro2023best,celli2022best,castiglioni2023online}. In online resource allocation, the resources are fixed and subject to known global supply constraints. The online arrivals are \textit{requests} to consume these resources and generate rewards. Results in this literature rely critically on the primal objective being time-separable, i.e., of the form $\sum_{t=1}^T f_t(x)$.
\footnote{\citet{balseiro2021regularized} allow an additional regularizer that is not time-separable, but this regularizer depends only on the resource consumption vector, rather than on correlated rewards. Moreover, the notion of fairness in that setting is different from ours: fairness in \citet{balseiro2021regularized} refers to consuming different types of resources at balanced rates, whereas our focus is on fairly distributing items among participating agents.}
In contrast, our online fair allocation problem features arriving items that are allocated upon arrival, together with a strongly time-correlated global objective that depends on the long-term utilities of the participating agents. Although both problems involve dual optimization, the algorithms and analyses do not directly generalize from one setting to the other. In online resource allocation, dual variables are naturally associated with the global resource constraints and are used to control depletion rates. In our fair allocation setting, there are no analogous resource constraints to dualize; instead, the dual multipliers are associated with the redundant constraints that track the utility level of each agent.

\paragraph{Pacing equilibria in first-price auctions.}
The name PACE is motivated by the fact that the dual interpretation of PACE closely resembles how online advertising platforms manage advertiser budgets through \emph{pacing algorithms}~\citep{balseiro2021budget,conitzer2022multiplicative,conitzer2022pacing,wang2023learning}. In particular, PACE can be interpreted as a form of pacing in a first-price auction setting. \citet{conitzer2022pacing} introduce an equilibrium notion for pacing in first-price auctions, known as \emph{first-price pacing equilibrium} (FPPE), and extensive follow-up work on the FPPE model has been conducted by researchers from several major industry labs~\citep{feng2024strategic,deng2024efficiency,gaitonde2023budget,lucier2024autobidders}. In \Cref{section: app}, we show that PACE naturally extends to the FPPE setting.

\paragraph{Online convex optimization.}
For stochastic input, PACE can be interpreted as a composite, unbounded, and non-strongly-convex version of dual averaging, with no external regularizer because the composite term itself plays the role of a regularizer. Dual averaging for online optimization was studied by~\citet{xiao2009dual}. With strongly convex regularizers, \citet{xiao2009dual} shows $O(\log t/t)$ mean-square convergence, whereas with non-strongly-convex regularizers, the rate degrades to $O(\log t/\sqrt{t})$. Our implicit bounding analysis shows that, with high probability, the strictly convex regularizer in our problem behaves as a strongly convex regularizer on the relevant region, which allows us to recover the $O(\log t/t)$ bound. Such a bound does not hold for general online optimization problems with non-strongly-convex objectives.

Our stochastic analysis is also related to the online convex programming framework of~\citet{agrawal2014fast}. Applying their algorithm to the Eisenberg--Gale program yields a multiplier-based decision rule, which resembles the dual interpretation of PACE. However, implementing their framework with a general online learning update does not provide last-iterate convergence, and thus does not lead to fairness and efficiency guarantees. In contrast, the dual interpretation of PACE aligns with the underlying KKT conditions and admits a closed-form primal-dual connection, thus enabling last-iterate convergence. Moreover, the analysis of~\citet{agrawal2014fast} relies on primal Lipschitzness and dual boundedness. Neither of these is satisfied by the Eisenberg-Gale program, so their analysis is not directly applicable. Their framework also does not provide results for the nonstationary or adversarial cases.

\paragraph{\textbf{Prior conference papers on PACE.}}
The present journal version builds on two earlier conference papers on PACE while substantially changing the scope of the results. For nonstationary inputs, \citet{liao2022nonstationary} analyzed the constrained version of PACE from~\citet{gao2021online}; here we replace constrained PACE with unconstrained PACE in the fair allocation setting, and establish nonstationary guarantees. \citet{liao2022nonstationary} also give broader online convex optimization results that are not related to fair allocation; these are not subsumed by the present paper. 

For adversarial input, \citet{yang2024greedy} studied the same unconstrained PACE algorithm, but only analyzed envy and NW for non-extreme inputs, and NW with seed utility. The present paper completes the guarantees for all fairness and efficiency metrics, with sharper dependence on the monopolist utility.
Moreover, the present paper provides an explicit upper bound for the NW competitive ratio under non-extreme inputs, where \citet{yang2024greedy} only gives $C \cdot n$ for some unspecified constant $C$.
The stochastic analysis of unconstrained PACE, the blood donation platform application, the extension to first-price pacing equilibrium, and the numerical simulations on Instagram notification are all new contributions by the present paper.

%% file: best-of-many-worlds/preliminaries.tex
\section{Preliminaries}
\label{section: preliminaries}

\paragraph{{Notations.}} {We let $\mathbb{R}_+^n$ denote the space of non-negative $n$-dimensional real vectors, and let $\mathbb{R}_{++}^n$ denote the space of strictly positive $n$-dimensional real vectors. For any integer $k \in \mathbb{N}_+$, let $[k]$ denote the set $\{1, \cdots, k\}$. For any $p>0$ or $p = \infty$, let $\|\cdot \|_p$ denote the $\ell_p$-norm. When $p$ is not specified, $\|\cdot \|$ is the $\ell_2$-norm by default. Let $\langle x, y\rangle:= \sum_{i=1}^n x_i y_i$ denote the inner product of vectors. We use $(\cdot)^+$ to denote the non-negative restriction $\max\{\cdot, 0\}$.}

\subsection{{Online Fair Allocation Setup}}

An online fair allocation instance is characterized by a $4$-tuple $(t,n,B,\gamma)$, where $t$ is the length of the finite horizon, unknown to the decision maker; $n$ is the number of agents; $B=(B_1,\ldots,B_n)\in \mathbb{R}_{++}^n$ is the vector of weights of the agents, encoding their relative priority or entitlement in the fairness and efficiency criteria that we introduce below; and $\gamma=(v_i^\tau)_{\tau\in[t],\, i\in[n]}$ encodes the agents' valuations for the items. At each time step $\tau\in[t]$, an item arrives and its value vector $v^\tau=(v_1^\tau,\ldots,v_n^\tau)$ is revealed, where $v_i^\tau\geq 0$ is the utility that agent $i$ would derive from receiving one unit of that item. We let $V\subseteq \mathbb{R}_+^n$ be the space of possible value vectors. 
We use $v_i=(v_i^1,\ldots,v_i^t)\in \mathbb{R}_+^t$ to capture the valuations of a single agent $i$. We will refer to $\gamma=(v_i^\tau)_{n\times t}$ as the value matrix, or simply the online input instance of the problem. We assume throughout that there are no trivial items (those with value $0$ for all agents) and no trivial agents (those assigning value $0$ to all items).\footnote{Our PACE algorithm, as will be clear later, is robust to inputs with trivial agents by simply never allocating to them. We exclude them only to simplify the presentation and to avoid the Nash welfare benchmark becoming trivially zero.}

In the online setting, the decision maker allocates the stream of items one at a time in an irrevocable manner without seeing the future items. At step $\tau$, after observing the value vector $v^\tau$ of the current item, the decision maker chooses an allocation $x^\tau = (x_1^\tau, \cdots, x_n^\tau)$ subject to the unit supply constraint $\sum_{i=1}^n x_i^\tau \leq 1$, and distributes the item accordingly. Here, $x_{i}^\tau \ge 0$ is the fraction of item $\tau$ allocated to agent $i$. Let $\mathcal X := \{x \in \mathbb{R}_+^{n\times t} : \sum_{i=1}^n x_i^\tau \le 1, \ \forall \tau \in [t]\}$ be the set of feasible allocations.\footnote{Our eventual PACE algorithm will only use integral allocations. We still allow the competing hindsight benchmarks to use fractional allocations.} We let $x_i=(x_i^1,\ldots,x_i^t)$ denote the bundle received by agent $i$. Upon receiving her fraction at step $\tau$, agent $i$ obtains instantaneous utility $u_i^\tau:=v_i^\tau x_i^\tau$. Utilities are linear and additive over time. In our online analysis, we write
$
U_i^\tau := \sum_{s=1}^{\tau} u_i^{s}
$ and $
\bar u_i^\tau := U_i^\tau/\tau
$
for the cumulative utility and time-averaged utility of agent $i$ up to time step $\tau$, respectively. We will use both cumulative and time-averaged utilities in the sequel, depending on the input model and the guarantee under study. A summary of the utility notations appears in \Cref{table: notation-1}. We also let $W_i = \sum_{\tau=1}^t v_i^\tau$ be the utility of agent $i$ for the grand bundle, \textit{a.k.a.} monopolist utility in the fair division literature.

\begin{table}[!htbp]
\centering
\renewcommand{\arraystretch}{1.5}
\begin{minipage}[t]{0.48\textwidth}
\caption{\small Different superscripts used in analysis.}
\label{table: notation-2}
\small
\centering
\begin{tabular}{c|c|c}
\Xhline{1.2pt}
superscript                  & Meaning & Examples \\ \Xhline{1.2pt}
$\gamma$ &  hindsight sol. \textit{w.r.t.} $\gamma$      &     $x_i^\gamma, u_i^\gamma, U_i^\gamma$       \\ \hline
$*$&  underlying sol. \textit{w.r.t.} $\bar{Q}$& $x_i^*, u_i^*, \beta_i^*$\\ \hline
$\tau$ or $t$& algorithmic result& $\Bar{u}_i^t, U_i^t$\\ \hline
\end{tabular}
\end{minipage}%
\hfill
\begin{minipage}[t]{0.48\textwidth}
\caption{\small Notations for algorithmic utilities.} 
\label{table: notation-1}
\small
\centering
\begin{tabular}{c|c|c}
\Xhline{1.2pt}
             & Meaning & Definition \\ \Xhline{1.2pt}
$(u_i^\tau)$ &  instantaneous        &     $u_i^\tau = x_i^\tau \cdot v_i^\tau$       \\ \hline
$ (U_i^\tau)$&  current cumulative & $U_i^\tau = \sum_{\tau^\prime \leq \tau} u_i^{\tau^\prime}$\\ \hline
$(\bar{u}_i^\tau)$&time-averaged & $\bar{u}_i^\tau = U_i^\tau / \tau$\\ \hline
\end{tabular}
\end{minipage}
\end{table}

In the subsequent analysis, we will consider three classes of online input models: (i) stationary, or\textit{ i.i.d.} input, (ii) nonstationary-yet-stochastic input, and (iii) adversarial input. For the stochastic input models, let $\Omega$ denote the sample space of $\gamma$, let $Q_\tau\in \Delta({V})$ be the marginal distribution of the value vector $v^\tau$, and let $\bar Q := \frac{1}{t}\sum_{\tau=1}^t Q_\tau$ denote the average marginal distribution; the assumption that there are no trivial agents implies that $\min_{i}\int_V v_i\,d\bar Q>0$, which we treat as a constant. We assume that values are finite, and, with a slight abuse of notation, write $\|v\|_\infty := \max_i \|v_i\|_\infty < \infty$ for a uniform upper bound on all values. We remark that these parameters only enter the theoretical guarantees, and are not required by our algorithm, which is prior-free. 

\subsection{Fairness and Efficiency Metrics}

Now we introduce the fairness and efficiency notions that we use to evaluate allocations. For any allocation $x \in \mathcal X$, let $U_i(x_j) := \sum_{\tau=1}^t v_i^\tau x_j^\tau$ denote the utility that agent $i$ would derive from agent $j$'s bundle, and in particular $U_i(x_i)$ is agent $i$'s actual utility under $x$. We state these notions in both their additive and multiplicative approximate forms. Throughout, we use $\eta \ge 0$ for additive approximation and $\rho \in [1, +\infty)$ for multiplicative competitiveness; the exact notions are recovered by setting $\eta=0$ or $\rho = 1$.

\paragraph{Fairness metrics.}
Our fairness discussion centers on two well-known notions: envy-freeness and proportionality. Envy-freeness is a local comparison notion which requires that no agent should prefer another agent's allocation to her own, after normalizing by priority weights. Proportionality is a global share notion which requires that each agent should receive at least as much utility as if they received a fraction of every good proportional to their priority weight. These two notions capture complementary aspects of fairness: envy-freeness rules out justified pairwise complaints, while proportionality ensures that no agent is left far below a natural baseline. 

\begin{definition}[Additive and multiplicative envy-freeness]
A feasible allocation $x \in \mathcal X$ is \emph{additive $\eta$-envy-free} (additive $\eta$-EF) if, for every pair of agents $i,j \in [n]$, we have $U_i(x_i)/B_i + \eta \ge U_i(x_j)/B_j$. It is \emph{multiplicative $\rho$-envy-free} (multiplicative $\rho$-EF) if, for every pair of agents $i,j \in [n]$, we have $U_i(x_i)/B_i \ge \rho^{-1} \, U_i(x_j)/B_j$.
\end{definition}
\begin{definition}[Additive and multiplicative proportionality]
A feasible allocation $x \in \mathcal X$ is \emph{additive $\eta$-proportional} if, for every agent $i \in [n]$, we have $U_i(x_i) + \eta \ge \frac{B_i}{\|B\|_1}\sum_{\tau=1}^t v_i^\tau$. It is \emph{multiplicative $\rho$-proportional} if, for every agent $i \in [n]$, we have $U_i(x_i) \ge \rho^{-1} \, \frac{B_i}{\|B\|_1}\sum_{\tau=1}^t v_i^\tau$.
\end{definition}

\paragraph{{Efficiency metrics.}}
We next turn to efficiency. Our primary efficiency benchmark is weighted Nash welfare, a well-studied social welfare function defined as the weighted geometric mean of agent utilities~\citep{nash1950bargaining}. 
It can be maximized by maximizing the sum of log-utilities.
Nash welfare is a more balanced social welfare function than the standard utilitarian welfare function, since it requires every agent to achieve a reasonable amount of utility.
\begin{definition}[Nash welfare suboptimality and competitiveness]
For a feasible allocation $x \in \mathcal X$, define its (weighted) Nash welfare by $\operatorname{NW}(x) := \prod_{i=1}^n U_i(x_i)^{B_i / \|B\|_1}$. It is \emph{$\eta$-suboptimal for weighted Nash welfare} if $\sup_{y \in \mathcal X} \operatorname{NW}(y) - \operatorname{NW}(x) \le \eta$. It is \emph{$\rho$-competitive for weighted Nash welfare} if $\operatorname{NW}(x) \ge \rho^{-1} \sup_{y \in \mathcal X} \operatorname{NW}(y)$.
\end{definition}

Another natural and well-studied efficiency notion is (strong) Pareto optimality, which asks whether the planner could reallocate the entire set of items and make all agents weakly better off, with at least one strict improvement. Here we consider a strictly stronger concept,\textit{core stability}, which further asks whether some coalition of agents, using only the endowment of resources proportional to their group weight, could reassign those resources among its members so that all become better off, with at least one strict improvement. It rules out not only global inefficiency but also the possibility that a subgroup is collectively underserved and prefers more efficient improvements.
\begin{definition}[Additive and multiplicative core stability]
A feasible allocation $x \in \mathcal X$ is \emph{additive $\eta$-core stable}, if there does not exist a nonempty coalition $S \subseteq [n]$ and allocation bundles $(y_i)_{i\in S}$ such that $\sum_{i\in S} y_i^\tau \le \frac{B(S)}{\|B\|_1}, \forall \tau \in [t]$, where $B(S):=\sum_{i\in S} B_i$, and $U_i(y_i) \geq U_i(x_i) + \eta , \forall i \in S$, with strict inequality for at least one $i\in S$. It is \emph{multiplicative $\rho$-core stable} if there do not exist a nonempty coalition $S \subseteq [n]$ and bundles $(y_i)_{i\in S}$ such that $\sum_{i\in S} y_i^\tau \le \frac{B(S)}{\|B\|_1}, \forall \tau \in [t]$, and $U_i(y_i) \geq \rho \  U_i(x_i), \forall i\in S$, with strict inequality for at least one $i\in S$.
\end{definition}

Let us briefly compare our choice of fairness and efficiency notions with existing works in the online fair allocation literature. One active line of work~\citep{benade2018make,zeng2020fairness,hosseini2024class,elkind2024temporal,choo2025approximate,neoh2025online} focuses primarily on achieving approximate envy-freeness and proportionality guarantees, and handles the tradeoff with Pareto optimality separately---whereas we tie envy-freeness, proportionality, Nash welfare, and core stability to the same offline target. A second line of literature evaluates allocations primarily through cardinal welfare objectives, rather than fairness criteria. Such welfare functions include Nash~\citep{banerjee2022online,huang2022online}, generalized-mean~\citep{balseiro2022uniformly,banerjee2022online,huang2025long}, maximin~\citep{kawase2022online}, and Gini-based ones~\citep{lejeune2019planning}. 

\subsection{Offline Benchmark Programs}

We now specify the hindsight offline benchmark that satisfies all the fairness and efficiency notions defined above. This is given by the optimal solution to the Eisenberg--Gale (EG) convex program.

\paragraph{Hindsight benchmark.}

For an input instance $\gamma=(v_i^\tau)_{\tau\in[t],\,i\in[n]}$, EG gives the offline allocation that maximizes the sum of weighted logarithmic utilities, which is equivalent to maximizing the weighted Nash welfare:
\begin{equation}
    \label{eq: EG-primal}
\max_{x\geq 0, U\geq 0} 
\left\{
    \left. \sum_{i=1}^n B_i \log U_i \ \right| 
    U_i\leq \langle v_i, x_i\rangle \ \forall i \in [n],  \ 
    \sum_{i=1}^n x_i^\tau \leq 1 \ \  \forall \tau \in [t]
\right\}.
\end{equation}

We let $x^\gamma$ be the optimal solution to \eqref{eq: EG-primal}, and denote the resulting utility vectors as $U^\gamma$ (total) and $u^\gamma$ (time-averaged). By associating a Lagrange multiplier $\beta_i\geq 0$ with each constraint $U_i \leq \langle v_i, x_i\rangle$ and dropping constants in the derivation, one can get the following dual formulation (normalized by time), which will be useful in the following analysis; see \Cref{section: review} for detailed steps.
\begin{equation}
    \label{eq: EG-dual}
\min_{\beta\geq 0} \left\{ \frac{1}{t}
    \sum_{\tau=1}^t \max_{i \in [n]}\beta_i v_i^\tau - \sum_{i=1}^n B_i \log \beta_i
\right\}
.
\end{equation}

The benchmark program \eqref{eq: EG-primal} is directly aligned with the fairness and efficiency metrics that we defined previously. By construction, it optimizes weighted Nash welfare. Moreover, the resulting solution is known to satisfy exact envy-freeness, proportionality, and core stability. 
\begin{proposition}
\label{prop:hindsight-benchmark-properties}
For any problem instance $(t,n,B,\gamma)$, the optimal solution $x^\gamma$ to \eqref{eq: EG-primal} exactly satisfies the following: 1) weighted Nash welfare maximization, 2) envy-freeness, 3) proportionality, and 4) core stability (and thus strong Pareto optimality).  
\label{proposition: EG-fairness}
\end{proposition}

\paragraph{Underlying benchmark for stochastic inputs.}
When the input is stochastic, the hindsight benchmark $x^\gamma$ fluctuates with the realization $\gamma$, and it is technically convenient to also consider a deterministic benchmark induced by the average input distribution. Let $\nu$ be the reference measure for the average item value distribution $\bar Q=\frac{1}{t}\sum_{\tau=1}^t Q_\tau$, and let $\langle v_i, x_i\rangle:= \int_{V} v_i x_i \mathrm{d}\nu$. The underlying benchmark is then the infinite-dimensional analogue of \eqref{eq: EG-primal} and \eqref{eq: EG-dual}.
\begin{align}
    \label{eq: inf-EG-primal}
    & \max_{x\in L^{\infty}_+(V), u\geq 0}
    \left\{
    \left.
     \sum_{i=1}^n B_i \log u_i \ \right| \ 
     u_i \leq \langle v_i, x_i\rangle \ \forall i\in [n], \
     \sum_{i=1}^n x_i\leq s
        \right\}
        ,
    \\
    & \quad \quad \quad \quad \quad  \min_{\beta \geq 0}
    \left\{
    \int_{V} \left(\max_{i\in [n]}\beta_i v_i\right) \mathrm{d}\Bar{Q} - \sum_{i=1}^n B_i \log \beta_i
        \right\}.
    \label{eq: inf-EG-dual}
\end{align}
We let $x^*, u^*, \beta^*$ denote the optimal primal-dual solutions to the underlying problems. Notice that $u^*$ is intrinsically a time-averaged benchmark. The underlying solutions also satisfy fairness and efficiency properties which are analogous to those in \Cref{proposition: EG-fairness}. A rigorous mathematical treatment of the infinite-dimensional programs can be found in \citet{gao2023infinite}. To avoid confusion, \Cref{table: notation-2} recaps the notations of algorithmic, hindsight, and underlying solutions for readers. 

In the stochastic settings, we will show that the average utility vector produced by PACE converges to the underlying benchmark $u^*$. Concretely, we will investigate the term 
$\mathbb E_\gamma \bigl[\|\bar u^t-u^*\|_2^2\bigr]$. Utility convergence suggests that each agent's attained level of benefit is stable, predictable, and well-calibrated in the long run. It is also technically useful in proving several fairness and efficiency guarantees.

\paragraph{Connection to competitive equilibria.}
The solution to the Eisenberg--Gale convex program \eqref{eq: inf-EG-dual} also coincides with the \textit{competitive equilibrium} of the underlying Fisher market when we interpret the agents as buyers and $B_i$ as their budgets. 
This connection is the foundation for the \emph{competitive equilibrium from equal incomes} solution concept for fair allocation~\citep{varian1974equity}.
Under this interpretation, by correctly defining the prices induced by the PACE algorithm, we show that PACE attains the competitive equilibrium under our stochastic input models, in the sense of clearing all agents' budgets and meeting their demands asymptotically. All setups and results for online market equilibria can be found in \Cref{section: fisher}.

%% file: best-of-many-worlds/algorithm.tex
\section{The PACE Algorithm}
\label{section: algorithm}
We next present the unconstrained PACE (Pace According to Current Estimated utility) algorithm; see \Cref{alg: PACE} for details. We then derive two complementary interpretations of the algorithm. 

PACE uses a simple allocation principle at each step $\tau\in [t]$: it allocates the item to the agent for whom the item creates the \textit{largest proportional utility increase} $v_{i}^\tau / U_{i}^{\tau-1}$, adjusted by the agents' priority weights $B$; see step \eqref{eq: pace-allocation-principle}.\footnote{For initial rounds where cumulative utilities could be zero, we use the convention $v_i^\tau / U_i^{\tau-1} = +\infty$ when $v_i^\tau >0$ and $U_i^{\tau-1} = 0$, and $v_i^\tau / U_i^{\tau-1} = 0$ when both are zero.} This encodes efficiency and fairness in an intuitive way: the numerator \(v_i^\tau\) captures the immediate value of assigning the current item to agent \(i\), while the denominator \(U_i^{\tau-1}\) captures how well that agent has already been served. This provides a balance between the efficiency-only extreme of solely maximizing the value increment $v_i^\tau x_i^\tau$ (as one would for utilitarian welfare maximization) and the conservative fairness extreme of maximizing the minimum utility by allocating to the currently-worst-off agent, who has the smallest $U_i^{\tau-1}$ value.

\begin{algorithm}
\caption{PACE (unconstrained)}
\label{alg: PACE}

    \SetKwInput{KwInit}{Initialization}
    \KwIn{number of agents $n$, agent weights $B$.}
    \KwInit{$U_i^0 \gets 0$.}
    \For{$\tau = 1, \cdots, t$}{
        Observe agent valuation $v^\tau$. 
        
        Find the agent who would get the largest proportional increment from the current item (adapted by weights; lexicographical tiebreaking):
        \begin{equation}
        \label{eq: pace-allocation-principle}
            i^\tau \gets \arg \max_{i \in [n]} B_i\cdot \frac{v_{i}^\tau}{U_{i}^{\tau-1}}.
        \end{equation}

        Allocate the item to $i^\tau$ and update the cumulative utilities,
        $$x_i^\tau \gets \bm{1}(i =i^\tau),  \ \ \ U_i^\tau \gets U_i^{\tau-1} + x_i^\tau\cdot v_i^\tau.$$
    }
\end{algorithm}

\Cref{alg: PACE} has a series of properties that are desirable for real-world applications: 1) \textit{{Transparency and} decentralization}. The platform only needs to compare agent-level ratios $v_i^\tau/U_i^{\tau-1}$ each round, and these ratios can be maintained and reported by the agents in a decentralized manner.
2) \textit{Integrality}. PACE allows each item to be fully allocated to a single agent, while being competitive against the hindsight metric which allows fractional allocations. While fractional allocations can be interpreted as randomized allocations in many settings, this may not always be desirable. 3) \textit{Free of prior knowledge and tuning}. It does not rely on any form of prior knowledge, requiring no information on the value space, the distributions, the horizon length $t$, or the type of the arrival model. Each agent has no tuning parameters whatsoever.

\subsection{{Dual Interpretation of PACE}}
\label{section: PACE-as-DA}

For subsequent analysis, it is convenient to associate an induced dual sequence $(\beta^\tau)_{\tau \in [t]}$ with \Cref{alg: PACE}, even though the algorithm itself does not explicitly maintain one. We let $\beta_{i}^{\tau}$ be inversely proportional to the time-averaged utility $\bar u_i^{\tau-1}$, so the decision rule in \Cref{alg: PACE} is exactly equivalent to finding the agent with the highest $\beta^\tau$-paced value. Concretely, 
$$
\beta_{i}^{\tau}:= B_i / \bar{u}_i^{\tau-1}, \ \  \quad i^\tau  \in  \arg\max_{i \in [n]} \beta_i^\tau v_i^\tau.
$$

We next interpret $(\beta^\tau)_{\tau \in [t]}$ as a decision sequence for a dual online learning problem. Given an instance $\gamma$, we define the dual online learning problem with loss function $F(\cdot;v^\tau)$ per step; $F(\beta;v) = f(\beta;v)+\Psi(\beta)$, where $f(\beta;v) =\max_i \beta_i v_i$ and $\Psi(\beta) = - \sum_{i=1}^n B_i\log \beta_i$. Denote $f^\tau(\cdot) = f(\cdot; v^\tau)$ for simplicity. 
For this dual loss, a key observation is that the subgradient $g^\tau \in \partial f^\tau(\beta^\tau)$ coincides with the instantaneous utility $u^\tau$, so $\bar{u}^\tau$ will be the time-average of dual subgradients, $\bar{g}^\tau$. Meanwhile, $\Psi(\beta)$ is a convex function that can be minimized as a regularizer-type composite term. A natural approach is then to apply \textit{dual averaging}, which recovers PACE's implicit update on $\beta^\tau$: 
\begin{equation*}
    \beta^{\tau+1} = \arg \min _{\beta \in \mathbb{R}_+^n} \langle \bar{g}^\tau, \beta\rangle + \Psi(\beta)  = \arg \min_{\beta\in \mathbb{R}_+^n}\langle \bar{u}^\tau, \beta\rangle -\sum_{i}B_i\log \beta_i = B/\bar{u}^\tau. 
\end{equation*}

A more detailed setup of dual averaging can be found in \Cref{appendix: PACE-as-DA}. Notice that regret bounds for the dual online learning problem alone cannot imply any of our desired fairness and efficiency results, which are properties tied to the primal space. To enable dual-to-primal translation, a key observation is that $\beta_i^{\tau+1} = B_i/\bar{u}_i^\tau$ is aligned with the KKT condition of the hindsight primal-dual programs, $\beta_i^* = B_i/u_i^*$~\citep{cole2017convex}. This alignment property is unique to our specific update rule, and does not hold for general online learning updates such as mirror descent.  

The PACE algorithm is substantially different from the standard regularized dual averaging algorithm~\citep{xiao2009dual} because here the regularizer $\Psi(\beta)$ is inherent to the loss function. Moreover, our problem satisfies neither boundedness nor strong convexity: we have an open domain $\beta\in \mathbb{R}_+^n$, and $\Psi(\beta)$ is unbounded and not strongly convex. This causes failure in directly applying the analysis of \citet{xiao2009dual}. 

\paragraph{Price and expenditure.} From the above dual interpretation we can define the \textit{price} of each item and agents' \textit{expenditures}, which are quantities that are independent of the fair allocation setup, but essential to PACE's Fisher market interpretations and extensions to multi-agent bidding. Notice that the decision rule can be read as allocating the item to the highest bidder when each agent bids $\beta_i^\tau v_i^\tau$. Under the first-price rule, the \textit{price} of the $\tau$th item is $p^\tau = \max_{i\in [n]} \beta_i^\tau v_i^\tau$. Only the winner is charged $p^\tau$ for the item, i.e., the expenditure of agent $i$ on item $\tau$ is $b_i^\tau=p^\tau x_i^\tau$. 

\paragraph{Constrained PACE.} Unlike our simple update rule, the previous constrained version of PACE by \citet{gao2021online} requires explicitly computing the dual and projecting to $[B_i / (1+\delta_0), B_i\cdot (1+ \delta_0)]$. This breaks the clean primal interpretations, and, more importantly, requires knowing the range of $u_i^*$ in advance to set $\delta_0$ correctly. This causes the constrained PACE algorithm to be prior-dependent and easily corrupted by simple adversarial instances; see constructions in \Cref{section: failure}.

\subsection{{Greedy Interpretation of PACE}}
\label{section: PACE-as-greedy}
\Cref{alg: PACE} can also be interpreted as a primal first-order greedy algorithm for online Nash welfare maximization. To show this connection, we first write down the pure greedy algorithm that finds the largest increment in Nash welfare myopically at each step:
\begin{equation}
\label{eq: pure-greedy}
    \max_{U_i^\tau \geq 0, x_i^\tau \in \{0,1\}}  \ \left\{ \sum_{i=1}^n B_i \log U_{i}^\tau  \ \right|\left. 
    \  U_{i}^\tau = x_{i}^\tau v_{i}^\tau + U_{i}^{\tau-1}  ,\ \forall i\in [n] ,
                  \ \ \sum_{i=1}^n x_{i}^\tau = 1 \right\}    
\end{equation}
The solution of \eqref{eq: pure-greedy} is to allocate the whole item to $i^\tau \in \arg \max_{i\in [n]} B_i \log\left( 1 + {v_i^\tau}/{U_i^{\tau-1}}\right)$. When $B_i$ are equal, this coincides with PACE's decision. For unequal weights, taking a first-order approximation of the logarithmic term gives the decision rule $i^\tau \in \arg\max_{i\in [n]} B_i{v_i^\tau}/{U_i^{\tau-1}}$, which recovers PACE's rule \eqref{eq: pace-allocation-principle}. The first-order approximation here can be justified by the fact that the ratio $v_i^\tau/U_i^\tau$ is small when the total utility becomes large. \Cref{alg: PACE} can thereby be viewed as a \textit{first-order, integral} greedy algorithm in the primal space.

%% file: best-of-many-worlds/stationary.tex
\section{Stationary Input}
\label{section: stationary}
In this section, we show performance guarantees for PACE under stationary (\textit{i.i.d.}) inputs. 
Our analysis begins with convergence results on the dual iterate $\beta^t$ and time-averaged utility $\bar u^t$. Using these convergence results as building blocks, we then derive fairness and efficiency guarantees. Missing proofs in this section and additional results under the market equilibrium interpretation can be found in \Cref{proof: iid}.

\subsection{Dual and Primal Convergence}
As discussed in \Cref{section: PACE-as-DA}, existing convergence results for dual averaging cannot be directly applied to our problem because $\Psi(\beta)$ is neither bounded nor strongly convex. Indeed, in some cases $\beta^t$ might not converge at all since the dual can be infinite with nonzero probability; see \Cref{example: unconditional}.

\begin{example}
    \label{example: unconditional}
    Consider the stationary input with $\Pr[v_1^\tau = 0] = 1/2$. Then $
        \Pr\left[\beta_1^{t+1} = \infty\right] = \Pr\left[\Bar{u}_1^t = 0\right] \geq \frac{1}{2^t}>0$. 
    Since $\beta_1^t$ is infinite with positive probability, the expectation of $\|\beta^t - \beta^*\|^2$ does not even exist.  
\end{example}
Despite the impossibility of unconditional convergence, we show that $\beta^t$ converges \textit{with high probability} under stationary input to both the hindsight dual $\beta^\gamma$ (optimal solution to \eqref{eq: EG-dual}) and the underlying dual $\beta^*$ (optimal solution to \eqref{eq: inf-EG-dual}).

\begin{restatable}[Convergence of pacing multipliers under stationary input]{theorem}{iidthmZ}
    \label{thm: convergence-multiplier-iid}
    Under stationary input, there exists a high-probability event $\Hat{A} \subseteq \Omega$ such that $\beta^t$ converges conditional on $\Hat{A}$. Specifically, 
    \begin{equation}
        \label{eq: conditional-multiplier-convergence}
    \mathbb{E} \left[ \|\beta^{\tau+1} - \beta^*\|_2^2 \mid \Hat{A} \right], \mathbb{E} \left[ \|\beta^{\tau+1} - \beta^\gamma\|_2^2 \mid \Hat{A} \right] = O\left( \frac{ n^4 \log n \log t}{t}\right), \Pr\left[(\Hat{A})^c\right]  =  O\left(\frac{1}{t}\right).
    \end{equation}
\end{restatable}

The $\widetilde{O}(1/t)$ mean-square convergence rate in \Cref{thm: convergence-multiplier-iid} is as strong as one could obtain for bounded and strongly convex objectives. For dual-averaging under general unbounded and non-strongly-convex functions like ours, previous analysis only achieves a substantially slower rate of $\widetilde{O}(1/\sqrt{t})$~\citep{xiao2009dual}. The proof of \Cref{thm: convergence-multiplier-iid} requires careful design of the high-probability event and decomposition of the dual online learning loss conditional on the event. We provide a sketch of the main proof components in \Cref{subsection: proof-sketch-iid}. 

We next show the convergence of average utilities. Since PACE's update rule $\beta^{t+1} = B/\bar u^t$ aligns with the KKT condition $\beta^* = B/u^*$, utility convergence can be obtained from \Cref{thm: convergence-multiplier-iid} using a conditional probability argument. Notice that the average utilities are bounded by $\|v\|_\infty$ even when $\Hat{A}$ does not hold. The utility convergence result is thus unconditional.

\begin{restatable}[Convergence of utilities under stationary input]{theorem}{iidthmY}
\label{thm: convergence-utility-iid}
    Under stationary input, it holds for \Cref{alg: PACE} that
    \begin{equation*}
        \mathbb{E} \left[\|\Bar{u}^t - u^*\|^2\right], \mathbb{E} \left[\|\Bar{u}^t - u^\gamma\|^2\right] = O\left(\frac{n^4\log n\log t}{t}\right). 
    \end{equation*}
\end{restatable}
\Cref{thm: convergence-utility-iid} shows that PACE attains the offline benchmarks asymptotically in agent-wise utility levels. The $\widetilde{O}(1/t)$ rate is as strong as that of constrained PACE~\citep{gao2021online}. By Cauchy--Schwarz, this further implies $\mathbb{E}[|\bar u_i^t-u_i^*|]$ and $\mathbb{E}[|\bar u_i^t-u_i^\gamma|]$ vanish at rate $\widetilde{O}(1/\sqrt{t})$ agent-wise. This will be useful for deriving the fairness and efficiency guarantees below.
 For the dependence on the number of agents $n$, the order is $\Omega(n^4)$ in \citet{gao2021online} even under prior knowledge and weight normalization; they did not show an upper bound. Here we provide an $\widetilde{O}(n^4)$ upper bound free of normalization.

\subsection{Fairness and Efficiency Metrics}

We now convert the utility and dual multiplier convergence results into guarantees for our fairness and efficiency notions. Intuitively, \Cref{thm: convergence-utility-iid} connects the online outcome to the hindsight benchmark; the latter is exactly envy-free, proportional, core stable, and maximizes Nash welfare.  We first present the results on the fairness metrics. 

\begin{restatable}[Fairness metrics under stationary input]{theorem}{iidFair}
\label{thm:stationary-fairness-metrics}
Under stationary inputs, the allocation of PACE is additive $\eta^t_{\mathrm{EF}}$-envy-free and additive $\eta^t_{\mathrm{Prop}}$-proportional, with
\[
\mathbb E\!\left[\eta^t_{\mathrm{EF}}\right]
=
O\!\left({n^2  \sqrt{\log n} \cdot \log t \,  \sqrt{t} }\right), \   \mathbb E\!\left[\eta^t_{\mathrm{Prop}}\right] = O\left (n^2 \sqrt{(\log n) \, (\log t)  t} \right).
\]
\end{restatable}

In Theorem~\ref{thm:stationary-fairness-metrics}, both additive fairness gaps are sublinear in $t$, and vanish in the time-average sense. This suggests that the online allocation is fair both locally and globally: no agent substantially prefers another agent's bundle or the weighted proportional share to what they actually receive, up to $\widetilde{O}(\sqrt{t})$ error. The proof of the envy part is not a straightforward transformation of \Cref{thm: convergence-utility-iid} and requires more refined treatment via Fisher market interpretations, since it is unclear if an agent's valuation on someone else's bundle also converges. 

We next show guarantees on the efficiency notions: Nash welfare maximization and core stability.

\begin{restatable}[Efficiency metrics under stationary input]{theorem}{iidEff}
\label{thm:stationary-efficiency-metrics}
Under stationary inputs, the allocation of PACE is $\eta^t_{\mathrm{NW}}$-suboptimal for weighted Nash welfare, and additive $\eta^t_{\mathrm{Core}}$-core stable, with
\[
\mathbb E\!\left[\eta^t_{\mathrm{NW}}\right], \ \mathbb E\!\left[\eta^t_{\mathrm{Core}}\right] = O\left (n^2 \sqrt{(\log n) \, (\log t)  t} \right).
\]
\end{restatable}

Theorem~\ref{thm:stationary-efficiency-metrics} captures both aggregate efficiency and efficiency in terms of robustness to coordinated deviations, up to vanishing per step errors. First, PACE computes a globally near-optimal solution to a group welfare objective--Nash welfare. To the best of our knowledge, this is the first no-regret type of guarantee for Nash welfare in the stationary setting. Second, approximate core stability rules out the possibility that a more efficient allocation for a group can be found by re-allocating their entitled resources. This provides a stronger efficiency certificate than approximate Pareto optimality.

\subsection{Proof of \Cref{thm: convergence-multiplier-iid}, Major Lemmas}
\label{subsection: proof-sketch-iid}
We consider the following class of \textit{implicit bound} events. For a given upper bound $h = (h_1, \cdots, h_n)\in \mathbb{R}_{++}^n$ and time step $s \in \mathbb{N}$, define
\begin{equation}
    \label{eq: event-definition}
    A_{h, s}:= \left\{\gamma:\ \beta_i^\tau \in [B_i/\|v_i\|_{\infty}, h_i], \forall i\in[n], \forall \tau \geq s \right\}.
\end{equation}

The event $A_{h, s}$ requires that the dual multipliers lie within bounded intervals after $s$ rounds: the lower bound is trivially $B_i/\|v_i\|_{\infty}$ and the upper bound $h_i$ is a parameter to be fixed later. We denote
$\mathcal{F}_h = [B_1/\|v_1\|_{\infty}, h_1] \times \cdots \times [B_n/\|v_n\|_{\infty}, h_n]$, and define $\Bar{F}(\beta) := \sup_v F(\beta, v)$, $G(h) = \max_{\beta \in \mathcal{F}_h} \Bar{F}(\beta).$

The proof of \Cref{thm: convergence-multiplier-iid} has three major components: 1) \emph{A deterministic bound implied by $A_{h,s}$} (\Cref{lem: deterministic-mean-square}), which relates the squared distance $\|\beta^{t+1}-\beta\|^2$ to the event $A_{h,s}$ and the \textit{dual online learning regret} term, 2) a dual online learning regret bound in conditional expectation (\Cref{lem: dual-regret-iid}), and 3) a high-probability proof for a selected $\hat{A} = A_{\hat h,\hat s}$, where $\hat h = O(n)$ and $\hat s = O(n\log t)$  (\Cref{lem: high-prob-bounds-iid}). 

\begin{restatable}[Deterministic convergence with implicit bounds]{lemma}{iidlemA}
    \label{lem: deterministic-mean-square}
    For any $\beta \in \mathbb{R}_+^n$, consider the regret of the dual online learning problem \textit{w.r.t.}\ $\beta$ starting from time step $s+1$,
    $
        \textnormal{Dual-Regret}(s,\beta):= \sum_{\tau = s+1}^t \left(F(\beta^\tau, v^\tau) - F(\beta, v^\tau)\right).
    $
    Then, for any $t>s$ and $h>1$, it holds deterministically that
    \begin{equation}
        \label{eq: deterministic-mean-square}
        \gamma \in A_{h, s} \implies \|\beta^{t+1}-\beta\|^2 \leq \frac{2}{\sigma t} \left[\frac{\max_i\|v_i\|_{\infty}^2}{2\sigma} \log t + \left(\Bar{F}(\beta) - \Psi(h)\right) \cdot s
    - \textnormal{Dual-Regret}(s,\beta)\right],
    \end{equation}
    where $\sigma = \min \frac{B_i}{h_i^2}$ is the strong convexity parameter of $\Psi$ on $\mathcal{F}_h$.
\end{restatable}
\Cref{lem: deterministic-mean-square} is a deterministic decomposition, and fits our analysis in both stationary and nonstationary settings. Unlike in \citet{xiao2009dual}, the gradients before round $s$ are treated separately, since they come from iterates that could be infinitely far away from the target $\beta$. 

\begin{restatable}[Dual regret bound, stationary]{lemma}{iidlemB}
\label{lem: dual-regret-iid}
    Under stationary input, for any $t>s$, 
    \begin{equation}
        \label{eq: dual-regret-iid}
        \begin{aligned}
          \mathbb{E}\left[\textnormal{Dual-Regret}(s,\beta^\gamma)\mid A_{h,s} \right]
         & \geq    \mathbb{E}\left[\textnormal{Dual-Regret}(s,\beta^*)\mid A_{h,s} \right] \\
         & \geq -4 G(h) \cdot \left(1-\Pr\left[A_{h,s}\right]\right)\cdot (t-s).
        \end{aligned}
    \end{equation}
\end{restatable}

\Cref{lem: dual-regret-iid} holds for any choice of parameters $s$ and $h$. We remark that, without conditioning on an implicit event $A_{h,s}$, it is straightforward to show that the expected dual online learning regret is non-negative. However, introducing the condition $A_{h,s}$ makes this challenging: as a posterior event, $A_{h,s}$ causes correlation across time steps and shifts the conditional distributions away from the underlying $\bar Q$. The key step in the proof of \Cref{lem: dual-regret-iid} is to relate the per step regret to the total variation (TV) distance between $Q_\tau$ and $Q_\tau|A_{h,s}$.

\begin{restatable}[High-probability implicit bounds]{lemma}{iidlemC}
    \label{lem: high-prob-bounds-iid}
    There exists $\hat h = \Hat{h}(n) = (\Hat{h}_i(n))_{i=1}^n$ and $\hat s = \Hat{s}(t,n)$ such that
    \begin{enumerate}
        \item For each $i\in [n]$, we have $\Hat{h}_i(n) >1 $, $\Hat{h}_i(n) = O(n)$. Also, $\Hat{h}_i(n)$ does not depend on $t$.
        \item $\Hat{s}(t,n) = O(n \log n+ n \log t)$. 
        \item Denote $\Hat{A}:=A_{\Hat{h}(n), \Hat{s}(t,n)}$. The failure probability satisfies $1-\Pr[\Hat{A}] = O(1/t)$.
    \end{enumerate}
\end{restatable}

\Cref{lem: high-prob-bounds-iid} is a nontrivial result: it is equivalent to requiring all time-averaged utilities to be lower bounded by a \textit{constant} after logarithmic iterations. It does not hold for general online learning updates, and cannot be guaranteed by using $O(\log t)$ naive allocations to ``warm up'' the algorithm. To prove \Cref{lem: high-prob-bounds-iid}, 
we use the fact that, with high probability, each agent sees sufficiently many items with non-trivial values. This is due to the assumption that $\mathbb{E}[v_i]>0$ for all $i$. We then leverage the greedy interpretation of PACE in the primal space (\Cref{section: PACE-as-greedy}) to argue that each agent receives sufficiently many allocations with high probability.

%% file: best-of-many-worlds/nonstationary.tex
\section{Nonstationary Input}
\label{section: nonstationary}

In this section, we generalize PACE's performance guarantees from \textit{i.i.d.} input to stochastic inputs that are not exactly stationary. In many operational settings, the arrival sequence is affected by time-varying supply, user activity patterns, or other external conditions. As a result, arrivals may be locally correlated, seasonal, or temporarily shifted away from their long-run average behavior. A robust online allocation algorithm should therefore not rely on the assumption that every item is independently drawn from exactly the same distribution. We study two input models that capture these nonstationary deviations: ergodic input, where the influence of past arrivals decays after a mixing window, and blockwise independent input, where arrivals may be arbitrarily correlated within each operational block but are independent across blocks.

\paragraph{Ergodic input.}
The \textit{ergodic} input model captures settings where arrivals may be strongly related locally, but the process gradually returns to its long-run average behavior. For example, Internet platform users arriving within a short time window may come from a similar population or be affected by the same external event, while users separated by a sufficiently long period become much less correlated. The parameter $\iota$ measures the operational mixing time: after $\iota$ steps, the conditional distribution of the next item is required to be close to the time-averaged distribution $\Bar Q$, up to a deviation $\delta$ in TV distance. Concretely, for any integer $\iota \in\{1, \cdots, t-1\}$, we measure the maximum $\iota$-step deviation from the time-averaged distribution $\Bar{Q}$ by the quantity
\begin{equation*}
\delta^{{\mathrm{E}}}(\iota) = \sup_{\tau = 1, \cdots, t-\iota} {\sup_{v^{1:\tau}}} \ \| Q^{\tau+\iota}(\cdot | v^{1:\tau})-\Bar{Q}\|_{\mathrm{TV}},
\end{equation*}
where $Q^{\tau+\iota}(\cdot | v^{1:\tau})$ is the conditional distribution of $v^{\tau+\iota}$ given $v^{1:\tau}:=(v^1, \cdots, v^\tau)$. The ergodic input distribution class is then defined as follows:
\begin{equation*}
    \mathsf{C}^{\mathrm{E}}(\delta, \iota):=
    \left\{
        (Q^1, \cdots, Q^t)\in \Delta(V)^t: {\delta^{\mathrm E}(\iota)}\leq \delta
    \right\}.
\end{equation*}

\paragraph{Blockwise independent input.}
\label{sec: blockwise}

Our second nonstationary model is blockwise independent input. In many scenarios, arrivals may be highly correlated by operational constraints within a block, but are independent across blocks. Formally, define a partition $\mathcal{P}$ of items by dividing the time horizon into $K$ blocks with $1= \tau_1\leq \cdots \leq \tau_{K+1} = t$. The $k$th block includes time steps $I_k = \{\tau_k, \cdots, \tau_{k+1}-1\}$, and has average value distribution ${Q}^{(k)}:= (\sum_{\tau\in I_k} Q^\tau )/|I_k|.$ Let $|\mathcal{P}|_{\infty} = \max_{k\in [K]} |I_k|$ be the maximum block length. The blockwise arrival distribution class is then defined as follows:
\begin{equation*}
    \mathsf{C}^{\mathrm{B}}(\mathcal{P},\delta):=
    \left\{
        (Q^1, \cdots, Q^t)\in \Delta(V)^t: \frac{1}{t}\sum_{k=1}^K |I_k|\cdot  \|{Q}^{(k)} - \Bar{Q}\|_{\mathrm{TV}}\leq \delta, \ \text{blocks are independent}
    \right\}.
\end{equation*}

This model also captures \textit{periodic} and \textit{mildly corrupted i.i.d.}\ inputs. For example, if arrivals follow a weekly periodic pattern, one can take each week as a block: arrivals within the week may follow different weekday and weekend patterns, while the block average remains close to the long-run average when the weekly cycles are similar. Also, mildly corrupted \textit{i.i.d.} input can be viewed as a blockwise input with block size $1$; the parameter $\delta$ captures the average degree of corruption. The periodic and mildly corrupted input models are formally defined in \Cref{appendix: corrupted-and-periodic}.

\subsection{Results}
We now state the convergence, fairness, and efficiency guarantees under the two nonstationary input models. The proof follows the same high-level structure as the stationary analysis: the main decomposition is still \Cref{lem: deterministic-mean-square}, but the distribution-dependent parts require nonstationary analogues---the dual regret bound and the high-probability implicit bounds now must account for nonstationarity. The full proofs are deferred to \Cref{proof: non}. 
\begin{restatable}[Convergence of utilities under nonstationary input]{theorem}{nonUtility}
\label{thm: convergence-utility-non}
    Under ergodic or blockwise independent inputs, it holds that
    \begin{equation*}
        \mathbb{E} \left[\|\Bar{u}^t - u^*\|^2\right], \mathbb{E} \left[\|\Bar{u}^t - u^\gamma\|^2\right] = \begin{cases}
            O\left(\dfrac{n^4 \iota \log n \log t}{t}+ n^3 \delta \right),& \gamma \sim Q \in \mathsf{C}^{\mathrm{E}}(\delta, \iota),\\
             O\left(\dfrac{n^4 |\mathcal{P}|_{\infty}^2 \log n \log t}{t}+ n^3 \delta \right),& \gamma \sim Q \in \mathsf{C}^{\mathrm{B}}(\mathcal{P},\delta).
        \end{cases} 
    \end{equation*}
\end{restatable}

\begin{restatable}[Fairness and efficiency metrics under nonstationary input]{theorem}{nonFE}
\label{thm:nonstationary-fairness-efficiency-metrics}
Under ergodic or blockwise independent inputs, the allocation of PACE is additive $\eta^t_{\mathrm{EF}}$-envy-free, additive $\eta^t_{\mathrm{Prop}}$-proportional, additive $\eta^t_{\mathrm{Core}}$-core stable, and $\eta^t_{\mathrm{NW}}$-suboptimal for weighted Nash welfare, with
\begin{align*}
    \mathbb E\!\left[\eta^t_{\mathrm{EF}}\right]
= 
\begin{cases}
    O\!\left({n^2  \sqrt{ \iota \,\log n} \cdot \log t \,  \sqrt{t} + \sqrt{n^3 \delta\, \log t} \cdot t  }\right), & \gamma \sim Q \in \mathsf{C}^{\mathrm{E}}(\delta, \iota),\\
    O\!\left({n^2  |\mathcal{P}|_\infty \sqrt{\log n} \cdot \log t \,  \sqrt{t} + \sqrt{n^3 \delta\, \log t} \cdot t  }\right)   , & \gamma \sim Q \in \mathsf{C}^{\mathrm{B}}(\mathcal{P},\delta).
\end{cases}
\\
\mathbb E\!\left[\eta^t_{\mathrm{Prop}}\right], \mathbb E\!\left[\eta^t_{\mathrm{NW}}\right], \mathbb E\!\left[\eta^t_{\mathrm{Core}}\right] = \begin{cases}
    O\!\left({n^2  \sqrt{ \iota \,(\log n) \, (\log t )\, t}+ \sqrt{n^3 \delta\,} \cdot t  }\right), & \gamma \sim Q \in \mathsf{C}^{\mathrm{E}}(\delta, \iota),\\
    O\!\left({n^2  |\mathcal{P}|_\infty\sqrt{ (\log n) \, (\log t )\, t}+ \sqrt{n^3 \delta\,} \cdot t  }\right), & \gamma \sim Q \in \mathsf{C}^{\mathrm{B}}(\mathcal{P},\delta).
\end{cases}
\end{align*}
\end{restatable}
The above bounds degrade smoothly as the amount of nonstationarity increases. In the ergodic model, larger $\iota$ slows convergence because the process mixes more slowly, providing fewer essentially independent observations. In the blockwise model, the bound degrades with larger $|\mathcal{P}|_\infty$ since long blocks create longer periods in which the algorithm may be exposed to a single correlated operational regime. For periodic inputs, this means that longer seasonal cycles lead to slower convergence.

%% file: best-of-many-worlds/adversarial.tex
\section{Adversarial Input}
\label{section: adversarial}

In this section, we assume that the online input $\gamma = (v_{i}^\tau)_{i\in [n], \tau \in [t]}$ is chosen by an adversary. Consequently, no-regret type of results are impossible, and instead we analyze the performance of \Cref{alg: PACE}  in terms of fairness and efficiency via competitive analysis. The missing proofs in this section can be found in \Cref{proof: adv}.

When the adversary is allowed to choose a completely arbitrary sequence, one cannot expect any nontrivial fairness and efficiency guarantees, and any online algorithm will suffer from a trivial Nash welfare competitive ratio~\citep{banerjee2022online}. However, these pessimistic results are restricted to inputs with extremely disparate values and tiny agent utilities, which may be more extreme than the type of unpredictable data encountered in practice. Once these cases are ruled out with certain mild and practical assumptions, we show that the PACE algorithm will attain competitive performance. To this end, we consider two input classes: inputs with non-extremity modulus $\varepsilon$ (\Cref{section: exterme}), and inputs where a seed utility $\xi$ is distributed to agents at the beginning (\Cref{section: seed}). We will develop multiplicative bounds that depend on these parameters.   

\subsection{Adversarial Input with Non-extreme Modulus}
\label{section: exterme}

We measure the extent of extremity of an input by the ratio between the minimum and maximum nonzero valuations that any single agent could give for different items. 

\begin{definition}
    \label{definition: eps}
   An input instance $\gamma = (v_i^\tau)_{i\in [n], \tau \in [t]}$ is \textit{non-extreme} with parameter $\varepsilon\in (0, 1]$ if 
    \begin{equation*}
        \frac{\min_{\tau \in [t]} \left\{ v_i^\tau: v_i^\tau >0\right\}}{\max_{\tau \in [t]}\left\{v_i^\tau\right\}} \geq \varepsilon, \ \forall i \in [n].
    \end{equation*}
\end{definition}

As defined above, $\varepsilon \in (0,1]$ is a lower bound on the minimum-to-maximum ratio of nonzero values for any single agent. Thus, a smaller $\varepsilon$ corresponds to greater within-agent dispersion in valuations across items. In many real-world allocation settings, however, the items belong to the same broad category and are therefore economically comparable or similar---for example, food, network bandwidth, or Internet content. In such environments, it is implausible that an agent's positive values for items differ by an arbitrarily large factor. A positive lower bound on $\varepsilon$ rules out these extreme instances of valuation heterogeneity, while still allowing zero valuations.

\paragraph{Fairness guarantees under non-extreme adversarial inputs.} Now we state the multiplicative envy-freeness and proportionality guarantees for non-extreme inputs. 

\begin{restatable}[Fairness metrics under non-extreme adversarial inputs]{theorem}{advthmA}
    \label{thm: mul-envy-upper}
    Under adversarial inputs that are non-extreme with parameter $\varepsilon$, \Cref{alg: PACE} gives an allocation that is multiplicative $\rho_{\mathrm{EF}}$-envy-free and multiplicative $\rho_{\mathrm{Prop}}$-proportional, with 
    \begin{equation}
        \label{eq: mul-envy-upper}
        \rho_{\mathrm{EF}}\leq 1+2 \log \frac{1}{\varepsilon} + O\left(\frac{1}{\min_i W_i}\right), \ \ \rho_{\mathrm{Prop}}  \leq 1 + 2 \left(1-\frac{ \min_{i\in [n]}B_i}{\|B\|_1} \right)\log \frac{1}{\varepsilon} +  O\left(\frac{1}{\min_i W_i}\right). 
    \end{equation}
\end{restatable}

In the asymptotic scenario, it is natural to assume that each agent's total value for all items is unbounded as the horizon expands, i.e., $W_i \to \infty$ as $t \to \infty$; otherwise, the time-averaged utility would be negligible even when receiving the grand bundle, and it is generally difficult to provide fair shares for such an agent. Under this growth condition, \Cref{thm: mul-envy-upper} shows that for \Cref{alg: PACE}, the multiplicative approximation factors for envy-freeness and proportionality are asymptotically bounded by $(1+2\log (1/\varepsilon))$ under non-extreme adversarial inputs. The dependence on the modulus $\varepsilon$ is logarithmic. When $\varepsilon \to 0$, the bound becomes infinite, corresponding to the fact that one cannot expect any guarantees on PACE when valuations are constructed in an arbitrarily extreme manner. When $\varepsilon$ is larger, i.e., the items are more economically similar, the multiplicative bound becomes tighter. 

The proof of \Cref{thm: mul-envy-upper} is intuitive and leverages the unique structure of PACE. We first exploit a \textit{consistency} property: re-running PACE on any subset of agents and their received bundles will result in the same allocation (\Cref{lem: reductive}), so to analyze envy, it suffices to consider only $2$-agent instances. Next, using a monotone instance transformation that keeps envy non-decreasing, we show an LP characterization of the $2$-agent input family that gives worst-case envy; see \eqref{eq: canonical}. We sketch the proof for the equal weight case as follows; the complete proof is deferred to the appendix.

\vspace{3mm}\noindent \textit{Proof sketch of \Cref{thm: mul-envy-upper}.}
Assume $\|v_i\|_\infty= 1$ without loss of generality. Then the non-extreme condition is equivalent to $v_i^\tau \in \{0\}\cup {[\varepsilon, 1]}$.  Let $T_i := \{\tau: x_i^\tau = 1\}$ be the set of items that agent $i$ receives from \Cref{alg: PACE}.

First, observe that the envy analysis between any pair of agents can be reduced to 2-agent instances by a consistency property of PACE; see \Cref{lem: reductive}. 

\begin{lemma}[Consistency on any agent subset] 
\label{lem: reductive}
    For an arbitrary input $\gamma$ and any subset $J \subseteq [n]$, define $\gamma_J$ as the restriction of $\gamma$ to $J$: 1) remove all agents that are not in $J$; 2) remove all items that are not allocated to $J$ by PACE. Then, when PACE is run on $\gamma_J$, it gives each agent in $J$ the same subset of items as when it is run on $\gamma$ with all agents.
\end{lemma}

For any $2$-agent instance, we consider the following transformation: 1) Set $v_1^\tau = 0$ for all $\tau \in T_2$; 2) Reorder the items by moving the columns in $T_2$ to the beginning, and $T_1$ to the end (the order is preserved for items within $T_1$). One can show that both agents' utilities under PACE are preserved after this transformation. It suffices to consider only transformed instances, i.e., inputs where agent $2$ receives their entire allocation in the first $R$ rounds, and agent $1$ receives their entire share in rounds $R+1$ and afterwards. It turns out that the worst-case multiplicative envy for transformed sequences can be identified by

\begin{equation}
\label{eq: canonical}
    \begin{aligned}
    \max_{v_1^\tau, v_2^\tau\in \{0\}\cup [\varepsilon,1]}  & \ \ \left\{\frac{1}{ U}\sum_{\tau=1}^{\infty} v_2^\tau \ \right|\left.\ \frac{v_2^\tau}{v_1^\tau} \leq \frac{U}{U_1^{\tau-1}}, \ U_1^{\tau} \geq \sum_{s=1}^{\tau} v_{1}^s, \  \forall \tau\right\}.
    \end{aligned}
\end{equation}
We observe that $v_2^\tau/v_1^\tau$ for each $\tau$ can be bounded by a function of $U_1^{\tau-1}$, defined as
\begin{equation*}
    Z(U_1^{\tau-1}) = \begin{cases}
        \min \left\{U/U_1^{\tau-1}, 1/\varepsilon\right\}, & U_1^{\tau-1} \in [0,U/\varepsilon] \\
        0, & U_1^{\tau-1} \in ( U/\varepsilon, \infty]
    \end{cases}.
\end{equation*}
We can then upper bound the objective value of \eqref{eq: canonical} by 
\begin{equation*}
    \frac{1}{U}\sum_{\tau=1}^\infty v_2^\tau = \frac{1}{U}\sum_{\tau=1}^\infty v_1^\tau \cdot \frac{v_2^\tau}{v_1^\tau} \leq \frac{1}{U}\sum_{\tau=1}^\infty v_1^\tau \cdot Z(U_1^{\tau-1})\rightarrow \frac{1}{U} \int_0^{U/\varepsilon} Z(u) \mathrm{d}u 
    = 1 +2 \log \frac{1}{\varepsilon}.
\end{equation*}

Intuitively, the integral upper bound follows from the observation that the increment $v_1^\tau \leq 1$ is infinitesimal when $U\rightarrow \infty$. Finally, the proportionality result follows by noticing that $\rho_{\mathrm{Prop}} \leq (B_i + \rho_{\mathrm{EF}} \sum_{j\neq i} B_j )/\|B\|_1$ for any algorithm that fully allocates all items. 

$\hfill \square$

We complement \Cref{thm: mul-envy-upper} with a lower bound, showing that our analysis on multiplicative envy and proportionality is asymptotically tight for the candidate algorithm.

\begin{restatable}[Lower bound for multiplicative envy and proportionality of PACE]{theorem}{advthmB}
\label{thm: mul-envy-lower}
    For any $\varepsilon \in (0, 1]$, there exists a sequence of inputs $(\gamma^t)_{t=1}^\infty$ that are non-extreme with parameter $\varepsilon$, such that 
    \begin{equation*}
        \limsup_{t \rightarrow \infty} \ \rho_{\mathrm{EF}}^t = 1+ 2\log\frac{1}{\varepsilon}, \ \limsup_{t \rightarrow \infty} \ \rho_{\mathrm{Prop}}^t = 1+ 2\left(1-\frac{\min_{i\in [n]}B_i}{\|B\|_1}\right)\log\frac{1}{\varepsilon}, 
    \end{equation*}
    where $\rho_{\mathrm{EF}}^t, \rho_{\mathrm{Prop}}^t$ are the multiplicative envy-freeness and proportionality metrics when \Cref{alg: PACE} is run under $\gamma^t$. 
\end{restatable}

\paragraph{Efficiency guarantees under non-extreme adversarial inputs.} 

For the multiplicative efficiency guarantees under non-extreme inputs, we have the following \Cref{thm: cr-upper}.

\begin{restatable}[Efficiency metrics under non-extreme adversarial inputs]{theorem}{advthmC}
    \label{thm: cr-upper}
    Under adversarial inputs that are non-extreme with parameter $\varepsilon$, \Cref{alg: PACE} gives an allocation that is $\rho_{\mathrm{NW}}$-competitive for Nash welfare and multiplicative $\rho_{\mathrm{Core}}$-core stable, with 
    \begin{equation}
        \label{eq: cr-upper}
            \rho_{\mathrm{NW}}, \, \rho_{\mathrm{Core}} \leq n \left(1+2\log \frac{1}{\varepsilon}\right) +   O\left( \frac{n}{\min_i W_i}\right).
    \end{equation}
\end{restatable}

For the dependence on the number of agents $n$ in the Nash welfare competitive ratio, we show that an $\Omega(n)$ lower bound is inevitable for any online algorithm, even when the input values are in $\{0,1\}$, i.e., the input has non-extreme modulus $\varepsilon = 1$. This shows that the Nash welfare approximation of PACE is asymptotically tight and optimal: it guarantees a constant fraction of the optimal online algorithm's Nash welfare up to a factor logarithmic in $\varepsilon$. 
\begin{restatable}[Lower bound for Nash welfare competitive ratio]{theorem}{advthmD}
    \label{thm: cr-lower}
    There exists a sequence of inputs $(\gamma^t)_{t=1}^\infty$ where all values are binary, i.e., $v_i^\tau \in \{0, 1\}$, such that any online algorithm must have $\limsup_{t\rightarrow \infty} \rho_{\mathrm{NW}}^t =  (n!)^{1/n} = \Omega(n),$
    where $\rho_{\mathrm{NW}}^t$ is the Nash welfare competitive ratio of the online algorithm when it is run under $\gamma^t$. 
\end{restatable}

\subsection{Adversarial Input with Seed Utility}
\label{section: seed}

We now analyze PACE under a \textit{seed utility} assumption: the cumulative utilities of agents start from a positive initial level of at least $\xi > 0$. This aligns with practical online systems where the horizon begins after the platform has already been running for some time, so agents may possess pre-existing usage rights or entitlements, or have accumulated service before the online procedure begins. From a technical point of view, this excludes degenerate early-stage effects caused by zero initial utilities, which might otherwise distort fairness or efficiency metrics.

\begin{definition}
    \label{definition: seed}
   An input instance $\gamma = (v_i^\tau)_{i\in [n], \tau \in [t]}$ has seed utility $\xi>0$ if $U_i^0 \geq  \xi$ for all $i\in[n]$.  
\end{definition}

To interpret a seeded instance as a standard one, one can equivalently consider the seed as a prefix containing $\xi$ copies of value-1 items for each agent (other agents have value 0); PACE will allocate these added items to the corresponding agents. A rigorous treatment of seed utility and PACE's implementation can be found in \Cref{alg: seed} in the appendix. 

\paragraph{Fairness guarantees under seed utility.} We now give the guarantees for multiplicative envy-freeness and proportionality with seed utility $\xi>0$. 
\begin{restatable}[Fairness metrics under adversarial inputs with seeds]{theorem}{advthmEnvyB}
    \label{thm: seed-envy}
    Under adversarial inputs with seed utility $\xi>0$, \Cref{alg: PACE} gives an allocation that is multiplicative $\rho_{\mathrm{EF}}$-envy-free and multiplicative $\rho_{\mathrm{Prop}}$-proportional, with 
    \begin{equation*}
        \begin{aligned}
            \rho_{\mathrm{EF}} &\leq \max\left\{ 1,\frac{\|v\|_\infty \log (1+\|v\|_\infty t/\xi)}{\xi \log (1+\|v\|_\infty/\xi)} \right\}, \\
            \rho_{\mathrm{Prop}} &\leq \max\left\{1, \frac{\min_{i\in [n]} B_i }{\|B\|_1} + \left(1-\frac{\min_{i\in [n]}B_i}{\|B\|_1}\right)  \frac{\|v\|_\infty \log (1+\|v\|_\infty t/\xi)}{\xi \log (1+\|v\|_\infty/\xi)}\right\}.
        \end{aligned}
    \end{equation*}
\end{restatable}
The above bounds on $\rho_{\mathrm{EF}}$ and $\rho_{\mathrm{Prop}}$ are decreasing in $\xi$. In practice, the platform can achieve sharper fairness guarantees by providing larger initial utility levels to agents. When $\xi$ is a constant independent of $t$, the bounds become logarithmic. When $\xi$ is at least a linear function of $t$, i.e., the provided initial utility is comparable with a constant factor of the entire bundle as the horizon scales, both approximation ratios $\rho_{\mathrm{EF}}$ and $\rho_{\mathrm{{Prop}}}$ become constant. 

\paragraph{Efficiency guarantees under seed utility.} We next give competitive ratio guarantees on Nash welfare and core stability. Similarly, the bounds are decreasing in $\xi$, so the efficiency guarantees become sharper as the initial seed utilities increase.

\begin{restatable}[{Efficiency metrics under adversarial inputs with seeds}]{theorem}{advthmE}
{Under adversarial inputs with seed utility $\xi>0$, \Cref{alg: PACE} gives an allocation that is $\rho_{\mathrm{NW}}$-competitive for Nash welfare and multiplicative $\rho_{\mathrm{Core}}$-stable, with}
    \label{thm: seed}
    \begin{equation*}
    \begin{aligned}
                \rho_{\mathrm{NW}}\leq 3+ \frac{4}{\xi}\|v\|_\infty +2\log \left(1+\frac{\|v\|_\infty\cdot t}{\xi}\right), \ 
                \rho_{\mathrm{Core}} \leq \frac{n\|v\|_\infty \log (1+\|v\|_\infty t / \xi)}{\xi\log (1+\|v\|_\infty /\xi)}.
    \end{aligned}
    \end{equation*}
\end{restatable} 

\paragraph{Connections to algorithms with predictions.} Our result for the seed utility setting provides meaningful extensions to the settings where no initial seed utility is available, but the decision maker knows the monopolist utility $W_i$ for each agent in advance. In this setting, a ``set-aside'' algorithm that guarantees $x_{i}^t \geq 1/(2n)$ (i.e., at least half of each item is divided uniformly) can be considered to have a seed utility of $W_i / (2n)$ effectively; the online decisionis then made on the other half of each item, which is not allocated uniformly. By running this set-aside version of PACE and applying \Cref{thm: seed}, we can get an $O(\log t)$ competitive ratio. This matches the asymptotic order obtained by \citet{banerjee2022online}, while using a simpler algorithm and enabling a richer set of fairness guarantees via \Cref{thm: seed-envy}. Detailed discussions on this extension can be found in \Cref{proof: adv-set-aside}.

%% file: best-of-many-worlds/applications.tex
\section{Applications}
\label{section: app}

\subsection{Fair Matching in Blood Donation Platforms}

We next show how PACE applies to an online fair matching problem motivated by blood donation platforms, such as Facebook Blood Donation~\citep{american2019new,jin2021million}, where potential donors visit the platform over time, and the platform recommends compatible donors to recipients such as hospitals, blood drives, or donation sites. Because blood donation is scarce and demand is often uneven across regions, blood types, and time periods, the platform's objective is not only to maximize the number of successful donations, but also to ensure that access to donor opportunities is distributed fairly across recipients with different demands and priorities. The online nature of the problem comes from the fact that donor availability, recipient demand, and compatibility information may change over time, while recommendations may need to be made promptly when donors visit the platform.

We use the same online matching setup as \citet{mcelfresh2020matching}. Let \(\mathcal I\) be the set of recipients and \(\mathcal J\) be the set of potential donors, i.e., users of the platform. On each day \(d \in [D]\), the platform observes a weighted bipartite graph \(\mathcal G^d=(\mathcal I,\mathcal J,\mathcal E^d,w^d)\), where an edge \((i,j)\in \mathcal E^d\) means that donor \(j\) is feasible for recipient \(i\) on day \(d\), and the weight \(w^d_{ij}\ge 0\) represents the donation probability, or more generally the operational score, obtained if donor \(j\) is recommended to recipient \(i\). Infeasible or unavailable pairs have weight \(w^d_{ij}=0\). A matching decision on day \(d\) is denoted by \(x^d=(x^d_{ij})_{i\in \mathcal I,j\in \mathcal J}\), where \(x^d_{ij}\in\{0,1\}\) and each active donor is assigned to at most one compatible recipient. The total matched weight of recipient \(i\) over \(D\) days is therefore \(W_i^D=\sum_{d=1}^D\sum_{j\in \mathcal J} w^d_{ij}x^d_{ij}\), which can be interpreted as the total expected donation value routed to that recipient. The recipient weights \(B_i\) encode priority or entitlement, and the goal is to obtain long-run fair and efficient total matched weights \(W^D=(W_i^D)_{i\in \mathcal I}\).

\paragraph{Applying PACE.}
The blood donation setting is more complicated than the standard online fair allocation model since each day's input is a dynamic weighted compatibility graph rather than a single item. However, we can apply PACE in a simple and direct way, and view the dynamic graph as a new structured arrival model for PACE, for which we develop specialized guarantees below. Concretely, {we treat each active donor as a single online item and the recipients as the agents, and process the donors within a day with PACE in arbitrary order; see Algorithm~\ref{alg: blood}.} This gives a simple ``look-once'' donor-by-donor rule: the platform does not need to solve a full-day matching problem over the graph or know the number of active donors in advance. It only needs to maintain each recipient's cumulative matched value and make an immediate recommendation when an active donor is processed. The resulting procedure is shown in Algorithm~\ref{alg: blood}.

\begin{algorithm}
\caption{PACE for Fair Matching in Blood Donation Platforms}
\label{alg: blood}
    \SetKwInput{KwInit}{Initialization}
    \KwIn{recipient set $\mathcal{I}$, donor set $\mathcal{J}$, number of days $D$, recipient weights $B=(B_i)_{i\in\mathcal{I}}$.}
    \KwInit{iteration count $\tau \leftarrow 1$, cumulative utilities $U_i^0 \leftarrow 0$ for all $i\in\mathcal{I}$.}
    \For{$d = 1, \cdots, D$}{
        Observe the daily donation graph $\mathcal{G}^d=(\mathcal{I},\mathcal{J},\mathcal{E}^d,w^d)$.\\
        Initialize $x^d_{ij}\leftarrow 0$ for all $i\in\mathcal{I}$ and $j\in\mathcal{J}$.\\
        \For{each donor $j \in \mathcal{J}$, in arbitrary order}{
            \If{$w^d_{ij} > 0$ for some $i \in \mathcal{I}$}{
                Set the current value vector by $v_i^\tau \leftarrow w^d_{ij}$ for all $i\in\mathcal{I}$.\\
                Choose the recipient according to one iteration of PACE:
                \[
                    i^\tau \in \min \arg\max_{i\in\mathcal{I}} B_i\frac{v_i^\tau}{U_i^{\tau-1}} .
                \]
                
                Match donor $j$ to recipient $i^\tau$ by setting $x^d_{i^\tau j}\leftarrow 1$, and update cumulative utilities for each $i\in\mathcal{I}$: \ $
                    U_i^\tau \leftarrow U_i^{\tau-1}+\mathbf{1}\{i=i^\tau\}v_i^\tau .$\\
                Set $\tau \leftarrow \tau+1$.
            }
        }
    }
\end{algorithm}

\paragraph{Fairness and Efficiency Results.} 

We now state the fairness and efficiency guarantees. Assume that the daily donation graphs are generated independently according to the same distribution across days, while allowing arbitrary correlation among donor-recipient edges within a day. Let \(\bar w^D\) be the recipient matched-weight vector produced by Algorithm~\ref{alg: blood}, averaged over the \(D\) days. Let \(w^\gamma\) be the corresponding day-averaged matched-weight vector of the hindsight Nash welfare benchmark for the realized sequence of daily graphs, and let \(w^*\) be the underlying day-level benchmark induced by the common distribution of daily graphs. The fairness and efficiency notions have the same interpretation as in our general online fair allocation problem, where we treat active donors as items and donation recipients as agents. For each day \(d\), let \(m_d:=|\{j\in\mathcal{J}:\exists i\in\mathcal{I},\, w^d_{ij}>0\}|\), and let \(\Delta_{\mathcal{G}}\) be an upper bound on the maximum degree of the daily donation graph after ignoring zero-weight edges. For compactness, write \(\Lambda_{\mathrm{BD}}:=\Delta_{\mathcal{G}}\log|\mathcal{I}|+\mathbb{E}[m_d]+\mathbb{E}[m_d^2]/|\mathcal{J}|\).

\begin{restatable}[Matched-weight convergence, fairness, and efficiency]{theorem}{appthmA}
\label{thm:blood-fairness-efficiency-metrics}
Under independently generated daily donation graphs, it holds for Algorithm~\ref{alg: blood} that
\begin{equation*}
    \mathbb{E}\!\left[\|\bar w^D-w^*\|^2\right],\mathbb{E}\!\left[\|\bar w^D-w^\gamma\|^2\right]
    =
    O\!\left(
        \frac{|\mathcal{I}|^4\log D}{|\mathcal{J}|D}\cdot \Lambda_{\mathrm{BD}}
    \right).
\end{equation*}
Moreover, the allocation of Algorithm~\ref{alg: blood} is additive $\eta^D_{\mathrm{EF}}$-envy-free, additive $\eta^D_{\mathrm{Prop}}$-proportional, additive $\eta^D_{\mathrm{Core}}$-core stable, and $\eta^D_{\mathrm{NW}}$-suboptimal for weighted Nash welfare, with
\begin{align*}
    \mathbb E\!\left[\eta^D_{\mathrm{EF}}\right]
    &=
    O\!\left(
        |\mathcal{I}|^2\log D\cdot
        \sqrt{\frac{D\Lambda_{\mathrm{BD}}}{|\mathcal{J}|}}
    \right),
    \mathbb E\!\left[\eta^D_{\mathrm{Prop}}\right],
    \mathbb E\!\left[\eta^D_{\mathrm{NW}}\right],
    \mathbb E\!\left[\eta^D_{\mathrm{Core}}\right]
    &=
    O\!\left(
        |\mathcal{I}|^2
        \sqrt{\frac{D(\log D)\Lambda_{\mathrm{BD}}}{|\mathcal{J}|}}
    \right).
\end{align*}
\end{restatable}

Theorem~\ref{thm:blood-fairness-efficiency-metrics} is a specialized counterpart of our earlier stochastic guarantees, and achieves $\widetilde{O}(1/D)$ convergence in the number of days. The daily compatibility graph model is not covered by the earlier \textit{i.i.d.} item arrival model because all donors and feasible edges within the same day may be correlated through certain operational conditions, such as location, donor availability, blood-type demand, and site capacity. The blockwise model could be applied by treating each day's donors as a large correlated block of items, but this would only result in a cruder dependence on $m_d$ and the daily graph size. The specialized analysis instead exploits the structure of the matching model and replaces this dependence by the graph complexity term \(\Lambda_{\mathrm{BD}}\), which depends on the maximum degree \(\Delta_{\mathcal{G}}\), the expected number of active donors \(\mathbb{E}[m_d]\), and the second moment \(\mathbb{E}[m_d^2]\). These quantities can be relatively small when the daily compatibility graph is sparse: even when a platform has a very large registered donor base, on most days only a small set of donors are active. For example, there are 85 million registered donors on Facebook, but most are rarely active, and donors are rate-limited in how often they may be notified~\citep{mcelfresh2020matching}. 

The case of adversarial arrivals for this graph model reduces to the general adversarial case, so there is no need for specialized analysis. %

\subsection{Multi-Agent Pacing Equilibrium in First-Price Auctions}
\label{section: fppe}

We next show how PACE can be interpreted as multi-bidder dynamics for repeated first-price auctions with budgets, a widely studied and practically important setting in online advertising markets. Consider \(n\) agents, usually advertisers, who participate in \(t\) repeated first-price auctions, each corresponding to an online item such as an impression opportunity. Bidder \(i\) has per-round budget \(B_i\), or equivalently a total budget constraint \(tB_i\) over the horizon. In each auction, bidders submit bids $(b_i^\tau)$, the highest bidder wins the item, and the winner pays the winning bid as the first-price payment $p^\tau = \max_{i\in [n]} b_i^\tau$. A central operational problem is to manage bids over time so that advertisers can compete for valuable items while keeping their long-run spending consistent with their budgets.

In this scenario, \citet{conitzer2022pacing} introduce the notion of First-Price Pacing Equilibrium (FPPE), an important offline solution concept that characterizes a static outcome when advertisers use multiplicative pacing to shade their bids. It describes a benchmark in which advertisers bid according to paced values, the auction outcomes are budget-feasible, and there is no unnecessary pacing~\citep{conitzer2022pacing,conitzer2022multiplicative}. The FPPE solution is captured by the following quasi-linear Eisenberg-Gale convex program developed by \citet{cole2017convex}. Let $\Bar Q$ be the time-averaged distribution of values as in previous sections. The infinite-dimensional versions of the primal and dual programs are
\begin{equation}
    \label{eq: inf-QL-primal}
    \max_{x\in L^{\infty}_+(V), u\geq 0, \delta\geq 0}
    \left\{
    \left.
     \sum_{i=1}^n \left(B_i \log (u_i+\delta_i) - \delta_i\right) \ \right| \ 
     u_i \leq \langle v_i, x_i\rangle  \  \ \forall i\in [n], \
     \sum_{i=1}^n x_i\leq s
        \right\}
        .
\end{equation}
\begin{equation}
    \label{eq: inf-QL-dual}
    \min_{\beta \in \mathbb{R}_+^n}
    \left\{\left.
    \int_{V} \left(\max_{i\in [n]}\beta_i v_i\right) \mathrm{d}\Bar{Q} - \sum_{i=1}^n B_i \log \beta_i
        \ \right|  \ 0\leq \beta_i\leq 1\right \}
        .
\end{equation}

The primal program describes the FPPE benchmark in terms of allocation $x$, gross value $u$, and leftover budget $\delta$; \(u_i\) denotes bidder \(i\)'s time-averaged gross value from the items they receive, while the new variable \(\delta_i\) denotes her time-averaged remaining budget. We let \((x^*, u^*, \delta^*)\) and \(\beta^*\) denote the primal-dual optimal solutions to \eqref{eq: inf-QL-primal} and \eqref{eq: inf-QL-dual}. 
Unlike in the fair allocation setting, the actual utility of each bidder is \textit{quasi-linear}, depending on the expenditure of their budget $B_i$. Concretely, in the optimal solution, the corresponding quasi-linear utility of bidder \(i\) is \(u_i^{\ql,*}=u_i^*+\delta_i^*\), which combines the gross value of items won with the leftover budget.\footnote{An alternative definition of quasi-linear utility is the \textit{profit}, \(\hat{u}_i^{\ql,*}=u_i^*-(B_i-\delta_i^*)\). The two definitions differ by the constant \(B_i\), and are therefore equivalent for measuring convergence. We use \(u_i^{\ql,*}=u_i^*+\delta_i^*\) because it is better behaved for competitive-ratio-type measures in the adversarial case.} The optimal dual solution \(\beta^*\) characterizes the equilibrium pacing multipliers.

\paragraph{Applying PACE.} The quasi-linear dual program \eqref{eq: inf-QL-dual} is highly similar to the standard EG dual program \eqref{eq: inf-EG-dual}: it has the same objective but with the additional constraints \(\beta_i\in[0,1]\). In the auction setting, this prevents bidders from bidding above their private value and thereby rules out negative utility from winning an item. Given the natural dual interpretation of PACE where $\beta_i^\tau$ can be considered as the bid-shading multiplier, we can describe a joint bidding dynamic by a capped version of PACE: each bidder submits the paced bid \(\beta_i^\tau v_i^\tau\), and the pacing multiplier is updated by projecting onto \([0,1]\), i.e., $\beta_i^{\tau+1} = \min \left\{{B_i}/{\Bar{u}_i^\tau}, 1\right\}$.
For each auction, the highest paced bid determines both the winner and the first-price payment. The detailed pseudocode is shown in \Cref{alg: QL-1} in the appendix.

\paragraph{Convergence under stochastic input.}
Under stochastic input, we show that the outcome utility levels under the adapted PACE dynamics converge to those described by the FPPE benchmark \eqref{eq: inf-QL-primal}; see \Cref{thm: convergence-ql-iid}. The analysis is technically similar to constrained PACE in \citet{gao2021online}. However, the constrained PACE algorithm is not directly applicable to the FPPE setting: its projection interval is introduced for technical boundedness and requires prior scale information, whereas the upper cap \(1\) here is imposed by the economic no-overbidding requirement in first-price auctions. The nonstationary extensions are deferred to the appendix. 

\begin{theorem}
\label{thm: convergence-ql-iid}
    Let $\bar{u}_i^{\ql, t}$ be the resulting (time-averaged) quasi-linear utility of agent $i$ under the dynamics of \Cref{alg: QL-1}. Then under stationary \textit{i.i.d.}\ input, 
    \begin{equation*}
        \mathbb{E} \left[\left\| \Bar{u}^{\ql,t}  - u^{\mathrm{\ql, *}}\right\|^2\right] = O\left(\frac{n^4\log n (\log t)^2}{t}\right). 
    \end{equation*}
\end{theorem}

\paragraph{Approximation guarantees under adversarial input.} 

We next consider the performance of \Cref{alg: QL-1} when the item value sequence is chosen adversarially. Here, it is technically challenging to derive approximation guarantees on the utility vector entry-wise. We therefore provide performance certificates for the PACE bidding dynamic through aggregate quasi-linear Nash welfare and envy metrics.\footnote{These metrics should not be directly interpreted as fairness and efficiency measures as in the fair allocation setup. In the auction setting, payments and leftover budgets are part of an agent's utility, so quasi-linear Nash welfare and envy notions are not as meaningful as in fair allocation. In addition, proportionality and core stability are not natural targets here because in the hindsight solution a bidder may keep part of their endowed budget rather than spend it on items.} 

We first consider Nash welfare with respect to quasi-linear utilities. Let \(U_i^{\ql,t}=U_i^t+\Delta_i^t\) be the total quasi-linear utility obtained by bidder \(i\) under \Cref{alg: QL-1}, where \(U_i^t\) is her total gross value derived from the items she obtains and \(\Delta_i^t\) is her remaining budget. The following \Cref{thm: ql-adv} gives a bound on the quasi-linear Nash welfare competitive ratio that does not depend on $t$.

\begin{restatable}[Competitive ratio for quasi-linear Nash welfare]{theorem}{qlthmmain}
\label{thm: ql-adv}
    We say that the allocation is $\rho$-competitive for quasi-linear Nash welfare if $\prod_i (U_i^{\ql,t})^{B_i/\|B\|_1} \geq \rho^{-1} \prod_{i}(\widetilde{U}_i +\widetilde{\Delta}_i)^{B_i/\|B\|_1}$, for any optimal solution $(\widetilde{U}, \widetilde{\Delta})$ of the following hindsight FPPE program,
\begin{equation*}
    \label{eq: EG-primal-hindsight-ql}
\max_{x\geq 0, U\geq 0} 
\left\{
    \left. \sum_{i=1}^n (tB_i \log (U_i+\Delta_i) - \Delta_i ) \ \right| 
    U_i\leq \langle v_i, x_i\rangle \ \forall i \in [n],  \ 
    \sum_{i=1}^n x_i^\tau \leq 1 \ \forall \tau \in [t], \Delta\geq 0
\right\}.
\end{equation*}
Under general adversarial value input, \Cref{alg: QL-1} is $\rho_{\mathrm{QLNW}}$-competitive with
\begin{equation*} \rho_{\mathrm{QLNW}} \leq \left(1+\frac{8}{t}\right) \sqrt{ \frac{n\|v\|_\infty}{\|B\|_1} +\frac{(n+1)\|v\|_\infty}{t} +5+4\log\left(1+\sqrt{\frac{n\|v\|_\infty}{\|B\|_1}}\right) } . \end{equation*} 
\end{restatable}  

The above guarantee in \Cref{thm: ql-adv} is assumption-free: unlike our previous analysis in the adversarial case, it holds without assumptions on non-extremity or seed utility parameters. The upper bound on the competitive ratio is free of $t$ and has an $O(\sqrt{n})$ dependence on the number of bidders.

Next, we consider envy-type metrics. The standard envy notion we considered in the fair allocation setting is based only on the gross value derived from items and does not involve expenditure. This is not an appropriate notion here since the bidder's payoff is quasi-linear. Also, even the hindsight FPPE allocation can exhibit large value-based envy if many items are priced above a bidder's value---such items would reduce that bidder's quasi-linear utility if the bidder were to receive them at the realized prices. This motivates two modified envy notions for the quasi-linear setting. The first notion, \textit{justified value envy}, compares bidder \(i\)'s allocation only to another agent's items where the price is lower than $v_i^\tau$; this is reasonable since bidder $i$ would not benefit from purchasing an item whose price exceeds their value, and thus would not be envious of that item's allocation to another bidder. The second notion, \textit{quasi-linear envy}, compares the quasi-linear gain that bidder \(i\) would obtain from another bidder's allocation; it also takes the prices given by PACE into account when comparing the bundles of two agents, and is thus more natural for the quasi-linear agent utility model.

\begin{definition}[Justified value envy and quasi-linear envy] Let \(p^\tau:=\max_{j\in[n]}\beta_j^\tau v_j^\tau\) denote the price of the \(\tau\)th auction under \Cref{alg: QL-1}. We say that the pricing and allocation are multiplicative \(\rho\)-justified envy-free if, for every bidder $i\neq k$, 
\[ U_i^t \geq \rho^{-1}\frac{B_i}{B_k}\cdot \sum_{\tau=1}^t v_i^\tau x_k^\tau \mathbb{I}\{p^\tau<v_i^\tau\}. \] 
We say that they are multiplicative \(\rho\)-quasi-linear envy-free if, for every bidder pair $i\neq k$, 
\[ U_i^{\ql,t} \geq \rho^{-1}\frac{B_i}{B_k}\cdot \sum_{\tau=1}^t (v_i^\tau-p^\tau)x_k^\tau . \] 
\end{definition}

For both multiplicative envy notions, we get asymptotic constant upper bounds that are similar to \Cref{thm: mul-envy-upper}, with the same condition that agent values are \textit{non-extreme} with parameter $\varepsilon$; see the following \Cref{thm: justified-envy}.
\begin{restatable}{theorem}{appthmB}
    \label{thm: justified-envy}
   Under adversarial inputs that are non-extreme with parameter $\varepsilon$, \Cref{alg: QL-1} gives prices and allocations that are multiplicative $\rho_{\mathrm{JVEF}}$-justified value envy-free and $\rho_{\mathrm{QLEF}}$-quasi-linear envy-free, with 
    \begin{align*}
       \rho_{\mathrm{JVEF}} \leq 1+ 2\log \frac{1}{\varepsilon} + O\left({n}/{W_i^\prime}\right), \quad 
       \rho_{\mathrm{QLEF}} \leq 1+ 2\log \frac{1}{\varepsilon} + \frac{\|v\|_\infty}{B_i}+O(1/t),
    \end{align*}
    where $W_i^\prime = \sum_{\tau=1}^t v_i^\tau \mathbb{I}\{p^\tau < v_i^\tau\}$ is bidder $i$'s total value on the items that could be won without overbidding.
\end{restatable}

%% file: best-of-many-worlds/experiments.tex
\section{Numerical Experiments}
\label{section: experiments}
We evaluate the performance of our PACE algorithm on the fair notification task, where items are users' attention, and agents are different notification categories or stakeholders. Fairness can be an important concern in this scenario~\citep{kroer2023fair}. We use a notification prioritization dataset from Instagram and a recommender system model derived from MovieLens. For each dataset, we experiment under three types of arrivals: 1) \textit{i.i.d.}, 2) periodic, and 3) true temporal arrival, i.e., the real-world time order according to the timestamps in the datasets. A summary of configurations is given in \Cref{table: configurations}. In \Cref{appendix: experiments}, we give the detailed configurations.   

\begin{table}[b]
\small
\caption{\small Experiment configurations.
\label{table: configurations}
}
\centering
\begin{tabular}{@{}ccccc@{}}
\toprule
Dataset                & $n$ & \#item types & \makecell{$t_0$ \\(true temporal)} & \makecell{$t_1$ \\(\textit{i.i.d.} \& periodic)} \\ \midrule
Instagram Notification & 4   & 41072        & 41072               & 200000           \\
MovieLens              & 10  & 610          & 100836              & 200000           \\ \bottomrule
\end{tabular}
\end{table}
\begin{figure*}[t] 
    \centering
    \begin{subfigure}
	\centering
\includegraphics[width=0.26\linewidth]{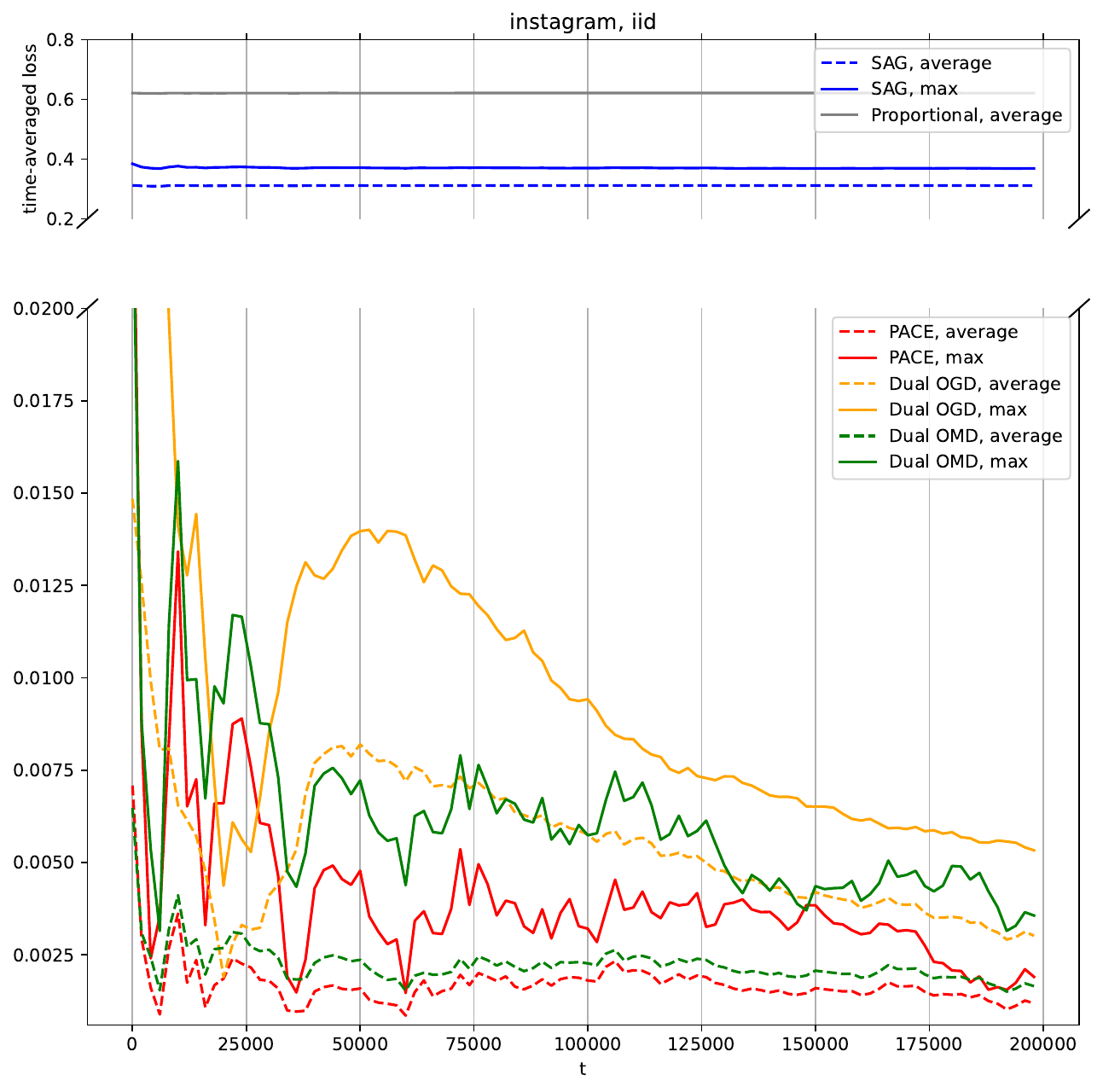}
	\end{subfigure}
     \begin{subfigure}
	\centering
\includegraphics[width=0.26\linewidth]{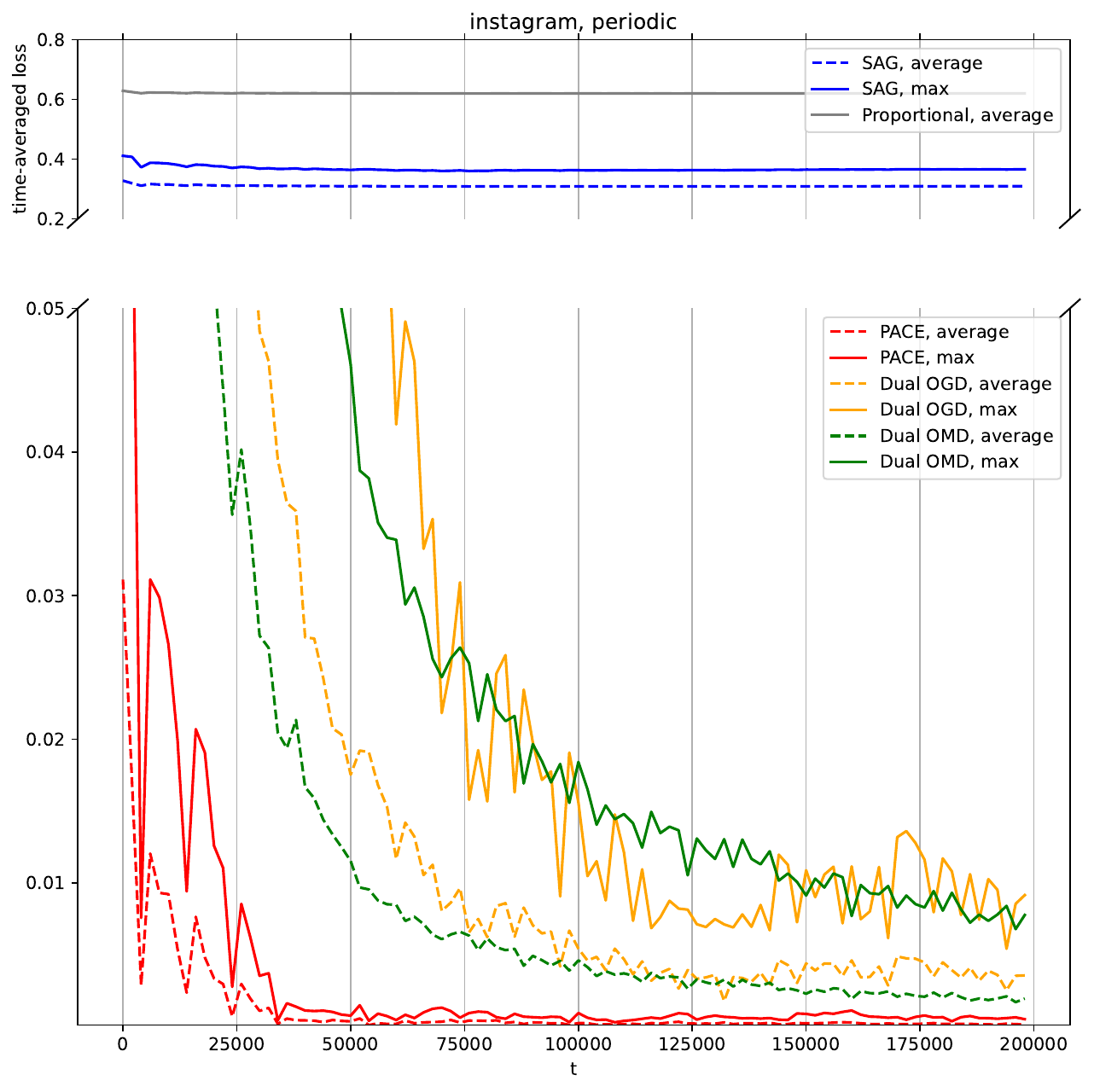}
	\end{subfigure}
     \begin{subfigure}
	\centering
\includegraphics[width=0.26\linewidth]{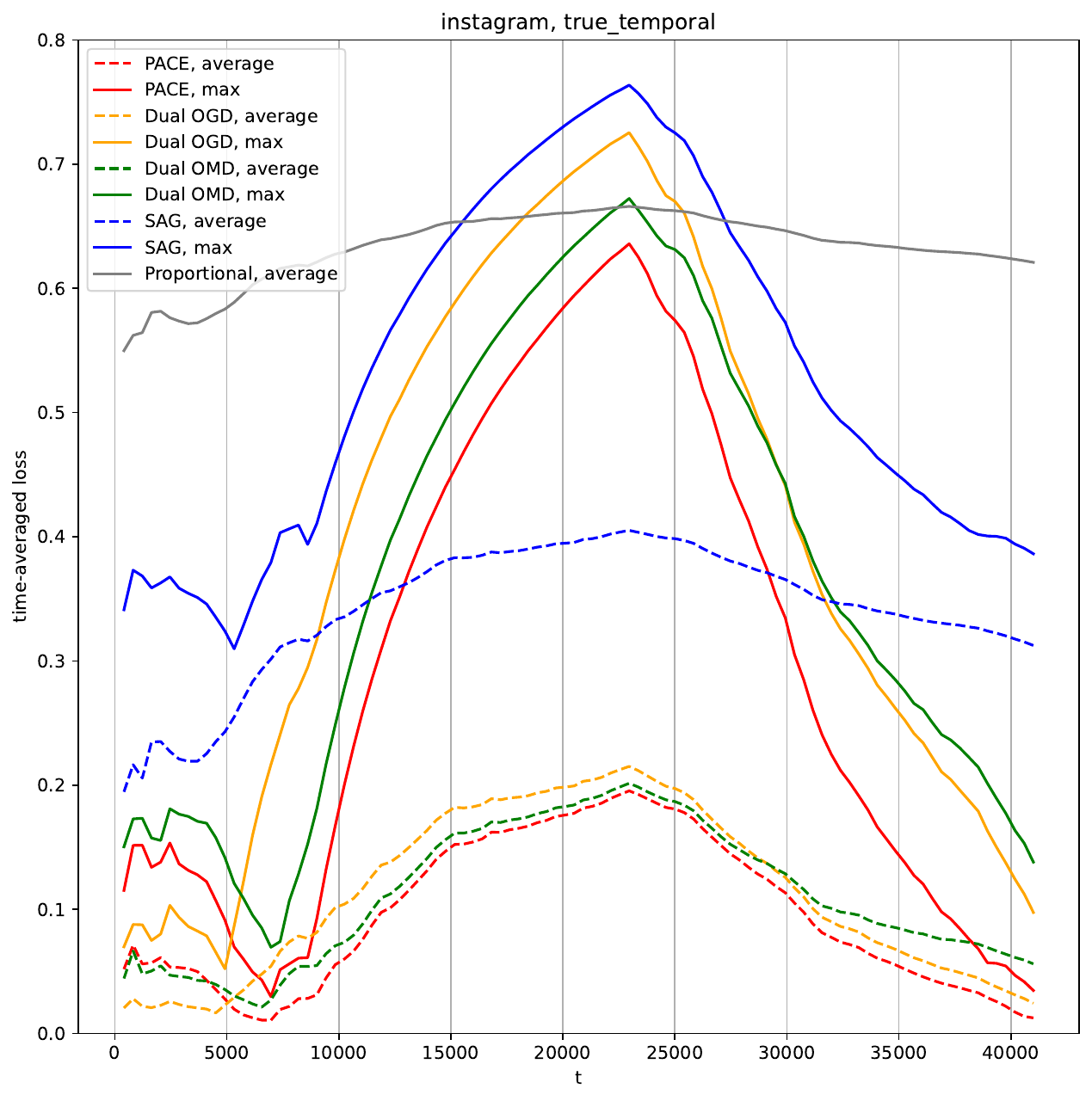}
	\end{subfigure}
     \begin{subfigure}
	\centering
\includegraphics[width=0.26\linewidth]{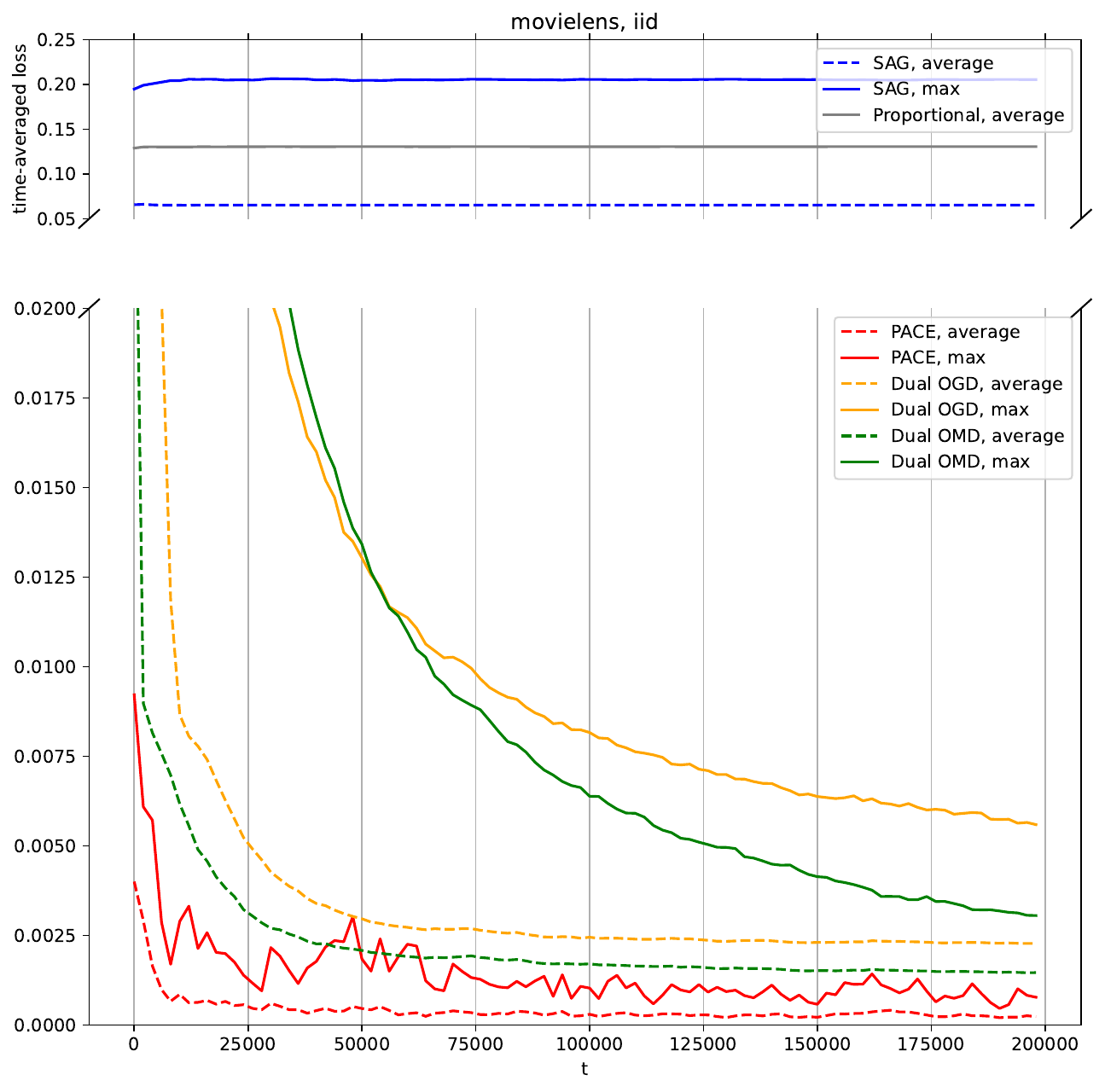}
	\end{subfigure}
     \begin{subfigure}
	\centering
\includegraphics[width=0.26\linewidth]{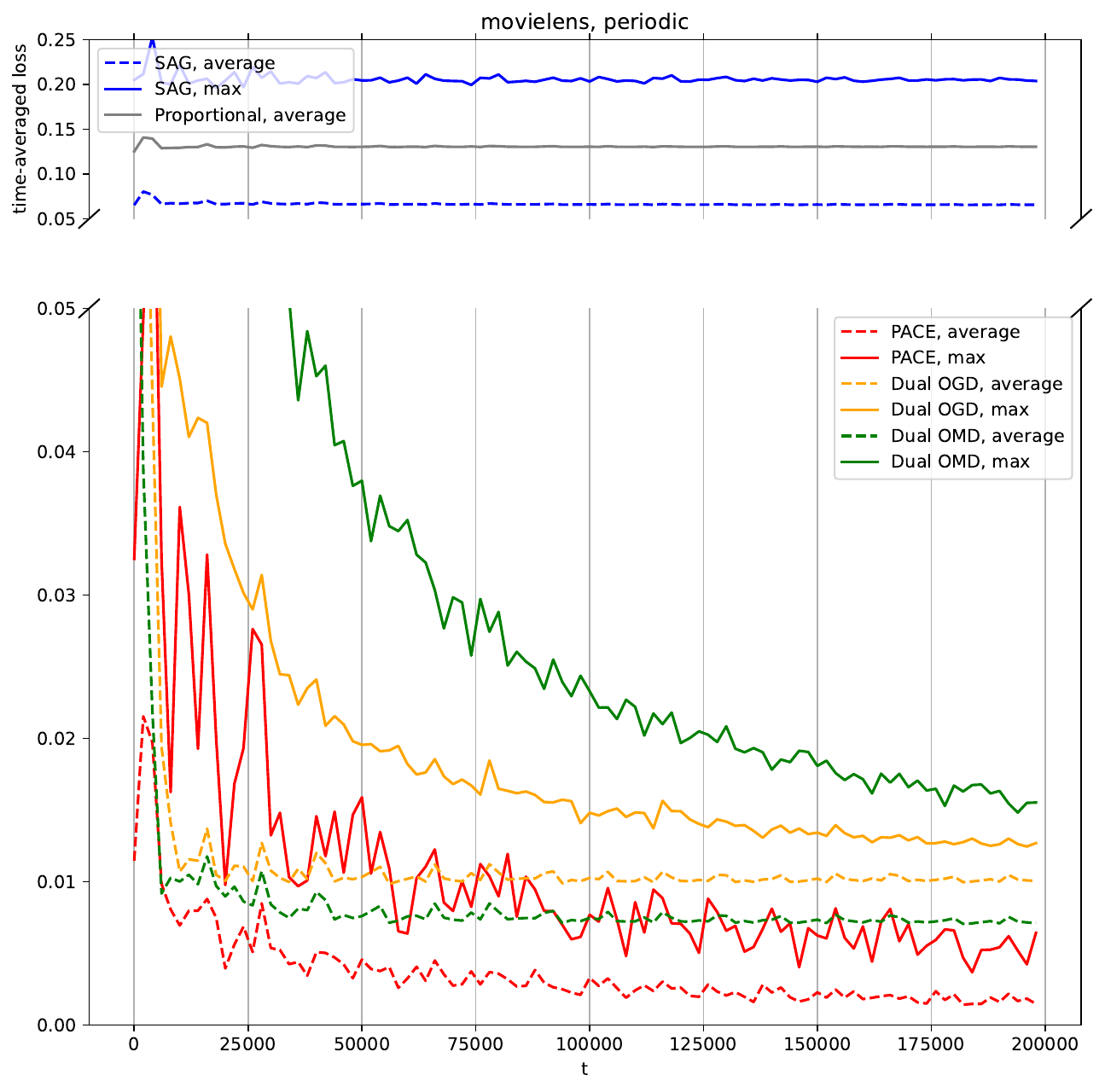}
	\end{subfigure}
     \begin{subfigure}
	\centering
\includegraphics[width=0.26\linewidth]{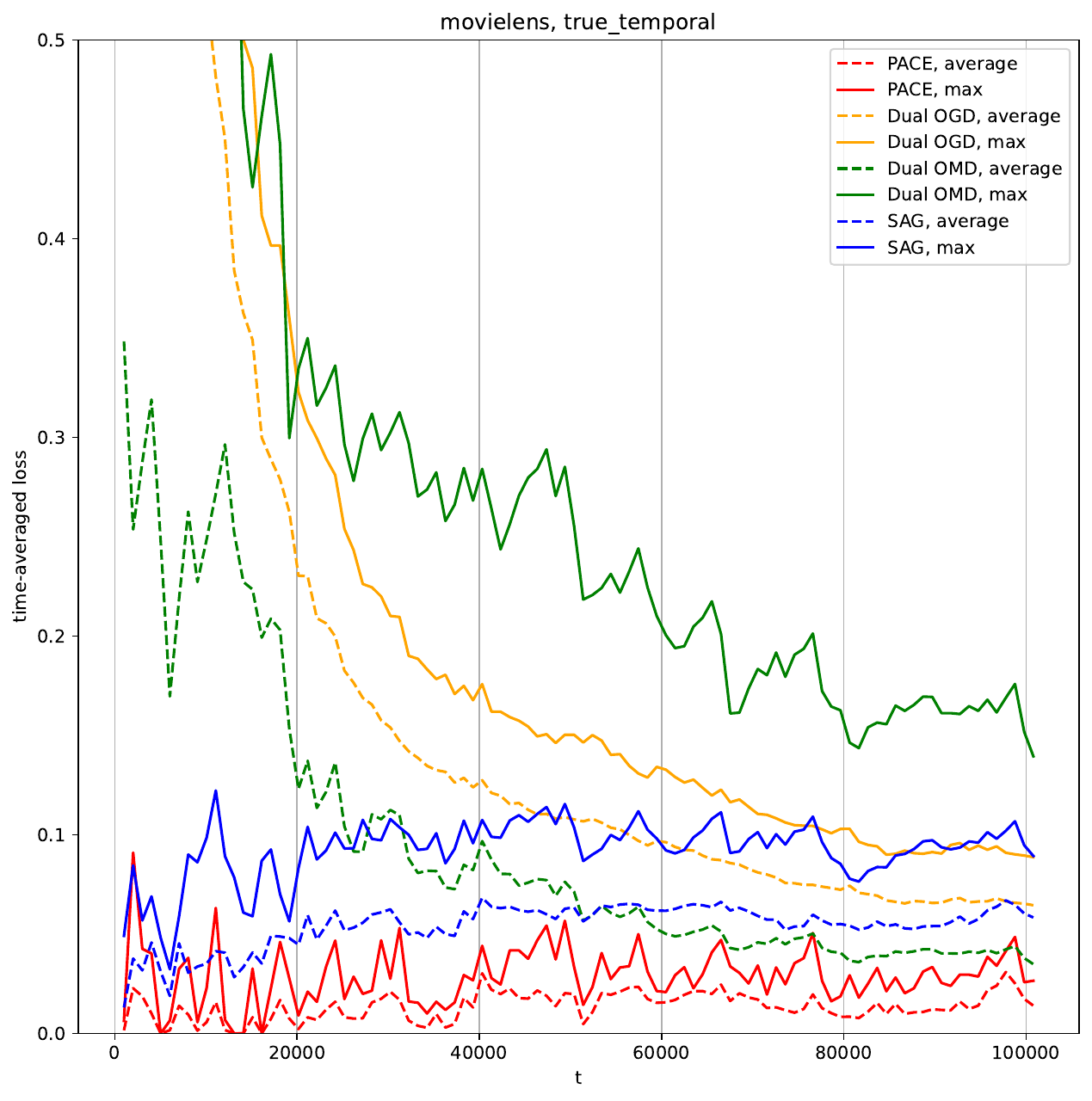}
	\end{subfigure}
    \caption{\small Simulations of PACE on the Instagram notification dataset \textit{(top)} and the MovieLens dataset \textit{(bottom)}, under three input models: stationary \textit{(left)}, periodic \textit{(middle)}, and true temporal \textit{(right)}.}
    \label{fig: experiments}
\end{figure*}

\paragraph{Benchmark Algorithms.} We compare PACE with three existing or potential algorithms for the online setting: 1) \textit{Set-Aside Greedy} (SAG), designed by \citet{banerjee2022online} for adversarial input, provided with perfect prediction and normalization as required by their algorithm; it can be considered as an enhanced version of \citet{huang2022online} for their balanced input model. 2) \textit{Online (sub)gradient descent} on the dual of EG (Dual OGD). This means keeping the same dual-to-primal rule as PACE ($i^\tau \in \arg \max \beta_i^\tau v_i^\tau$), but using online (sub)gradient descent to update the pacing multipliers. 3) \textit{Online mirror descent} on the dual of EG (Dual OMD), with entropic Bregman divergence. We tune the parameters in these benchmark algorithms instance-wise and the results shown are under optimal configurations.

\paragraph{Results.}
For each algorithm and input model, we track the loss of the algorithm in terms of the relative difference between the outcome utility $\bar u_i^\tau$ and the Eisenberg-Gale utility on the first $\tau$ items. Concretely, we plot $(1- \bar{u}_i^\tau/u_i^{\gamma, 1:\tau})^+$ as a function of $\tau$, where ${u_i^{\gamma,1:\tau}}$ is from solving EG on the first $\tau$ items. When $\tau = t$, this shows how close the eventual output utilities are to the offline benchmark and reflects the $\|\bar u^t - u^\gamma\|$ measure we considered in \Cref{section: stationary} and \Cref{section: nonstationary}. In the plot, we show both the maximum (worst agent, labeled as ``max'') and the arithmetic mean across agents (labeled as ``average'' and shown as a dotted line). The result shown is the average of $10$ independent repetitions; see \Cref{fig: experiments}.

{PACE has robust experimental performance under all input models.} For \textit{i.i.d.} inputs, we see that PACE converges quickly in the experiments. Within a few iterations, its deviation falls within 1\% of the current hindsight optimal utilities. After $t_1=200,000$ iterations, the relative average loss falls within 0.2\% for all agents. For periodic inputs, despite the fluctuations caused by nonstationarity, the loss is still low. For true temporal input, PACE suffers from a larger deviation than for stochastic and periodic inputs. However, at the end of the true temporal arrivals, it still attains an allocation where every agent attains at least $95\%$ of their hindsight equilibrium utility. Remarkably, the true temporal arrivals in the Instagram dataset have an adversarial structure; we provide a detailed discussion on this in \Cref{section: true-temporal-extra}.

{The benchmark algorithms are outperformed by PACE in each set of simulations. The SAG algorithm fails to approach the optimum in all cases and suffers from a large loss on the worst agent. This is not surprising due to its conservative design, which allocates $50\%$ of the items uniformly. The alternative online convex optimization algorithms (OGD and OMD) applied to the dual have decreasing loss in the stochastic cases, but are outperformed by PACE in convergence rate even for their optimal parameter configuration. In the true temporal simulations, both OGD and OMD suffer considerable performance degradation, on both datasets.}

\section{Future Work}
{
Our PACE algorithm achieves the first best-of-many-worlds type of guarantee in online fair allocation, in the sense of achieving convergence in stochastic cases and $O(1)$ competitive ratio results in the adversarial case \textit{w.r.t.} horizon length $t$. However, in the stochastic setting it is unknown whether the $\widetilde{O}(1/t)$ mean-square convergence rate is optimal. It will be interesting to either show an improvement, or an impossibility result.  }

%% file: best-of-many-worlds/appendix-related-work.tex
\section{Additional Related Work}
\label{appendix: related-work}

\paragraph{Online envy minimization. }\citet{bogomolnaia2022fair} assume stochastic input and enforce envy-freeness as a soft constraint, while maximizing social welfare. \citet{he2019achieving} allow reallocating previous items, and show that $O(t)$ re-allocations is enough to achieve envy-freeness up to an item. \citet{benade2018make} considers envy minimization in the stochastic, indivisible setting, and shows that allocating each item to each agent uniformly at random is near-optimal. \citet{zeng2020fairness} considers the indivisible setting with a non-adaptive adversary, showing that nontrivial approximation of envy-freeness and Pareto-optimality is hard to achieve simultaneously. While we also give approximate envy-freeness results, we remark that the focus on strict envy-freeness seems to remove the connection to market equilibrium in this line of work. 
By focusing on CEEI-style allocation rather than directly constraining envy, we get asymptotic envy guarantees, but also achieve other fairness guarantees such as approximate NW maximization, and get stronger efficiency guarantees.

\paragraph{Fairness in different online models.} We also comment on lines of work that consider online fair allocation under models that are more significantly different from ours, either in utilities or in the input setup.
\citet{liu2015dynamic} studies online allocation for Leontief utilities where each agent wants a bundle with items of fixed proportions, and shows how to balance various properties for this setting.
\citet{aleksandrov2015online} studies a simple mechanism where agents can declare if they like an item, and then a coin is flipped to determine which of the agents that desired the item will get it. 
\citet{bateni2022fair} models item arrival with Gaussian valuations and proposes a stochastic approximation scheme based on frequently resolving the EG convex program, which ensures a constant approximation ratio in terms of a proportional fairness metric. 
\citet{cheung2018tracing} considers an evolving market environment and shows that the PR dynamics generates iterates that are close to the changing equilibrium. As in PR dynamics, they assume prior knowledge of future value distributions, while our algorithm does not rely on such input. 
\citet{gkatzelis2021fair} assumes that the valuations are normalized a priori, and studies the setting where items arrive online and with two agents. They focus on satisfying the no-envy condition while maximizing social welfare, and show that one can do this approximately by allocating items proportionally to valuations. 
\cite{manshadi2021fair} studies the problem of rationing a social good and proposes simple, implementable algorithms that promote fairness and efficiency. In their setting, it is the agents’ demands rather than the supply that are sequentially realized and possibly correlated over time. Finally, there is also a recent line of online fair allocation literature that considers sequentially arriving agents, instead of items~\citep{jalota2022stochastic,sinclair2022sequential}. 
\paragraph{Convex optimization for market equilibria.}
Convex optimization algorithms and their theory have been applied to computing competitive market equilibria, which has been of interest in economics and computation for a long time~\citep{nisan2007algorithmic}. The computation of static Fisher market equilibrium with divisible items is captured by the Eisenberg-Gale convex program~\citep{eisenberg1959consensus}, which also has an infinite-dimensional generalization with stochastic items~\citep{gao2023infinite}. Specific cases of finite-dimensional static Fisher markets are studied through various convex optimization formulations~\citep{birnbaum2011distributed,kroer2019computing,shmyrev2009algorithm,gao2020first}. Applying first-order methods to equilibrium-capturing convex programs often leads to interpretable market dynamics that emulate real-world market-behaviors, such as the proportional response (PR) dynamic~\citep{birnbaum2011distributed,gao2020first,zhang2011proportional,cheung2018dynamics}, and tâtonnement~\citep{cheung2020tatonnement}. The constrained~\citep{gao2021online} and unconstrained PACE algorithms are no exception: they can be interpreted as applying a first-order method (dual averaging) to the dual of Eisenberg-Gale convex program, but can also be interpreted as a natural bidding dynamic in a sequence of first-price auctions.

The Eisenberg-Gale program is equivalent to maximizing Nash welfare (NW)~\citep{nash1950bargaining} subject to supply feasibility constraints, which is known as a strong proxy for balancing fairness and efficiency~\citep{kaneko1979nash}. In the offline setting, there is a line of literature on approximating NW with indivisible items. The problem is APX-hard ~\citet{lee2017apx}, though constant competitive-ratio algorithms are known ~\citet{cole2018approximating, anari2016nash,barman2018finding}. Beyond additive utilities, there are also competitive-ratio-type results for budget-additive utilities~\citep{garg2018approximating}, separable, piecewise-linear and concave utilities~\citep{anari2018nash}, and submodular utilities~\citep{garg2020approximating,li2022constant, barman2020tight, chaudhury2021fair}. Discrete variants of EG (i.e. NW maximization) have been applied to real-world fair division 
on the \emph{spliddit} website~\citep{caragiannis2019unreasonable}. In the discrete case, NW maximization still yields attractive efficiency and fairness guarantees, though the connection to CEEI no longer holds.

%% file: best-of-many-worlds/review.tex
\section{More on Dual Averaging and Constrained PACE}
\label{section: review}

\subsection{The Dual of EG and its infinite-dimensional analogue}
We next show how to derive the dual problem \eqref{eq: EG-dual} from the primal problem \eqref{eq: EG-primal}. Introduce the dual variables $\beta_i\geq 0$ with $i\in[n]$ for each constraint of the first type and variables $p^\tau\geq 0$ with $\tau \in [t]$ for constraints of the second type. The Lagrangian $L: \mathbb{R}_+^{n\times t} \times  \mathbb{R}_+^{n}\times  \mathbb{R}_+^{n}\times \mathbb{R}_+^{t}\rightarrow \mathbb{R}$ is given by
\begin{align*}
    L(x,U,\beta,p) &= \sum_{i=1}^n B_i \log U_i + \sum_{i=1}^n \beta_i\left(\langle v_i, x_i\rangle - U_i\right) +\sum_{\tau=1}^t p^\tau \left(1 - \sum_{i=1}^n x_i^\tau\right)\\
    &= \sum_{\tau=1}^t p^\tau + \sum_{i=1}^n \left(B_i\log U_i -\beta_iU_i\right) + \sum_{i=1}^n\langle\beta_i v_i - p, x_i\rangle.
\end{align*}
Maximizing over the primal variables $(x, U)$ gives the dual program
\begin{equation*}
    \min_{p\geq 0, \beta\geq 0} \left\{
    \left.
        \sum_{\tau=1}^t p^\tau - \sum_{i=1}^n B_i \log \beta_i + \sum_{i=1}^n(B_i\log B_i - B_i)\right| p\geq \beta_i v_i, \forall i \in [n]
    \right\}.
\end{equation*}
Ignoring the constants and moving the constraint $p\geq \beta_i v_i$ to the criterion, we obtain the following equivalent optimization problem, which is exactly \eqref{eq: EG-dual}.
\begin{equation*}
    \min_{\beta\geq 0} \left\{
    \sum_{\tau=1}^t \max_{i \in [n]}\beta_i v_i^\tau - \sum_{i=1}^n B_i \log \beta_i
\right\}
.
\end{equation*}
Consider the infinite-dimensional analogue of Eisenberg-Gale program \eqref{eq: inf-EG-primal},
\begin{equation*}
    \max_{x\in L^{\infty}_+(V), u\geq 0}
    \left\{
    \left.
     \sum_{i=1}^n B_i \log u_i \ \right| \ 
     u_i \leq \langle v_i, x_i\rangle \ \forall i\in [n], \
     \sum_{i=1}^n x_i\leq s
        \right\}
        .
\end{equation*}
We can derive its dual problem, which is the infinite-dimensional analogue of \eqref{eq: EG-dual},
\begin{equation*}
    \min_{\beta \geq 0}
    \left\{
    \int_{V} \left(\max_{i\in [n]}\beta_i v_i\right) \mathrm{d}\Bar{Q} - \sum_{i=1}^n B_i \log \beta_i
        \right\}
        .
\end{equation*}
The relationship between the two versions of the dual program is that we replaced the uniform averaging over samples in \eqref{eq: EG-dual} with an integral \textit{w.r.t.} the value distribution $\Bar{Q}$ in \eqref{eq: inf-EG-dual}. When the value space is continuous, the supply, allocation and pricing functions are measurable functions over the space in \eqref{eq: inf-EG-dual}, which can be infinite-dimensional. In the stochastic case where item arrivals are drawn from a distribution with density $s$, they correspond to an \textit{underlying market} with item supply $s$; the underlying dual program will serve as the reference programs that facilitate the analyze of dual convergence and regret. The rigorous mathematical treatment of infinite-dimensional markets can be found in \citet{gao2023infinite,gao2021online}.

\subsection{Interpretation from PACE to Dual Averaging}
\label{appendix: PACE-as-DA}
We review the stochastic dual averaging setup by \citet{xiao2009dual}. Consider the stochastic optimization problem:
\begin{equation}
    \label{eq: da-optimization}
    \min_{w \in \mathcal{F}} \left\{
    \phi(w) := \mathbb{E}_{z\sim \Pi}\left[F(w,z)\right] = \mathbb{E}_{z\sim \Pi}\left[f(w,z)\right]+\Psi(w)
    \right\},
\end{equation}
where $\Pi$ is distribution over $\mathcal{Z}$, and $f(\cdot, z)$ and $\Psi$ are both convex functions on domain $\mathcal{F}$ for any $z \in \mathcal{Z}$. In a more general online optimization setting, at each time $\tau$ we choose an action $w^\tau$ before an unknown convex loss function $f^\tau(\cdot) := f(\cdot, z^\tau)$ arrives, where $z^\tau$ is generated in some stochastic or adversarial manner. The goal is to minimize \textit{regret} when comparing the action sequence to any fixed action $w\in \mathcal{F}$ in hindsight. The regret with respect to a static solution $w$ is defined as 
\begin{equation*}
    R_t(w) := \sum_{\tau=1}^t \left(f^\tau(w^\tau ) + \Psi(w^\tau)\right) - \sum_{\tau=1}^t \left(f^\tau(w ) + \Psi(w)\right).
\end{equation*}
The dual averaging algorithm (DA) assumes access to the subgradients, i.e., given any $f^\tau$ and $w\in \mathcal{F}$, we can compute a subgradient $g^\tau\in \partial f^\tau(w)$ efficiently. DA initializes with $w_1 \in \mathcal{F}$ and $\Bar{g}^0 = 0$, and for each $\tau = 1, 2, \cdots$ it performs the following steps:
\begin{enumerate}
    \item [(1)] Observe $f^\tau$ and compute a gradient $g^\tau \in \partial f^\tau(w^\tau)$.
    \item [(2)] Update the time-averaged subgradient (the \textit{dual average}) via $\Bar{g}^\tau = \frac{\tau-1}{\tau}\Bar{g}^{\tau-1} + \frac{1}{\tau}g^\tau$.
    \item [(3)] Compute the next iterate $w^{\tau+1} = \arg \min_{w\in\mathcal{F}}\left\{ \langle
    \Bar{g}^\tau, w \rangle + \Psi(w)\right\}$.
\end{enumerate}

Let us see how PACE, i.e. \Cref{alg: PACE}, can be cast as dual-averaging on the EG dual \eqref{eq: EG-dual}. Let
\begin{align*}
f^{\tau}(\beta ) &= f(\beta, v^\tau) = \max_i \beta_i v_i, \\
    \Psi(\beta) &= - \sum_{i=1}^n B_i \log \beta_i,\\
    u^\tau &= \left(u_1^\tau, \cdots, u_n^\tau \right) \in \partial f^{\tau}(\beta^\tau).
\end{align*}
The interpretation is shown in detail in \Cref{section: PACE-as-DA}. A key observation for PACE is that the subgradient in the dual space coincides with the primal utility vector, so the time-averaged utility is exactly the dual average for the dual problem. It is also worth noting that PACE does not deploy an auxiliary regularizer; $\Psi(\beta)=- \sum B_i \log \beta_i$ is a part of the dual objective itself and acts as a regularizer. 

We stress that, despite the interpretation from \Cref{alg: PACE} to DA, the existing theoretical results for dual averaging are in no way applicable to our problem. The results by \citet{xiao2009dual} that allow \citet{gao2021online} to derive bounds for PACE strictly rely on the assumptions that $\mathcal{F}$ is closed, $\Psi$ is bounded and strongly convex. We remark that none of these assumptions holds in our setting:
\begin{itemize}
    \item In our problem, the domain for $\beta$ is $\mathbb{R}_{++}^n$, which is \textit{open}. 
    \item In our problem, the regularizer $\Psi(\beta) = - \sum_i B_i \log \beta_i$ is strictly convex, but \textit{not strongly convex} on the domain $\mathbb{R}_{++}^n$. $\Psi$ is not bounded on $\mathbb{R}_{++}^n$ either.
\end{itemize}

The analysis in \citet{xiao2009dual} suffers a direct breakdown with either of these problems, and is technically challenging to adapt to our problem. \citet{gao2021online} tried to handle the misalignment by enforcing a projection of $\beta$ to a fixed bounded domain. This projection solves the above issues, as $\Psi(\beta)$ becomes bounded and strongly convex when the domain becomes bounded.
However, that approach only works with normalized agent budgets, and must be set with knowledge of the valuation range of each agent. Moreover, the projection is extremely vulnerable against adversarial input, which makes it impossible to obtain any guarantees without introducing restrictive assumptions; see further discussions in \Cref{section: failure}. 
In order to avoid projection, we develop a new dual-averaging-like algorithm, which initially sets some variables to positive infinity. We then leverage the boundedness of the primal problem in order to show that, after a logarithmic number of rounds, we have \textit{high-probability implicit bounds}. We then show that the dual averaging analysis from \citet{xiao2009dual} can be applied to the remaining $t-O(\log t)$ rounds, in order to achieve similar guarantees as in the constrained setting.
Our ``implicit bounds'' approach may be of independent interest for dealing with unbounded online optimization problems; see the detailed analysis in \Cref{section: stationary}.
Next we show how to cast PACE as dual averaging applied to the problem \eqref{eq: EG-dual}.

\subsection{Inputs that Cause Failure of Constrained PACE}
\label{section: failure}
We show that the constrained version of PACE algorithm~\citep{gao2021online,liao2022nonstationary} is vulnerable either when 1) the input value is adversarially chosen, or 2) agent values and budgets are not normalized in the stochastic case. It turns out that simple adversarial construction result in the constrained algorithm allocating nothing to one of the agent at all, even when the hindsight optimal allocation is a equal division.

The constrained PACE algorithm is similar to \Cref{alg: PACE}, the only difference being it projects the pacing multiplier to a fixed interval at each time-step:
\begin{equation*}
    \beta_i^{\tau+1} = \mathrm{Proj}_{\left[\frac{B_i}{1+\delta_0}, (1+\delta_0)\right]} \frac{B_i}{\Bar{u}_i^\tau}.
\end{equation*}

We show that, the projection causes the failure of the algorithm for any fixed (independent of $t$) target interval $[\ell_i, r_i]$, where $0\leq \ell_i < r_i < \infty$ . 
\begin{example}
\label{example: fail}
    Consider 2-agent instances with equal weights $B_1 = B_2 = 1$. For constrained PACE algorithm that projects agent multipliers to $[\ell_1, r_1]$, $[\ell_2, r_2]$ (assuming $r_2 \leq r_1$ without loss of generality), consider the following input instance:
    \begin{equation*}
        v_1^\tau = v_2^\tau = \min \left\{\frac{1}{r_2}, \|v\|_{\infty}\right\}, \ \forall \tau \in [t].
    \end{equation*}
    Then for constrained PACE, clearly we have for every time step,
    \begin{equation*}
        \beta_1^{\tau+1} = \beta_2^{\tau+1} =  r_2, \ \forall \tau \in [t].
    \end{equation*}
    Every item is allocated to agent $1$; agent $2$ has final utility $0$.
\end{example}

\Cref{example: fail} can be regarded both as an adversarially constructed input, or as a stationary input where each round is \textit{i.i.d.} with the distribution being degenerate. In fact, a fixed upper bound in the projection interval of constrained PACE requires prior knowledge or normalization on $B_i$ and agent values; otherwise it is easily corrupted otherwise by such simple instances as \Cref{example: fail}.

\section{Results for Online Fisher Market}
\label{section: fisher}

\subsection{Setup}
We review the linear Fisher market model~\citep{varian1974equity}; the corresponding convex optimization programs are discussed extensively in \citet{cole2017convex}.

For input \(\gamma=(v_i^\tau)_{i\in[n],\tau\in[t]}\), the associated linear Fisher market has the same \(n\) agents as buyers and the same \(t\) items as goods. Each item has unit supply. The vector \(B\) now has the interpretation of a per-round budget vector in faux-currency, so buyer \(i\)'s total budget over the \(t\) items is \(tB_i\). Additionally, each item has a \textit{price} $p^\tau \in \mathbb{R}_+$. Given item prices \(p=(p^1,\ldots,p^t)\in\mathbb R_+^t\), the expenditure on bundle \(z_i\) is \(\langle p,z_i\rangle\). The \textit{demand} of an agent under $p$ is the set of all utility maximizing bundles with expenditure not exceeding the budget constraint $t B_i$.  

\begin{definition}[Demand]
Given item prices \(p\in\mathbb R_+^t\), the demand of buyer \(i\) is the set of utility-maximizing affordable bundles:
\begin{equation}
    \label{eq: demand}
    D_i(p)
    =
    \arg\max_z
    \left\{
        \langle v_i,z_i\rangle:
        0\le z_i\le \mathbf{1}_t,\ 
        \langle p,z_i\rangle\le tB_i
    \right\}.
\end{equation}
\end{definition}

\begin{definition}[Market equilibrium]
A market equilibrium for the Fisher market induced by \(\gamma\) is an allocation-price pair \((x,p)\in\mathcal X\times\mathbb R_+^t\) such that the following conditions hold:
\begin{itemize}
    \item Supply feasibility: \(\sum_{i=1}^n x_i\le \mathbf{1}_t\).
    \item Buyer optimality: \(x_i\in D_i(p)\) for all \(i\in[n]\).
    \item Market clearance: \(\left\langle p,\mathbf{1}_t-\sum_{i=1}^n x_i\right\rangle=0\).
\end{itemize}
\end{definition}

This Fisher market interpretation is closely connected to fair allocation. In the equal-weight case, the market equilibrium is the classical \textit{competitive equilibrium from equal incomes} (CEEI): every agent is endowed with the same artificial purchasing power, and the resulting allocation is determined by preferences and market-clearing prices.

The market equilibrium allocation is computed by the same hindsight Eisenberg--Gale program \eqref{eq: EG-primal}~\citep{eisenberg1959consensus,cole2017convex}. Moreover, if \(\beta^\gamma\) is an optimal solution to the normalized dual program \eqref{eq: EG-dual}, then the corresponding equilibrium price of item \(\tau\) is
\begin{equation}
    \label{eq: hindsight-equilibrium-price}
    p^{\gamma,\tau}
    =
    \max_{i\in[n]}\beta_i^\gamma v_i^\tau .
\end{equation}
Together with an EG-optimal allocation \(x^\gamma\), these prices form a market equilibrium. Similarly, the underlying programs \eqref{eq: inf-EG-primal} and \eqref{eq: inf-EG-dual} define the competitive equilibrium of the infinite-dimensional Fisher market induced by \(\bar Q\).

\subsection{Performance Metrics for Online Market Equilibrium}

To interpret PACE in the online Fisher market setting where items arrive sequentially, we need to define its induced prices of the items. Recall the dual interpretation in \Cref{section: algorithm} gives us $p^\tau
    =
    \max_{i\in[n]}\beta_i^\tau v_i^\tau$. The expenditure of buyer $i$  on item \(\tau\) is $ b_i^\tau
    =
    p^\tau x_i^\tau $. 
We write \(b^\tau=(b_i^\tau)_{i\in[n]}\). Since PACE uses integral allocations, only the winner of item \(\tau\) has positive expenditure in that round.

The first performance notion we study is approximate budget clearance. In an exact Fisher market equilibrium, each buyer with at least one non-zero valuation spends the budget. In our online setting, buyer \(i\)'s cumulative benchmark budget is \(tB_i\), so we measure whether the total expenditure vector \(\sum_{\tau=1}^t b^\tau\) is close to \(tB\).

The second market-equilibrium condition is approximate buyer optimality, or \textit{buyer regret}. After observing the realized PACE prices \(p=(p^1,\ldots,p^t)\), buyers can compare the received bundles from PACE with their respective demand, i.e., the best bundles that could have been purchased in hindsight given these prices and budgets. Concretely, the buyer regret of the buyer $i$ can be defined as
\begin{equation*}
    \label{eq: buyer-regret}
    \text{Buyer-Regret}_{i} = \left(
    \max
    \left\{
        \langle v_i,z_i\rangle:
        0\le z_i\le \mathbf{1}_t,\ 
        \langle p,z_i\rangle\le tB_i
    \right\}
    -
    U_i^t
    \right)^+ .
\end{equation*}

\subsection{Results for Online Market Equilibrium}
The missing proofs in this subsection can be found in \Cref{proof: iid}. We first show that PACE approximately clears buyers' budgets under stationary input. The guarantee is conditional on the same high-probability event used in the multiplier convergence analysis, and it excludes the initial \(\Hat{s}\) rounds where the induced prices may be unbounded due to zero cumulative utilities. 

\begin{restatable}[Budget clearance under stationary input]{theorem}{iidthmC}
    \label{thm: convergence-spending-iid}
    Under stationary input, let \(\Hat{s}\) and \(\Hat{A}\) be specified as in \Cref{lem: high-prob-bounds-iid}. It holds that
    \begin{equation*}
        \mathbb{E}
        \left[
        \left\|
        \sum_{\tau=\Hat{s}+1}^t b^\tau
        -
        tB
        \right\|
        \mid \Hat{A}
        \right]
        =
        O
        \left(
        n^2\sqrt{\log n}\,\log t\cdot\sqrt{t}
        \right).
    \end{equation*}
\end{restatable}

Theorem~\ref{thm: convergence-spending-iid} shows that the cumulative spending vector is close to the cumulative budget vector, up to a sublinear error. Thus, in the time-averaged sense, PACE asymptotically clears buyers' budgets.

We next show that PACE also approximately satisfies buyer optimality. Given the realized price sequence generated by PACE, each buyer's received utility is close to the best utility that could have been obtained from any affordable hindsight bundle.

\begin{restatable}[Buyer regret under stationary input]{theorem}{iidthmD}
    \label{thm: regret-and-envy}
    Under stationary input, for every buyer \(i\in[n]\), it holds that
    \begin{equation*}
        \mathbb{E}
        \left[
        \textnormal{Buyer-Regret}_i
        \right]
        =
        O
        \left(
        n^2\sqrt{\log n}\,\log t\cdot\sqrt{t}
        \right).
    \end{equation*}
\end{restatable}

Theorem~\ref{thm: regret-and-envy} gives a sublinear additive buyer regret guarantee. Equivalently, after normalizing by the horizon length \(t\), every buyer asymptotically attains her demand utility level under the PACE-induced prices. Together with \Cref{thm: convergence-spending-iid}, this shows that PACE converges to the competitive equilibrium interpretation of the Eisenberg--Gale benchmark under stationary input.

The same market equilibrium guarantees extend to the nonstationary input models in Section~5. The proof follows the same route as the previous nonstationary analysis: the deterministic implicit bounds argument is unchanged, while the conditional-regret step is adjusted using the corresponding total variation estimates for ergodic and blockwise independent inputs.

\begin{theorem}[Online market equilibrium under nonstationary input]
    \label{thm: market-equilibrium-nonstationary}
    Under ergodic or blockwise independent inputs, there exist \(\Hat{A}\subseteq\Omega\) and \(\Hat{s}=O(\log t)\), with \(\Pr[(\Hat{A})^c]=O(1/t)\), such that the following holds. For budget clearance,
    \begin{equation*}
        \mathbb{E}
        \left[
        \left\|
        \sum_{\tau=\Hat{s}+1}^t b^\tau
        -
        tB
        \right\|
        \mid \Hat{A}
        \right]
        =
        \begin{cases}
        O
        \left(
        n^2\sqrt{\iota\log n}\,\log t\cdot\sqrt{t}
        +
        \sqrt{n^3\delta}\cdot t
        \right),
        & \gamma\sim Q\in \mathsf{C}^\mathrm{E}(\delta,\iota),\\[0.6em]
        O
        \left(
        n^2|\mathcal P|_\infty\sqrt{\log n}\,\log t\cdot\sqrt{t}
        +
        \sqrt{n^3\delta}\cdot t
        \right),
        & \gamma\sim Q\in \mathsf{C}^\mathrm{B}(\mathcal P,\delta).
        \end{cases}
    \end{equation*}
    Moreover, for every buyer \(i\in[n]\), the buyer regret satisfies
    \begin{equation*}
        \mathbb{E}
        \left[
        \textnormal{Buyer-Regret}_{i}
        \right]
        =
        \begin{cases}
        O
        \left(
        n^2\sqrt{\iota\log n}\,\log t\cdot\sqrt{t}
        +
        \sqrt{n^3\delta}\cdot t
        \right),
        & \gamma\sim Q\in \mathsf{C}^\mathrm{E}(\delta,\iota),\\[0.6em]
        O
        \left(
        n^2|\mathcal P|_\infty\sqrt{\log n}\,\log t\cdot\sqrt{t}
        +
        \sqrt{n^3\delta}\cdot t
        \right),
        & \gamma\sim Q\in \mathsf{C}^\mathrm{B}(\mathcal P,\delta).
        \end{cases}
    \end{equation*}
\end{theorem}

%% file: best-of-many-worlds/proof-iid.tex
\section{Missing Results and Proofs in \Cref{section: stationary}}
\label{proof: iid}
\subsection{Proof of \Cref{thm: convergence-multiplier-iid}}
\iidthmZ*

\begin{proof}
We show how to combine \Cref{lem: deterministic-mean-square}, \Cref{lem: dual-regret-iid}, and \Cref{lem: high-prob-bounds-iid} to prove \Cref{thm: convergence-multiplier-iid}. The proofs of these main lemmas can be found in subsequent subsections.

With the specification of a high-probability event $\Hat{A}$ in \Cref{lem: high-prob-bounds-iid}, we apply \Cref{lem: deterministic-mean-square} with $h = \Hat{h}(n)$ and $s = \Hat{s}(t,n)$ defined as in \Cref{lem: high-prob-bounds-iid}. For time step $r$ with $\Hat{s}(t,n)<r \leq t$, we have
\begin{align}
    \gamma \in \Hat{A}  &\implies \nonumber\\ 
    \|\beta^{r+1}-\beta^*\|^2
    &\leq \frac{2}{\Hat{\sigma}(n) \cdot r} \left[\frac{\max_i\|v_i\|_{\infty}^2}{2\Hat{\sigma}(n)} \log r + \left(\Bar{F}(\beta^*) - \Psi(\Hat{h}(n))\right) \cdot \Hat{s}(t,n)
    -\textnormal{Dual-Regret}(\Hat{s}(t,n),\beta^*)
    \right],\label{eq: deterministic-mean-square-specified}
\end{align}

We investigate the asymptotic order of each item,
\begin{align}
    \Hat{\sigma}(n) & = \min_{i\in[n]} \left\{B_i / (\Hat{h}_i(n))^2\right\}  = \Omega(1/n^2), \nonumber \\
    \Bar{F}(\beta^*)-\Psi(\Hat{h}(n)) &= \max_i \beta_i^* \|v_i\|_{\infty} + \sum_{i}B_i\log \frac{\Hat{h}_i}{\beta_i^*}  = O(n \log n), \label{eq: asymptotic-items} \\
    G(\Hat{h}(n)) &= \max_{\beta\in \mathcal{F}_{\Hat{h(n)}}} \left\{\max_i \beta_i\|v_i\|_\infty - \sum_{i}B_i \log \beta_i \right\} = O(n). \nonumber
\end{align}

For the dual regret term, by \Cref{lem: dual-regret-iid},
\begin{equation}
    \label{eq: dual-regret-iid-specified}
       - \mathbb{E}\left[\textnormal{Dual-Regret}(\Hat{s}(t,n),\beta^*)\mid \Hat{A} \right]  \leq 4\cdot \left(1-\Pr\left[\Hat{A}\right]\right)\cdot t \cdot G(\Hat{h}(n)) = O(n).
\end{equation}

Taking conditional expectation on \eqref{eq: deterministic-mean-square-specified}, and noticing that the second term is dominant asymptotically, we get
\begin{equation}
    \label{eq: deterministic-mean-square-r}
    \mathbb{E}\left[ \|\beta^{t+1}-\beta^*\|_2^2 \mid \Hat{A}\right ] = O\left(\frac{n^4 \cdot \log n \cdot (\log n + \log t )}{r}\right), \ \forall r \in \{\Hat{s}(t,n)+1, \cdots, t\}.
\end{equation}
Setting $r = t$ in \eqref{eq: deterministic-mean-square-r} and noticing that $t\geq n$ by assumption, we prove the desired bound for $\mathbb{E}\left[ \|\beta^{t+1}-\beta^*\|_2^2 \mid \Hat{A}\right ]$. The proof for $\mathbb{E}\left[ \|\beta^{t+1}-\beta^*\|_2^2 \mid \Hat{A}\right ]$ is similar. 

\end{proof}
\subsection{Proof of \Cref{thm: convergence-utility-iid}}
\iidthmY*
\begin{proof}
By the KKT conditions $u_i^* = B_i / \beta_i^*$ and the algorithm's update rule $\bar u_i^t  = B_i / \beta_i^{t+1}$, we have
\begin{equation}
    \label{eq: multiplier-utility-translation}
    \|\Bar{u}^t - u^*\|^2 = \sum_{i=1}^n B_i^2 \left(\frac{1}{\beta_i^{t+1}} - \frac{1}{\beta_i^*}\right)^2 
    \leq \sum_{i=1}^n \frac{\|v_i\|_\infty^2}{B_i^2}\left(\beta_i^{t+1} -\beta_i^*\right)^2
    \leq \max_i \frac{\|v_i\|_\infty^2}{B_i^2} \|\beta^{t+1} - \beta^*\|^2.
\end{equation}    
The first inequality holds uses the fact that bot $\beta_i$ and $\beta_i^*$ are no smaller than $B_i / \|v_i\|_\infty$. 
Let $\Hat{A}$ be given as in \Cref{lem: high-prob-bounds-iid}. By \Cref{thm: convergence-multiplier-iid},
\begin{align*}
    \mathbb{E}\left[ \|\Bar{u}^t - u^*\|^2 \right]
    &= \mathbb{E}\left[ \|\Bar{u}^t - u^*\|^2 \mid \Hat{A}\right] \Pr\left[\Hat{A}\right] + \mathbb{E}\left[ \|\Bar{u}^t - u^*\|^2 \mid ({\Hat{A}})^c \right] \Pr\left[({\Hat{A}})^c\right]\\
    &\leq \max_i \frac{\|v_i\|_\infty^2}{B_i^2} \cdot \mathbb{E}\left[ \|\beta^{t+1} - \beta^*\|_2^2\mid \Hat{A}\right] + n\max_i\|v_i\|_{\infty}^2 \cdot (1-\Pr[\Hat{A}]) \\ 
    &= O\left(\frac{n^4 \log n \log t}{t}\right).
\end{align*}
The proof for $\mathbb{E}\left[ \|\Bar{u}^t - u^\gamma\|^2 \right]$ is similar. 

\end{proof}
\subsection{Proof of \Cref{thm:stationary-fairness-metrics}}
\iidFair*
\begin{proof}

We begin with asymptotic proportionality. By \Cref{prop:hindsight-benchmark-properties} we know that the hindsight program gives exact proportionality, 
\begin{equation*}
    U_i^\gamma\geq \frac{B_i}{\|B\|_1} \sum_{\tau=1}^t v_i^\tau. 
\end{equation*}
Therefore, we have
\begin{equation*}
    \eta_{\mathrm{Prop}}^t \leq  \max \left\{0, U_i^t - \frac{B_i}{\|B\|_1} \sum_{\tau=1}^t v_i^\tau \right\} \leq U_i^t -  U_i^\gamma  = t \cdot  (\bar u_i^t - u_i^\gamma). 
\end{equation*}
Taking expectations and applying Jensen's inequality, we have
\begin{equation}
    \label{eq: proportionality-decomposition}
    \mathbb{E}[\eta_{\mathrm{Prop}}^t] \leq \sqrt{\mathbb{E}[(\eta_{\mathrm{{Prop}}}^t)^2]} \leq \sqrt{t^2 \mathbb{E}(\bar u_i^t - u_i^\gamma)^2} = O\left(n^2 \sqrt{(\log n) (\log t) t}\right),
\end{equation}
where the last step is by the utility convergence result in \Cref{thm: convergence-utility-iid}. 

Next we prove asymptotic envy-freeness. We have the following deterministic upper-bound on the squared average envy term; the proof of \Cref{lem: envy} will be shown later. 
\begin{lemma}
Let $E_{ik}:= \left(\frac{\Bar{u}_{ik}^t}{B_k} - \frac{\Bar{u}_{i}^t}{B_i}\right)^2$ denote the squared time-averaged envy of agent $i$ to $k$. For any $s<t$ and $i, k \in [n]$ it holds that
\label{lem: envy}
    \begin{equation*}
 E_{ik} \leq \frac{4}{B_i^2}\left|\Bar{u}_i^t - u_i^*\right|^2 + 
\frac{4s^2 \left(\beta_i^*\right)^2 \|v_i\|_{\infty}^2}{t^2 B_i^2 B_k^2}
+\frac{4\|v_i\|_{\infty}^2}{B_i^2B_k^2}\left|\frac{1}{t}\sum_{\tau = s+1}^{t} b_k^\tau - B_k\right|^2 + \frac{4\|v_i\|_{\infty}^2\|v\|_{\infty}^2}{B_i^2B_k^2}\cdot \frac{1}{t} \cdot \sum_{\tau = s+1}^t \|\beta^\tau - \beta^*\|^2,
\end{equation*}
where $b_k^\tau = x_k^\tau \cdot\beta_k^\tau \cdot  v_k^\tau$. 
\end{lemma}

We bound each term in \Cref{lem: envy} under conditional expectation. Notice that by \eqref{eq: deterministic-mean-square-r}, we have for any $\tau \in \{\Hat{s}(t,n)+1, \cdots, t\}$,
\begin{equation*}
    \mathbb{E}\left[ \|\beta^\tau - \beta^*\|^2 \mid \Hat{A}\right] = O\left(\frac{n^4 \log n \log t}{\tau}\right).
\end{equation*}
Hence, the final summation term satisfies,
\begin{equation}
    \label{eq: sum-multipliers}
    \sum_{\Hat{s}({t,n})+1}^t \mathbb{E}\left[ \|\beta^\tau - \beta^*\|^2 \mid \Hat{A}\right] = O\left({n^4 \log n (\log t)^2}\right).
\end{equation}

From \Cref{thm: convergence-utility-iid} and \Cref{thm: convergence-spending-iid} we know that conditional on $\Hat{A}$, there is an $O(n^4 \log n (\log t)^2/t)$ bound on the other terms. Hence, 
\begin{equation*}
    \mathbb{E}\left[\max_{i, k \in [n]}E_{ik}\mid \Hat{A}\right] = O\left(\frac{n^4 \log n (\log t)^2}{t}\right).
\end{equation*}
Therefore, we know that for any $i, k\in [n]$,
\begin{equation}
\label{eq: envy-decomposition-0}
\begin{aligned}
    \mathbb{E}\left[\max_{i, k \in [n]}E_{ik}\right] &= \mathbb{E}\left[\max_{i, k \in [n]}E_{ik}\mid \Hat{A}\right] \cdot \Pr[\Hat{A}] + \mathbb{E}\left[\max_{i, k \in [n]}E_{ik}\mid \left(\Hat{A}\right)^c\right]\cdot (1-\Pr[\Hat{A}])\\
    &\leq \mathbb{E}\left[\max_{i, k \in [n]}E_{ik}\mid \Hat{A}\right] + \frac{\|v_i\|_{\infty}^2}{\min_{j}B_j^2} \cdot (1-\Pr[\Hat{A}])\\
    &= O\left(\frac{n^4 \log n (\log t)^2}{t}\right).
\end{aligned}
\end{equation}

By the definition of additive $\eta_{\mathrm{EF}}^t$-envy-freeness and Jensen's inequality, we have
\begin{equation}
\label{eq: envy-decomposition-1}
    \mathbb{E}[\eta_{\mathrm{EF}}^t] \leq \sqrt{\mathbb{E}[(\eta_{\mathrm{{EF}}}^t)^2]} \leq t \cdot \sqrt{\mathbb{E}\left[\max_{i, k \in [n]}E_{ik}\right]} = O(n^2 \sqrt{\log n} \, \log t \sqrt{t}).  
\end{equation}

This gives the desired envy-freeness approximation. To complete the proof, it suffices to prove \Cref{lem: envy}. We show \Cref{lem: envy} as follows. 

\paragraph{{Proof of \Cref{lem: envy}}.}
For each $\tau$, decompose
\begin{equation}
    \label{eq: envy-1}
    p^{*,\tau}x_k^\tau = b_k^\tau + \left(p^{*, \tau}-\beta_k^\tau v_k^\tau \right) x_k^\tau,
\end{equation}
where $p^{*, \tau} = \max_{i\in [n]} \beta_i^* v_i^\tau$ for each $\tau \in [t]$. We then have
\begin{align*}
    \beta_i^* \Bar{u}_{ik}^t 
    &=
    \frac{1}{t}\sum_{\tau=1}^t \beta_i^* v_i^\tau x_i^\tau \\
    &=
    \frac{1}{t}\sum_{\tau=1}^{s} \beta_i^* v_i^\tau x_i^\tau
    +
    \frac{1}{t}\sum_{\tau = s+1}^{t} \beta_i^* v_i^\tau x_i^\tau \\
    &\overset{\text{(a)}}{\leq} \frac{s \beta_i^* \|v_i\|_\infty}{t}
    +
    \frac{1}{t}\sum_{\tau = s+1}^{t} p^{*,\tau}x_k^\tau \\
    &\overset{\text{(b)}}{=} \frac{s \beta_i^* \|v_i\|_\infty}{t}
    +
    \frac{1}{t}\sum_{\tau = s+1}^{t} b_k^\tau + \frac{1}{t}\sum_{\tau = s+1}^{t} \left(p^{*,\tau} - \beta_k v_k^\tau \right)x_k^\tau\\
    &\leq \frac{s \beta_i^* \|v_i\|_\infty}{t}
    +
    B_k + \left|\frac{1}{t}\sum_{\tau = s+1}^{t} b_k^\tau - B_k\right| + \frac{1}{t}\sum_{\tau = s+1}^{t} \left(p^{*,\tau} - \beta_k v_k^\tau \right)x_k^\tau,
\end{align*}
where step (a) is by the definition of $p^{*, \tau}$, and step (b) is by \eqref{eq: envy-1}. Using the KKT condition $\beta_i^*u_i^* = B_i$, we further have
\begin{align*}
    \frac{\Bar{u}_{ik}^t}{B_k}&\leq
    \frac{u_i^*}{B_i}\left(1+ \frac{s\beta_i^*\|v_i\|_{\infty}}{t\cdot B_k}+\frac{1}{B_k}\left|\frac{1}{t}\sum_{\tau = s+1}^{t} b_k^\tau - B_k\right| + \frac{1}{tB_k}\sum_{\tau = s+1}^{t} \left(p^{*,\tau} - \beta_k v_k^\tau \right)x_k^\tau\right)\\
    &\leq
    \frac{\Bar{u}_i^t}{B_i}+\frac{1}{B_i}\left|\Bar{u}_i^t - u_i^*\right| + 
    \frac{s\beta_i^* \|v_i\|_{\infty}}{t\cdot B_i B_k} + 
    \frac{\|v_i\|_{\infty}}{B_iB_k}\left|\frac{1}{t}\sum_{\tau = s+1}^{t} b_k^\tau - B_k\right| + \frac{\|v_i\|_{\infty}}{B_iB_k}\cdot\frac{1}{t}\sum_{\tau = s+1}^{t} \left(p^{*,\tau} - \beta_k v_k^\tau \right)x_k^\tau.
\end{align*}

Rearranging terms, and using the inequality $(x+y+z+w)^2 \leq 4(x^2+y^2+z^2+w^2)$ for any $x, y, z, w \in \mathbb{R}$, 
\begin{equation}
\label{eq: envy-3}
E_{ik} \leq \frac{4}{B_i^2}\left|\Bar{u}_i^t - u_i^*\right|^2 + 
\frac{4s^2 \left(\beta_i^*\right)^2 \|v_i\|_{\infty}^2}{t^2 B_i^2 B_k^2}
+\frac{4\|v_i\|_{\infty}^2}{B_i^2B_k^2}\left|\frac{1}{t}\sum_{\tau = s+1}^{t} b_k^\tau - B_k\right|^2 + \frac{4\|v_i\|_{\infty}^2}{B_i^2B_k^2} \left[\frac{1}{t}\sum_{\tau = s+1}^{t} \left(p^{*,\tau} - \beta_k v_k^\tau \right)x_k^\tau\right]^2.
\end{equation}
For the last term, notice that
\begin{align*}
    \left(\frac{1}{t}\sum_{\tau = s+1}^{t} \left(p^{*,\tau} - \beta_k v_k^\tau \right)x_k^\tau \right)^2
    &\leq  \left(\frac{1}{t}\sum_{\tau = s+1}^{t} \left|\left(p^{*,\tau} - \beta_k v_k^\tau \right)x_k^\tau\right|\right)^2\\
    &\leq \left(\frac{1}{t}\sum_{\tau = s+1}^{t} \left|p^{*,\tau} - \beta_{i^\tau} v_{i^\tau}^\tau \right|\right)\\
    &\leq \left(\frac{1}{t}\sum_{\tau = s+1}^{t} \|v\|_{\infty}\|\beta_i^\tau - \beta^*\|_\infty\right)^2\\
    &\leq \frac{\|v\|_{\infty}^2}{t} \sum_{\tau=s+1^t}\|\beta_i^\tau-\beta^*\|^2,
\end{align*}
where the last step is by the convexity of $(\cdot)^2$. Combining with \eqref{eq: envy-3}, this proves \Cref{lem: envy}.

\end{proof}

\subsection{Proof of \Cref{thm:stationary-efficiency-metrics}}
\iidEff*

\begin{proof}
    We first bound the expectation of Nash welfare suboptimality. Since $\hat h =\hat h (n) $ is independent of $t$, we know that conditional on $\hat A$, each $\bar u_i^t$ is lower-bounded by a $t$-independent constant, and upper-bounded by $\|v\|_\infty$. Let $L_{\hat h}$ be the Lipschitzness parameter (\textit{w.r.t.} the $\ell_2$-norm) of the weighted gemetric-mean function on this bounded area. We then have
    \begin{equation*}
    \gamma \in \hat A \implies \eta_{\mathrm{NW}}^t \leq L_{\hat h} \cdot \|U^t - U^\gamma\| = t L_{\hat h} \|\bar u^t - u^\gamma\|. 
    \end{equation*}
    By a conditional expectation argument and Jensen's inequality, we have
    \begin{equation}
        \begin{aligned}
    \mathbb{E}[\eta_{\mathrm{NW}}^t] &\leq t \|v\|_\infty \cdot \mathrm{Pr}[(\hat A)^c] + \mathbb{E}\left[ \eta_{\mathrm{NW}}^t \cdot \mathbb{I}\{\hat A\}\right] \\
    &\leq t \|v\|_\infty \cdot \mathrm{Pr}[(\hat A)^c] +  \sqrt{\mathbb{E}\left[ (\eta_{\mathrm{NW}}^t)^2 \cdot \mathbb{I}\{\hat A\}\right]}\\
    &\leq t \|v\|_\infty \cdot \mathrm{Pr}[(\hat A)^c] +  t L_{\hat h} \mathbb E\left[\|\bar u^t - u^\gamma\|^2\right]\\
    & = O\left(n^2 \sqrt{(\log n) (\log t) \, t}\right),
        \end{aligned}
            \label{eq: efficiency-nw-decomposition}
    \end{equation}
    where we applied \Cref{thm: convergence-utility-iid} in the last step. 

    We next show the asymptotic core stability guarantee. We argue that the allocation is \(\eta^t_{\mathrm{Core}}\)-core stable for $\eta^t_{\mathrm{Core}}(\gamma)=\|\bar u^t-u^\gamma\|_\infty$. Suppose not. Then there exist a nonempty coalition $S\subseteq [n]$ and re-allocation $(y_i)_{i\in S}$, such that
\[
\sum_{i\in S} y_i^\tau \le \frac{B(S)}{\|B\|_1} \ \  \ \  \forall \tau\in [t],  \ \ U_i(y_i) > U_i^t + \| U^t-U^\gamma\|_\infty, \ \ \ \forall i\in S,
\]
with at least one strict inequality for some $i\in S$. 
But for every $i\in[n]$, we have $U_i^\gamma \le U_i^t + \|U^t-U^\gamma\|_\infty$. Therefore, the above re-allocation satisfies $U_i(y_i)\geq U_i^\gamma, \forall i\in S$, with at least one strict inequality. This contradicts the exact core stability of the hindsight solution. 

By norm monotonicity,
\begin{equation}
    \bigl(\eta^t_{\mathrm{Core}}(\gamma)\bigr)^2
=
\|U^t-U^\gamma\|_\infty^2
\le
\|U^t-U^\gamma\|_2^2 = t^2\|\bar u^t - u^\gamma\|_2^2.
\label{eq: core-stability-decomposition}
\end{equation}

Taking expectation and applying Jensen's inequality and \Cref{thm: convergence-utility-iid}, we get the desired asymptotic core stability result. 

\end{proof}
\subsection{Proof of \Cref{lem: deterministic-mean-square}}
\label{proof: deterministic-mean-square}
\iidlemA*

\noindent{\textit{Proof.}}\paragraph{Conjugate Function $V^t(w)$ and Its Properties.}
Notice that $\Psi(\beta)$ is monotonically decreasing on each $\beta_i$. We have
\begin{equation*}
    h = (h_1, \cdots, h_n) = \arg \min_{\beta\in \mathcal{F}_h} \Psi(\beta).
\end{equation*}
In this proof, we let $w^t:= \sum_{\tau=1}^t g^\tau$ be the sum of gradients up to time step $t$. For each $t\geq s$, we consider the following conjugate-type function:
\begin{equation*}
    V^t(w)=  \max_{\beta \in \mathcal{F}_h} \left\{\langle w, \beta-h\rangle - t \Psi(\beta)\right\}.
\end{equation*}
By comparing the definition $V^t(w)$ and the update rule $\beta^{t+1}_i = B_i/\bar u_i^t$, we have for any $t \geq s-1$, 
\begin{equation}
    \label{eq: update-rule-after-log-rounds}
    \beta^{t+1} = \arg \min_{\beta> 0} \left\{\langle w^t, \beta\rangle + t\Psi(\beta) \right\} \overset{\text{(a)}}{=} \arg \min_{\beta \in \mathcal{F}_h} \left\{\langle w^t, \beta\rangle + t\Psi(\beta) \right\} = \arg \max_{\beta \in \mathcal{F}_h} V^t(-w^t),
\end{equation}
where (a) is by the definition of the implicit bound event $A_{h,s}$.

For any $t\geq s$, $\Psi(\beta)$ is strongly convex on the domain $\mathcal{F}_h$, which is non-empty and closed. Therefore, the maximum is always achieved, and the maximizer is unique. Also, by $\Psi(h)\leq 0$, the function $V^t(w)$ is always non-negative. 

Let $\pi^t(w)$ denote the unique maximizer in the definition of $V^{t}(w)$. We have
\begin{align*}
    \pi^t(w) &=  \arg \max_{\beta\in \mathcal{F}_h}\left\{\langle w, \beta-h\rangle - t \Psi(\beta)\right\}\\
             &=  \arg \min_{\beta\in \mathcal{F}_h}\left\{\langle -w, \beta\rangle + t \Psi(\beta)\right\}
\end{align*}
Combined with \eqref{eq: update-rule-after-log-rounds}, we have
\begin{equation*}
    \beta^{t+1} = \pi^t(-w^t), \ \ \forall t \geq s-1.
\end{equation*}
\begin{lemma}
    \label{lem: gradient-of-conjugate}
    The function $V^t(w)$ is convex and differentiable. Its gradient is given by
    \begin{equation}
        \label{eq: gradient-of-conjugate}
        \nabla V^t(w) = \pi^t(w) - h.
    \end{equation}
    Moreover, the gradient is Lipschitz continuous with parameter $1/\sigma t$, i.e., 
    \begin{equation*}
        \|\nabla V^t(w_1)-\nabla V^t(w_2)\|\leq \frac{1}{\sigma t} \|w_1 - w_2\|_*, \ \ \forall w_1, w_2 \geq 0.
    \end{equation*}
\end{lemma}

The proof of \Cref{lem: gradient-of-conjugate} follows from classical results of convex analysis, see, for example, \citet{boyd2004convex}. 

A direct consequence of the Lipschitz gradients result in \Cref{lem: gradient-of-conjugate} is the following inequality:
\begin{equation}
    \label{eq: lipschiz-contiuous}
    V^t(w+g)\leq V^t(w) + \langle g, \nabla V^t(w)\rangle + \frac{1}{2 \sigma t} \|g\|_*^2, \ \ \forall w, g \geq 0.
\end{equation}
\begin{lemma}
    \label{lem: conjugate-inequality}
    For each $t\geq s$, we have
    \begin{equation*}
        V^t(-w^t) + \Psi(\beta^{t+1}) \leq V^{t-1}(-w^t).
    \end{equation*}
\end{lemma}
To see \Cref{lem: conjugate-inequality}, notice that by the definition of $V^{t-1}(-w^t)$,
\begin{align*}
    V^{t-1}(-w^t) &= \max_{\beta \in \mathcal{F}_h}\left\{\langle-w^t, \beta - h \rangle- (t-1)\Psi(\beta)\right\}\\
    &\geq \langle -w^t, \beta^{t+1}-h\rangle - (t-1) \Psi(\beta^{t+1})\\
    &= \left\{\langle -w^t, \beta^{t+1}-h\rangle - t \Psi(\beta^{t+1})\right\}+ \Psi(\beta^{t+1})\\
    &= V^{t}(-w^t)+ \Psi(\beta^{t+1}),
\end{align*}
where the last step is due to \eqref{eq: update-rule-after-log-rounds}. 

Now we have proved \Cref{lem: gradient-of-conjugate} and \Cref{lem: conjugate-inequality}, we continue the proof of \Cref{lem: deterministic-mean-square}.

\paragraph{Bounding the Dual Iterates.}
By the optimality condition for the decision rule of PACE , there exists a subgradient $d^{t+1} \in \partial \Psi(\beta^{t+1})$ such that
\begin{equation}
    \label{eq: optimality-condition}
    \langle w^t + t d^{t+1}, \beta - \beta^{t+1}\rangle \geq 0, \ \forall \beta > 0.
\end{equation}
Consider $t \geq s$. By the strong convexity of $\Psi(\beta)$ on $\mathcal{F}_h$, 
\begin{equation}
    \label{eq: strong-convexity-condition}
    \Psi(\beta) \geq \Psi(\beta^{t+1}) + \langle d^{t+1}, \beta - \beta^{t+1}\rangle + \frac{\sigma^2}{2}\|\beta - \beta^{t+1}\|^2, \ \ \forall \beta >0, t \geq s.
\end{equation}
By multiplying both sides of \eqref{eq: strong-convexity-condition} by $t$ and rearranging terms, we have for $t\geq s$, 
\begin{equation*}
    \frac{\sigma t}{2}\|\beta - \beta^{t+1}\|^2 \leq - \langle t d^{t+1}, \beta - \beta^{t+1}\rangle + t\Psi(\beta) - t \Psi(\beta^{t+1}).
\end{equation*}
Using the optimality condition \eqref{eq: optimality-condition}, we have for $t\geq s$,
\begin{align*}
    \frac{\sigma t}{2}\|\beta - \beta^{t+1}\|^2 &\leq \langle w^t, \beta - \beta^{t+1}\rangle +t\Psi(\beta) - t \Psi(\beta^{t+1})\\
    &= \left\{\langle -w^t, \beta^{t+1}-h\rangle - t \Psi(\beta^{t+1})\right\} + t\Psi(\beta) + \langle w^t, \beta - h\rangle\\
    &= V^t(-w^t) + t\Psi(\beta) + \langle w^t, \beta - h\rangle.
\end{align*}
Now we expand the last term on the right hand side, and split the horizon at round $s$,
\begin{align*}
    \frac{\sigma t}{2}\|\beta - \beta^{t+1}\|^2 &\leq V^t(-w^t) + t\Psi(\beta) + \sum_{\tau=1}^t \langle g^\tau, \beta - h\rangle\\
    &= V^t(-w^t) + t\Psi(\beta)+\sum_{\tau=1}^{s}\langle g^\tau, \beta - h\rangle  +\sum_{\tau=s+1}^{t} \langle g^\tau, \beta -h\rangle  \\
    &= V^t(-w^t) + t\Psi(\beta)+\sum_{\tau=1}^{s}\langle g^\tau, \beta - h\rangle +\sum_{\tau=s+1}^{t} \langle g^\tau, \beta -\beta^{\tau}\rangle +\sum_{\tau=s+1}^{t} \langle g^\tau, \beta^\tau -h\rangle
\end{align*}
By further adding and subtracting $\sum_{\tau = s+1}^t \Psi(\beta^\tau)$ to the right-hand side, the inequality becomes
\begin{align*}
    \frac{\sigma t}{2}\|\beta - \beta^{t+1}\|^2  \leq &
    \underbrace{\sum_{\tau=s+1}^t\langle g^\tau, \beta - \beta^\tau \rangle+(t-s) \Psi(\beta) -\sum_{\tau = s +1}^t\Psi(\beta^\tau)}_{\text{(I)}} \\ + &
    \underbrace{\sum_{\tau=1}^{s}\langle g^\tau, \beta - h\rangle + s \Psi(\beta)}_{\text{(II)}} \\ + & 
    \underbrace{V^t(-w^t)+\sum_{\tau=s+1}^t\left(  \langle g^\tau, \beta^\tau - h\rangle +\Psi(\beta^\tau) \right)}_{\text{(III)}}.
\end{align*}
Next, we bound the three terms (I), (II), and (III), respectively.

\paragraph{Bounding Term (I).} Term (I) can be rearranged into the negative regret over rounds $s+1,\ldots,t$. By convexity of $f( \cdot , v^\tau)$, 
\begin{align*}
    \text{(I)} &= \sum_{\tau=s+1}^t\langle g^\tau, \beta - \beta^\tau \rangle+(t-s) \Psi(\beta) -\sum_{\tau = s +1}^t\Psi(\beta^\tau)\\
    &\leq \sum_{\tau = s+1}^t \left(f(\beta, v^\tau)-f(\beta^\tau, v^\tau)\right) +(t-s) \Psi(\beta) -\sum_{\tau = s +1}^t\Psi(\beta^\tau)\\
    &= - \textnormal{Dual-Regret}(s, \beta).
\end{align*}

\paragraph{Bounding Term (II).} Recall that in our problem $g_i^\tau = u_i^\tau \leq \|v_i\|_{\infty}$ for each $i$. Then,
\begin{align*}
    \text{(II)} = \sum_{\tau=1}^{s}\langle g^\tau, \beta - h\rangle + s \Psi(\beta)
    \leq -\sum_{\tau=1}^{s}\langle g^{\tau}, h\rangle+ s \cdot \left(\max_i \beta_i \|v_i\|_{\infty} - \sum_{i}B_i\log \beta_i\right)
\end{align*}

\paragraph{Bounding Term (III).}
For any $\tau \geq s+1$, we have
\begin{align*}
    V^\tau(-s^\tau) + \Psi(\beta^{\tau+1}) &\overset{\text{(a)}}{\leq} V^{\tau-1}(-s^\tau)\\
    &= V^{\tau-1}(-s^{\tau-1}-g^\tau)\\
    &\overset{\text{(b)}}{\leq} V^{\tau-1}(-s^{\tau-1})+ \langle - g^{\tau}, \nabla V^{\tau-1}(-s^{\tau-1})\rangle +\frac{\|g^{\tau}\|^2_{*}}{2\sigma(\tau-1)}\\
    &\overset{\text{(c)}}= V^{\tau-1}(-s^{\tau-1})+ \langle - g^{\tau}, \beta^{\tau}-h\rangle +\frac{\|g^{\tau}\|^2_{*}}{2\sigma(\tau-1)},
\end{align*}
where the steps (a), (b), and (c) are due to \Cref{lem: conjugate-inequality}, \eqref{eq: lipschiz-contiuous}, and \eqref{eq: gradient-of-conjugate}, respectively. Rearranging the terms, we have
\begin{equation*}
    \langle g^{\tau}, \beta^{\tau}-h\rangle+ \Psi(\beta^{\tau+1}) \leq V^{\tau-1}(-s^\tau)-V^\tau(-s^\tau)+\frac{\|g^{\tau}\|^2_{*}}{2\sigma(\tau-1)}, \ \ \forall \tau \geq s+1.
\end{equation*}
Summing the inequalities from $\tau = s +1, \cdots, t$, we have
\begin{align*}
    \text{(III)} &= \sum_{\tau=s+1}^t \left(\langle g^{\tau}, \beta^{\tau}-h\rangle+ \Psi(\beta^{\tau+1})\right) + V^{t}(-w^t)\\
    &\leq  V^{s}(-w^{s})+ \frac{1}{2}\sum_{\tau=s+1}^{t}\frac{\|g^{\tau}\|^2_{*}}{\sigma(\tau-1)}\\
    &= \langle -w^s, \beta^{s+1}-h\rangle - s \Psi(\beta^{s+1}) + \frac{1}{2}\sum_{\tau=s+1}^{t}\frac{\|g^{\tau}\|^2_{*}}{\sigma(\tau-1)}\\
    &\leq \sum_{\tau=1}^{s} \langle g^\tau, h \rangle + s \sum_{i=1}^n B_i \log h_i +  \frac{1}{2}\sum_{\tau=s+1}^{t}\frac{\|g^{\tau}\|^2_{*}}{\sigma(\tau-1)}
\end{align*}

\paragraph{Putting Everything Together.}
Combining the bounds for term (I), (II), and (III), 
\begin{equation}
    \label{eq: put-everything-together}
    \frac{\sigma t}{2}\|\beta - \beta^{t+1}\|^2  \leq 
    \frac{1}{2}\sum_{\tau=s+1}^{t}\frac{\|g^{\tau}\|^2_{*}}{\sigma(\tau-1)}+
     \left(\max_i \beta_i \|v_i\|_{\infty} + \sum_{i}B_i\log \frac{h_i}{\beta_i} \right) \cdot s
    -\textnormal{Dual-Regret}(s, \beta).
\end{equation}
We also have
\begin{equation}
    \label{eq: sum-to-log}
    \frac{1}{2}\sum_{\tau=s+1}^{t}\frac{\|g^{\tau}\|^2_{*}}{\sigma(\tau-1)} 
    < \frac{\max_{\tau}\|g^\tau\|_{*}^2}{2\sigma} \log t 
    \leq \frac{\max_i\|v_i\|_{\infty}^2}{2\sigma} \log t.
\end{equation}
Combining \eqref{eq: put-everything-together} and \eqref{eq: sum-to-log}, we have
\begin{equation*}
    \frac{\sigma t}{2}\|\beta - \beta^{t+1}\|^2 \leq \frac{\max_i\|v_i\|_{\infty}^2}{2\sigma} \log t + \left(\max_i \beta_i \|v_i\|_{\infty} + \sum_{i}B_i\log \frac{h_i}{\beta_i} \right) \cdot s
    -\textnormal{Dual-Regret}(s, \beta). 
\end{equation*}
Multiplying both sides by ${2}/{\sigma t}$, we arrive at
\begin{equation*}
    \|\beta - \beta^{t+1}\|^2 \leq \frac{2}{\sigma t}\left(\frac{\max_i\|v_i\|_{\infty}^2}{2\sigma} \log t + \left(\max_i \beta_i \|v_i\|_{\infty} + \sum_{i}B_i\log \frac{h_i}{\beta_i} \right) \cdot s
    -\textnormal{Dual-Regret}(s, \beta)\right).
\end{equation*}
This proves \Cref{lem: deterministic-mean-square}.

$\hfill \square$

\subsection{Proof of \Cref{lem: dual-regret-iid}}
\label{proof: dual-regret-iid}
\iidlemB*

\begin{proof}
For $\tau > s$, we decompose the dual regret at time $\tau$ as follows,

\begin{align}
\label{eq: dual-regret-decomposition-1}
    F(\beta^\tau, v^\tau) - F(\beta^*, v^\tau) = 
    \underbrace{\left(F(\beta^\tau, v^\tau) - \phi(\beta^\tau)\right)}_{\text{(I)}} + 
    \underbrace{\left(\phi(\beta^*) - F(\beta^*, v^\tau)\right)}_{\text{(II)}}  +  
    \underbrace{\left(\phi(\beta^\tau) - \phi(\beta^*)\right)}_{\text{(III)}}.
\end{align}

By the optimality of $\beta^*$ we have $\text{(III)}\geq 0$ always. Next, we bound (I) and (II). 

Let $\Hat{Q}$ be the distribution of $v|A_{h,s}$, $\Hat{Q}^{1:\tau}$ be the marginal distribution of $\Hat{Q}$ on the first $\tau$ rounds, and $\Hat{Q}^\tau(\cdot | v^{1:\tau-1})$ be the distribution of $v^\tau$ conditioned on $A_{h,s}$ and the information of previous rounds. 

Note that
\begin{equation*}
    \mathbb{E}\left[\text{(I)}|A_{h,s}\right] = \mathbb{E}_{v^{1:\tau-1} \sim \Hat{Q}^{1:\tau-1}}\left[ \mathbb{E}_{v^\tau \sim \Hat{Q}^\tau}\left[\left(F(\beta^\tau, v^\tau) - \phi(\beta^\tau)\right) \mid v^{1:\tau-1}\right]  \right].
\end{equation*}

Let us investigate the inner expectation. Conditional on $v^{1:\tau-1}$, the pacing multiplier $\beta^\tau$ is deterministic.
\begin{align}
    \left|\mathbb{E}_{v^\tau \sim \Hat{Q}^\tau}\left[\left(F(\beta^\tau, v^\tau) - \phi(\beta^\tau)\right) \mid v^{1:\tau-1}\right]\right|
&\leq \mathbb{E}_{v^\tau \sim \Hat{Q}^\tau}\left[\left|F(\beta^\tau, v^\tau) - \phi(\beta^\tau)\right| \mid v^{1:\tau-1}\right]  \nonumber \\
&=  \left| \int_{0}^{\|v\|_{\infty}}
F(\beta^\tau, v) \Hat{Q}^\tau(\mathrm{d}v|v^{1:\tau-1}) - 
\int_{0}^{\|v\|_{\infty}} F(\beta^\tau, v) \mathrm{d}\Bar{Q}(v)
\right| \nonumber \\
&\leq G(h) \int_{0}^{\|v\|_{\infty}} 
\left| \mathrm{d}\Hat{Q}^\tau(\cdot|v^{1:\tau-1}) - \mathrm{d}\Bar{Q}(v) \right| \nonumber \\
&= 2 G(h) \cdot \|\Hat{Q}^\tau(\cdot|v^{1:\tau-1}) - \Bar{Q}(\cdot)\|_{\mathrm{TV}}. \label{eq: dual-regret-1}
\end{align}

Next, we deal with the outer expectation. Let $\Hat{\mathcal{V}}^\tau$ be the support of $\Hat{Q}^{\tau}(\cdot \mid v^{1:\tau-1})$. Since $\Hat{Q}^{\tau}$ is also conditioned on $A_{h,s}$, we have
\begin{equation*}
    v^{1:\tau-1} \not \in \Hat{\mathcal{V}}^\tau \implies \Pr[ A_{h,s}|v^{1:\tau-1}] = 0.
\end{equation*}
Hence,
\begin{align}
    \label{eq: dual-regret-1.5}
    \Pr[A_{h,s}] \leq \mathbb{E}_{v^{1:\tau-1} \sim \Hat{Q}^{1:\tau-1}}\left[\Pr[A_{h,s} \mid v^{1:\tau-1}\right] + 
    \Pr[A_{h,s} \mid v^{1:\tau-1} \not \in \Hat{\mathcal{V}}^\tau]
    &=\mathbb{E}_{v^{1:\tau-1} \sim \Hat{Q}^{1:\tau-1}} \left[\Pr[A_{h,s} \mid v^{1:\tau-1}\right]
\end{align}
    
We can use a failure probability term to bound (I) as follows,

\begin{align}
\mathbb{E}\left[|\text{(I)}||A_{h,s}\right] &\leq  
    \mathbb{E}_{v^{1:\tau-1} \sim \Hat{Q}^{1:\tau-1}} \left[
        \left|\mathbb{E}_{v^\tau \sim \Hat{Q}^\tau}\left[\left(F(\beta^\tau, v^\tau) - \phi(\beta^\tau)\right) \mid v^{1:\tau-1}\right]\right|
    \right] \nonumber \\
    &\leq 2 G(h) \cdot 
    \mathbb{E}_{v^{1:\tau-1} \sim \Hat{Q}^{1:\tau-1}} \left[
        \|\Hat{Q}^\tau(\cdot|v^{1:\tau-1}) - \Bar{Q}(v)\|_{\mathrm{TV}}
    \right] \nonumber \\
    &= 2 G(h) \cdot 
    \mathbb{E}_{v^{1:\tau-1} \sim \Hat{Q}^{1:\tau-1}} \left[
        \|Q^\tau\left(\cdot|A_{h,s}, v^{1:\tau-1}\right) - \Bar{Q}(v)\|_{\mathrm{TV}}
    \right] \nonumber \\
    &\overset{\text{(a)}}{\leq} 2 G(h) \cdot 
     \mathbb{E}_{v^{1:\tau-1} \sim \Hat{Q}^{1:\tau-1}} \left[
        1- \Pr\left[A_{h,s} \mid v^{1:\tau-1}\right]
    \right] \nonumber \\
    &\overset{\text{(b)}}{\leq} 2 G(h) \cdot
    \left(1-\Pr\left[{ A_{h,s}}\right]\right), \label{eq: dual-regret-2}
\end{align}
where (a) is by the property of total variation distance that $\|Q^\tau(\cdot|A) - \Bar{Q}(\cdot)\|_{\mathrm{TV}}\leq (1-\Pr[A])$ for any measurable event $A$, and (b) is by \eqref{eq: dual-regret-1.5}.

Next we bound $\mathbb{E}\left[|\text{(II)}||A_{h,s}\right] $. Notice that similar analysis goes through without the outer expectation.
\begin{equation}
\label{eq: dual-regret-2-4}
    \mathbb{E}\left[|\text{(II)}||A_{h,s}\right] \leq 2 \Bar{F}(\beta^*) (1-\Pr[A_{h,s}])
\end{equation}

Combining (I), (II), and $\text{(III)}>0$ and summing over $\tau = s+1, \cdots, t$, we prove \Cref{lem: dual-regret-iid}.

\end{proof}
\subsection{Proof of \Cref{lem: high-prob-bounds-iid}}
\iidlemC*
\label{proof: high-prob-bounds-iid}

\noindent\textit{Proof.\,}
For agent $j$, let $T_j:=\{\tau\in [t]: x_j^\tau = 1\}$ be the index set of items that are allocated to agent $j$ in PACE. 
For \textit{i.i.d.} inputs, we denote $\mu_j:= \mathbb{E}[v_i^\tau]$. We are going to show that, for each $j\in [n]$ there exists $u_j^\prime = \Omega(1/n)$ (which is independent of $t$), and constant $c_1, c_2>0$ (which are independent of both $t$ and $n$), such that 
\begin{equation}
    \label{eq: high-prob-1}
    \Pr[\Bar{u}_j^\tau < u_j^\prime] = O\left(e^{-c_1\tau/n}\right), \ \forall \tau > c_2n.
\end{equation}
We first show how \eqref{eq: high-prob-1} leads to \Cref{lem: high-prob-bounds-iid}. By \eqref{eq: high-prob-1} and the fact that $c_1$ is independent of $n$, we know that the sum of $\Pr[\Bar{u}_j^\tau < u_j^\prime]$ over $\tau\geq s$ can be bounded by the sum of geometric series, where $s>c_2n$,
\begin{equation*}
    \sum_{\tau\geq s}\Pr[\Bar{u}_j^\tau < u_j^\prime] = O\left(e^{-c_1 s/n}\right).
\end{equation*}
By union bound over $j\in [n]$ and $\tau\geq s$,
\begin{equation*}
    \Pr\left[\exists \tau \geq s, j\in [n], \ \ \Bar{u}_j^\tau < u_j^\prime\right] = O\left(n^2 e^{-c_1 s/n}\right), \ s>c_2n.
\end{equation*}
Since we have $\beta_j^\tau \geq B_j/\|v_j\|_{\infty} $ always, 
\begin{equation*}
    \forall \tau\geq s, j \in [n], \Bar{u}_j^\tau \geq u_j^\prime \implies A_{h,s}, \ \left(h_j = B_j/u_j^\prime = O(n)\right).
\end{equation*}
Hence for $h = (h_1, \cdots, h_n), $ where $h_j = B_j/u_j^\prime$,
\begin{equation*}
    1-\Pr\left[A_{h,s}\right] = O\left(n^2 e^{-c_1 s/n}\right), \ s>c_2n.
\end{equation*}
Choosing $\Hat{s}(t,n) = \max\{c_2n+1, c_1^{-1}n(\log n+\log t)\}$ will result in an $O(n/t)$ failure probability, which shows \Cref{lem: high-prob-bounds-iid}. 

In the rest of the proof, we show \eqref{eq: high-prob-1}.
\paragraph{Bounding the monopolist utility.} 
Consider $j\in[n]$, and the following event 
$$
X:= \left\{ \gamma : \sum_{t^\prime=1}^\tau v_j^{t^\prime} > \left(\mu_j - {w}_j\right) \tau\right\},
$$
which characterizes a lower bound of his monopolist utility (sum of values for all items so far) at time step $\tau.$ By Hoeffding bounds, we can show a high probability bound for $X$, 
\begin{equation}
    \label{eq: concentration-monopolist}
    \Pr\left[ X^c\right] \leq \exp\left(- \frac{2 {w}_j^2 \tau}{\|v_j\|_{\infty}^2}  \right), \ \forall {w}_j>0.
\end{equation}
Notice that 
\begin{equation*}
    X \implies \sum_{t^\prime \not \in T_j} v_j^{t^\prime} > \left(\mu_j - {w}_j - \Bar{u}_j^\tau\right) \tau.
\end{equation*}
For $\lambda>0$, let $M_j(\lambda)$ be the index set of items which agent $j$ has valuation at least $\lambda$ and was not allocated to agent $j$ by PACE. Define $m_j(\lambda) = |M_j(\lambda)| $. Then, by
$$
\sum_{t^\prime\not \in T_j} v_j^{t^\prime} \leq m_j(\lambda) \|v_j\|_{\infty} + (\tau-m_j(\lambda)) \lambda ,
$$
we have
\begin{equation}
    X 
    \implies m_j(\lambda) > \frac{\mu_j - {w}_j - \Bar{u}_j^\tau - \lambda}{\|v_j\|_{\infty} - \lambda}\cdot \tau.
    \label{eq: lower-bound-y}
\end{equation}

\paragraph{Bounding competing agents' utility.} 
Let $i = \min \arg \max_{k\neq j}|M_j(\lambda) \cap T_k|.$ Then, we have 
\begin{equation*}
    |M_j(\lambda)\cap T_i| \geq \frac{m_j(\lambda)}{n}.
\end{equation*}
Let $r = \max \{M_j(\lambda)\cap T_i \}$. We consider the event
\begin{equation*}
    Y:= \left\{\gamma: U_i^r > (\mu_i - {w}_i) \frac{m_j(\lambda)}{n}\right\}. 
\end{equation*}
By the definition of $T_i$, we have
\begin{equation}
\label{eq: concentration-1}
    U_i^r \geq \sum_{t^\prime \in M_j(\lambda)\cap T_i} v_i^{t^\prime}.
\end{equation}
We know from the decision rule of PACE that for $t^\prime \in M_j(\lambda) \cap T_i$, 
\begin{equation*}
    v_i^{t^\prime} \geq \max_{k\neq i}\left\{ \frac{v_j^{t^\prime} U_i^{t^\prime-1}}{U_j^{t^\prime-1}} \right\}.
\end{equation*}
Therefore, the right hand side of \eqref{eq: concentration-1} stochastically dominates $\sum_{t^\prime \in M_j(\lambda)\cap T_i} S_i^{t^\prime}$, where random variable $S_i^{t^\prime}:= v_i^{t^\prime} \mid v_i^{t^\prime}>0$ is defined as the per-round agent value conditioned on itself being nonzero. By Azuma-Hoeffding inequality, we have
\begin{equation}
    \label{eq: concentration-competitive}
    \Pr\left[ Y^c \right] \leq \Pr\left[\sum_{t^\prime \in M_j(\lambda)\cap T_i} v_i^{t^\prime} \leq \left(\mu_i - {w}_i\right) \frac{m_j(\lambda)}{n} \right] \leq \exp\left(- \frac{2{w}_i^2 \cdot m_j(\lambda)}{n\cdot \|v_i\|_{\infty}^2}\right) , \ {w}_i>0.
\end{equation}

\paragraph{Lower Bound for $\Bar{u}_j^\tau$.}
By the PACE decision rule on item $r$, it holds that
\begin{equation}
    \label{eq: last-round-decision}
    U_i^r-\|v_i\|_{\infty} \leq U_i^{r-1} \leq \frac{B_iU_j^{r-1} v_i^r}{B_j v_j^r} \leq \frac{B_i \|v_i\|_\infty  \Bar{u}_j^\tau}{B_j \lambda} \cdot \tau.
\end{equation}

Combining \eqref{eq: last-round-decision} and the definition of $X$ and $Y$, we have
\begin{equation}
    \label{eq: high-prob-bounds-iid-max}
    X \cap Y \implies \Bar{u}_j^\tau \geq \frac{B_j\left(\mu_j - {w}_j - \lambda - \frac{n\|v_i\|_{\infty}(\|v_j\|_{\infty}-\lambda)}{(\mu_i - {w}_i)\tau}\right)\lambda}{B_j\lambda + \frac{ \|v_i\|_\infty B_i}{(\mu_i - {w}_i)} \cdot (\|v_j\|_\infty -\lambda) \cdot n}.
\end{equation}

\paragraph{Fixing ${w}$ and $\lambda$.}
We now determine the values for ${w}_i, {w}_j, \lambda$, such that the right hand side of \eqref{eq: high-prob-bounds-iid-max} gives us a meaningful bound. By choosing ${w}_i = \mu_i/2$ and ${w}_j = \lambda = \mu_j/4$, we have that for $\tau \geq \frac{8n\max_k\|v_k\|_{\infty}\|v_j\|_{\infty}}{\mu_k \mu_j}$,
\begin{equation*}
    \Bar{u}_j^\tau \geq \Bar{u}_j^\prime:= \min\left\{\frac{\mu_j}{4}, \frac{B_j \mu_j^2}{4 B_j \mu_j + 32 n\cdot(\|v_j\|_\infty -\mu_j/4)\cdot \max_k\left\{B_k\|v_k\|_{\infty} / \mu_k\right\}}\right\} \in \Omega\left(\frac{1}{n}\right) .
\end{equation*}
The failure probability of above is at most $\Pr[X^c] + \Pr[Y^c]$, which is  
\begin{equation*}
    \exp\left(-\frac{\mu_j^2}{8\|v_j\|_\infty^2}\tau\right) + 
    \exp \left(-\frac{\min_k \mu_k^2 \cdot \mu_j}{n\|v_i\|_{\infty} \cdot \left(8\|v_j\|_{\infty} - 2\mu_j\right)}\cdot \tau\right)= O(e^{-c_1\tau/n}),
\end{equation*}
where $c_1$ is a constant independent of $n$ and $\tau$. This proves \eqref{eq: high-prob-1}.

$\hfill \square$

\subsubsection{Proof of \Cref{thm: convergence-spending-iid}} 
\iidthmC*

\begin{proof}
We show the following lemma, which reduces the question of spending convergence into a convergence question on time-averaged utilities and multipliers, thereby allowing application of our earlier results. We will prove \Cref{lem: convergence-spending} later. 
\begin{lemma}
    \label{lem: convergence-spending}
    For any $s<t$, it holds that
    \begin{align*}
        \left\| \frac{1}{t}\sum_{\tau= s+1}^t b^\tau - B\right\|^2 
        \leq  3\left[ \max_i(\beta_i^*)^2 \cdot 
    \|\Bar{u}^t - u^*\| + n \cdot \|v\|_{\infty}^2 \cdot \left(\frac{s}{t}\right)^2  + \frac{\|v\|_{\infty}^2}{t} \sum_{\tau= s+1}^t \|\beta^\tau - \beta^*\|^2\right],
\end{align*}
where $b_i^\tau = p_i^\tau x_i^\tau$ is the expenditure of agent $i$ under price $p^\tau$ at step $\tau$. 
\end{lemma}

\Cref{lem: convergence-spending} helps us to decompose the target term into previous mean-square convergence results for dual multipliers and time-averaged utilities. Let $\Hat{s}(t, n)$ and $\Hat{A}$ be defined as in \Cref{lem: high-prob-bounds-iid}, and taking conditional expectations, 
\begin{equation}
    \label{eq: convergence-spending-decomposition}
    \begin{aligned}
        &\mathbb{E}\left[\left\| \frac{1}{t}\sum_{\tau= \Hat{s}(t, n)+1}^t b^\tau - B\right\|^2 \mid  \Hat{A} \right]
        \\
        \leq & \  \ 3\left[ \max_i(\beta_i^*)^2 \cdot 
    \mathbb{E}\left[\|\Bar{u}^t - u^*\|\mid  \Hat{A}\right] + n \cdot \|v\|_{\infty}^2 \cdot \left(\frac{\Hat{s}(t, n)}{t}\right)^2  + \frac{\|v\|_{\infty}^2}{t} \sum_{\tau= \Hat{s}(t, n)+1}^t \mathbb{E} \left[\|\beta^\tau - \beta^*\|^2\mid  \Hat{A}\right]
    \right].
    \end{aligned}
\end{equation}

Notice that $\Hat{s}({t, n}) = O(n \log n + n \log t)$. Combining \eqref{eq: sum-multipliers} with \Cref{thm: convergence-utility-iid} and applying Jensen's inequality, we get the desired bound. Now we prove \Cref{lem: convergence-spending}. 

\paragraph{Proof of \Cref{lem: convergence-spending}.} We decompose each $b_i^\tau$ as follows.
\begin{align*}
    b_i^\tau = \beta_i^\tau v_i^\tau \mathbb{I}\{i = i^\tau\} 
    = \beta_i^\tau u_i^\tau 
    = \beta_i^* u_i^\tau + \left(\beta_i^\tau - \beta_i^*\right) u_i^\tau.
\end{align*}
Summing up over time steps,
\begin{align*}
     \frac{1}{t}\sum_{\tau= s+1}^t b^\tau_i
    &= \beta_i^*\left( \Bar{u}_i^t - \frac{1}{t}\sum_{\tau = 1}^{s} u_i^\tau \right) +  \frac{1}{t}\sum_{\tau= s+1}^t (\beta_i^\tau - \beta_i^*) u_i^\tau.
\end{align*}
By the boundedness of $u_i^\tau$, 
\begin{align*}
     \frac{1}{t}\sum_{\tau= s+1}^t b^\tau_i - B_i
    \leq \left(\beta_i^*  \Bar{u}_i^t - B_i \right) + \frac{s}{t} \cdot {\beta_i^*}{\|v_i\|_{\infty}}+  \frac{1}{t}\sum_{\tau= s+1}^t (\beta_i^\tau - \beta_i^*) u_i^\tau.
\end{align*}
Then, we bound the square difference between expenditure and budgets as follows, using $(x+y+z)^2\leq 3(x^2+y^2+z^2)$ for any $x, y,z \in \mathbb{R}$,
\begin{align}
    \label{eq: expenditure-1}
    \left( \frac{1}{t}\sum_{\tau= s+1}^t b^\tau_i - B_i  \right)^2 
    &\leq 3 \left[ (\beta_i^*\Bar{u}_i^t - B_i)^2 + 
    \left(\frac{s}{t} \cdot \beta_i^* \|v_i\|_{\infty}\right)^2 + \left(\frac{1}{t} \sum_{\tau= s+1}^t (\beta_i^\tau - \beta_i^*) u_i^\tau \right)^2\right].
\end{align}
Using the convexity of $(\cdot)^2$ and the boundedness of $u_i^\tau$, we bound the final term as follows,
\begin{align}
    \label{eq: expenditure-2}
   \left(\frac{1}{t} \sum_{\tau= s+1}^t (\beta_i^\tau - \beta_i^*) u_i^\tau \right)^2 \leq \frac{1}{t}\sum_{\tau= s+1}^t(\beta_i^\tau - \beta_i^*)^2\|v_i\|_\infty^2.
\end{align}
Combining \eqref{eq: expenditure-1} and \eqref{eq: expenditure-2}, and using $B_i = \beta_i^* {u}_i^*$, 
\begin{align*}
    \left( \frac{1}{t}\sum_{\tau= s+1}^t b^\tau_i - B_i  \right)^2 
    \leq 3 \left[ \beta_i^*(\Bar{u}_i^t - u_i^*)^2 + 
    \left(\frac{s}{t} \cdot \beta_i^* \|v_i\|_{\infty}\right)^2 + \frac{1}{t} \sum_{\tau= s+1}^t (\beta_i^\tau - \beta_i^*)^2\|v_i\|_\infty^2\right].
\end{align*}
Summing up across $i\in [n]$, we prove \Cref{lem: convergence-spending}.

\end{proof}

\subsubsection{Proof of \Cref{thm: regret-and-envy}.}
\iidthmD*
\begin{proof}
We give the following decomposition of the squared buyer-regret term. We will prove \Cref{lem: regret} later. 
\begin{lemma}
    \label{lem: regret}
    For any $s<t$, it holds deterministically that
    \begin{equation*}
        \textnormal{Buyer-Regret}_{i}^2 \leq \frac{2\|v_i\|_{\infty}^2}{(\beta_i^*)^2}\cdot  \left({s^2} + t \sum_{\tau = s+1}^t \|\beta^\tau - \beta^*\|^2\right).
    \end{equation*}
\end{lemma}

Let $\Hat{s}(t,n)$ and $\Hat{A}$ be defined as in \Cref{lem: high-prob-bounds-iid}. Then, by \eqref{eq: sum-multipliers}, 
\begin{equation*}
    \sum_{\Hat{s}({t,n})+1}^t \mathbb{E}\left[ \|\beta^\tau - \beta^*\|^2 \mid \Hat{A}\right] = O\left({n^4 \log n (\log t)^2}\right).
\end{equation*}
Combining with \Cref{lem: regret} and the fact that $\Hat{s}(t,n) = O(n+\log t)$, we have
\begin{equation*}
    \mathbb{E}\left[ \textnormal{Buyer-Regret}_{i}^2\mid \Hat{A}\right] = O\left({n^4 \log n\; (\log t)^2}\; t\right).
\end{equation*}
Notice that the buyer regret is at most $t \|v\|_\infty$. By a conditional expectation argument, 
\begin{align*}
    & \ \quad\mathbb{E}\left[\textnormal{Buyer-Regret}_{i}^2\right]\\ 
    &\leq \mathbb{E}\left[\textnormal{Buyer-Regret}_{i}^2\mid \Hat{A}\right] + t^2\|v\|_{\infty}^2 \cdot (1-\Pr[\hat A])\\
    &= O\left({n^4 \log n (\log t)^2} \; t\right).
\end{align*}
Applying Jensen's inequality gives us the desired inequality. Now it remains for us to prove \Cref{lem: regret}. 
\paragraph{Proof of \Cref{lem: regret}.} 

Let $(z_i^\tau)_{\tau \in [t]}$ be any bundle attaining the maximum in the definition of $\textnormal{Buyer-Regret}_i$. Denote its time-averaged utility by $\hat{u}_i := (1/t)\sum_{\tau=1}^t z_i^\tau v_i^\tau$ for simplicity. Then $\textnormal{Buyer-Regret}_{i}=t(\hat u_i-\bar u_i^t)^+$. By the KKT conditions of the underlying problem~\citep{gao2023infinite}, we have $p^*\geq \beta_i^* v_i^\tau$ for all $\tau \in [t]$. Combined with the fact that $z_i^\tau \leq 1$ for all $\tau\in[t]$, we have 
\begin{equation}
    \label{eq: regret-1}
    \beta_i^* {\hat{u}_i}
    =     \beta_i^* \left( \frac{1}{t}\sum_{\tau = 1}^t v_i^\tau z_i^\tau \right) \leq \frac{1}{t}\sum_{\tau=1}^t\min\left\{v_i^\tau, p^{*, \tau} \cdot z_i^\tau\right\}.
\end{equation}

Meanwhile, we have for each $\tau$,
\begin{align}
    p^{*, \tau} \cdot z_i^\tau 
    &= p^\tau z_i^\tau + (p^{*,\tau}-p^\tau)z_i^\tau \nonumber \\
    &\leq p^\tau z_i^\tau + \|v_i\|_{\infty} \cdot \|\beta^\tau - \beta^*\|_{\infty}
    \label{eq: regret-2}
\end{align}
Combining \eqref{eq: regret-1} and \eqref{eq: regret-2}, 
\begin{align*}
    \beta_i^* {\hat{u}_i}
    &\leq \frac{1}{t} \sum_{\tau=1}^{s} v_i^\tau + \frac{1}{t}\sum_{\tau=s+1}^t p^\tau z_i^\tau
    + \frac{\|v_i\|_{\infty}}{t}\sum_{\tau=s+1}^t \|\beta^\tau - \beta^*\|_{\infty} \\
    &\leq \frac{s}{t} \cdot \|v_i\|_{\infty} + \frac{1}{t}\sum_{\tau = 1}^t p^\tau z_i^\tau
    + \frac{\|v_i\|_{\infty}}{t}\sum_{\tau=s+1}^t   \|\beta^\tau - \beta^*\|_{\infty} \\
    &\leq \frac{s}{t} \cdot \|v_i\|_{\infty} + B_i
    + \frac{\|v_i\|_{\infty}}{t}\sum_{\tau=s+1}^t  \|\beta^\tau - \beta^*\|_{\infty}.
\end{align*}
We notice that $B_i/\beta_i^* = u_i^*$. Then the demanded time-averaged utility satisfies
\begin{align*}
   {\hat{u}_i} \leq u_i^* +\frac{s}{t} \cdot \frac{1}{\beta_i^*}\cdot\|v_i\|_{\infty} +  \frac{1}{\beta_i^*}
     \frac{\|v_i\|_{\infty}}{t}\sum_{\tau=s+1}^t  \|\beta^\tau - \beta^*\|_{\infty}.
\end{align*}

Therefore,
\begin{align*}
   \frac{1}{t^2}\textnormal{Buyer-Regret}_{i}^2
    & \leq \frac{\|v_i\|_{\infty}^2}{(\beta_i^*)^2}\cdot  \left(\frac{s}{t} + \frac{1}{t} \sum_{\tau = s+1}^t \|\beta^\tau - \beta^*\|_{\infty}\right)^2\\
    &\leq \frac{2\|v_i\|_{\infty}^2}{(\beta_i^*)^2}\cdot  \left(\frac{s^2}{t^2} + \left(\frac{1}{t} \sum_{\tau = s+1}^t \|\beta^\tau - \beta^*\|^2\right)^2\right)\\
    &\leq \frac{2\|v_i\|_{\infty}^2}{(\beta_i^*)^2}\cdot  \left(\frac{s^2}{t^2} + \frac{1}{t} \sum_{\tau = s+1}^t \|\beta^\tau - \beta^*\|^2\right),
\end{align*}
where the last step is by the convexity of $(\cdot)^2$. This proves \Cref{lem: regret}.

\end{proof}

%% file: best-of-many-worlds/proof-non.tex
\section{Missing Results and Proofs in \Cref{section: nonstationary}}
\label{proof: non}

\subsection{Mildly Corrupted Independent Input and Periodic Input}
We introduced two subclasses of the blockwise independent inputs: independent inputs with mild adversarial corruption and periodic inputs. Each captures some practical scenarios of interest. The convergence, fairness, and efficiency results can be directly derived from \Cref{thm: convergence-utility-non} and \Cref{thm:nonstationary-fairness-efficiency-metrics}. 

\label{appendix: corrupted-and-periodic}
\paragraph{Independent input with mild adversarial corruption.} We consider the scenario where items generally behave in a predictable manner, but at some time steps behave erratically, due to malicious manipulation or unpredictable surges. or example, \citet{esfandiari2018allocation} study online allocation under such a model, in order to model unpredictable traffic spikes or malicious activity in online advertising allocation. We model this scenario by setting $|\mathcal{P}|_{\infty}=1$ in the blockwise independent model: the item distribution at each time step might be corrupted and deviate from $\bar Q$, but the distributions at different time steps are independent of each other. The average corruption is captured in parameter $\delta$ as measured in TV distance:
\begin{equation*}
     \mathsf{C}^{\mathrm{ID}}(\delta):=
     \left\{
         Q\in \Delta(V)^t: \frac{1}{t}\sum_{\tau=1}^t \|Q^\tau - \Bar{Q}\|_{\mathrm{TV}}\leq \delta, \ (v^\tau)_{\tau=1}^t \text{ are independent}
     \right\}.
\end{equation*}

\begin{corollary}[Independent data with adversarial corruption]
\label{corollary: corrupted}
    Under adversarially corrupted and independent distributions, $\gamma \sim Q \in \mathsf{C}^{\mathrm{ID}}(\delta)$, it holds that
\begin{align*}
\mathbb{E} \left[ \|\Bar{u}^t - u^*\|^2\right] = O\left(\frac{n^4 \log n\log t}{t}+ n^3 \delta \right).
\end{align*}
The allocation of PACE is additive $\eta^t_{\mathrm{EF}}$-envy-free, additive $\eta^t_{\mathrm{Prop}}$-proportional, additive $\eta^t_{\mathrm{Core}}$-core stable, and $\eta^t_{\mathrm{NW}}$-suboptimal for weighted Nash welfare, with
\begin{align*}
    \mathbb E\!\left[\eta^t_{\mathrm{EF}}\right]
    =
    O\!\left({n^2  \sqrt{  \,\log n} \cdot \log t \,  \sqrt{t} + \sqrt{n^3 \delta\, \log t} \cdot t  }\right),
    \\
    \mathbb E\!\left[\eta^t_{\mathrm{Prop}}\right], \mathbb E\!\left[\eta^t_{\mathrm{NW}}\right], \mathbb E\!\left[\eta^t_{\mathrm{Core}}\right]
    =
    O\!\left({n^2  \sqrt{ \,(\log n) \, (\log t )\, t}+ \sqrt{n^3 \delta\,} \cdot t  }\right).
\end{align*}
\end{corollary}

\paragraph{Periodic input.} When each block is \textit{i.i.d.} with $\delta = 0$ and has the same length $q$, we get a model for \textit{periodic}, or \textit{seasonal} input dat. For example, online traffic varies in the morning and evening, but daily patterns tend to repeat over time. Let $Q^{\tau_1:\tau_2}$ be the joint value distribution from time step $\tau_1$ to $\tau_2$, then the class of periodic input distribution is as follows:
\begin{equation*}
     \mathsf{C}^{\mathrm{P}}(q):=
    \left\{
        Q\in \Delta(V^q)^K: Q^{1:q} = Q^{q+1:2q} = \cdots = Q^{t-q+1:t}, \ \text{blocks are independent}
    \right\}.
\end{equation*}
\begin{corollary}[Periodic data]
    \label{corollary: periodic}
    Under adversarially corrupted and independent distributions, $\gamma \sim Q \in \mathsf{C}^{\mathrm{P}}(q)$, it holds that
\begin{align*}
\mathbb{E} \left[ \|\Bar{u}^t - u^*\|^2\right] = O\left(\frac{n^4 \log n\log t}{t}+ n^3 \delta \right).
\end{align*}
The allocation of PACE is additive $\eta^t_{\mathrm{EF}}$-envy-free, additive $\eta^t_{\mathrm{Prop}}$-proportional, additive $\eta^t_{\mathrm{Core}}$-core stable, and $\eta^t_{\mathrm{NW}}$-suboptimal for weighted Nash welfare, with
\begin{align*}
    \mathbb E\!\left[\eta^t_{\mathrm{EF}}\right]
    =
    O\!\left({n^2  q \sqrt{  \,\log n} \cdot \log t \,  \sqrt{t}   }\right),
    \\
    \mathbb E\!\left[\eta^t_{\mathrm{Prop}}\right], \mathbb E\!\left[\eta^t_{\mathrm{NW}}\right], \mathbb E\!\left[\eta^t_{\mathrm{Core}}\right]
    =
    O\!\left({n^2 q \sqrt{ \,(\log n) \, (\log t )\, t} }\right).
\end{align*}
\end{corollary}

\subsection{Missing Results and Proofs for Ergodic and Blockwise Independent Input}

We first state the complete set of results for the nonstationary input models, including dual and primal convergence, fairness and efficiency metrics, and online market equilibria results. 

\begin{theorem}[Convergence of pacing multipliers under nonstationary input]
\label{thm: convergence-multiplier-non}
Under ergodic and blockwise independent inputs respectively, such that 
\begin{align*}
    \mathbb{E}\left[\|\beta^{t+1}-\beta^*\|^2 \mid \Hat{A}\right]
    =
    \begin{cases}
        O\left(\dfrac{n^4 \iota \log n \log t}{t}+ n^3 \delta \right), 
        & \gamma \sim Q \in \mathsf{C}^{\mathrm{E}}(\delta, \iota),\\[1ex]
        O\left(\dfrac{n^4 |\mathcal{P}|_{\infty}^2 \log n \log t}{t}+ n^3 \delta \right), 
        & \gamma \sim Q \in \mathsf{C}^{\mathrm{B}}(\mathcal{P},\delta).
    \end{cases}
\end{align*}
\end{theorem}

\nonUtility*
\nonFE*

\begin{theorem}[Spending convergence under nonstationary input]
Under ergodic and blockwise independent inputs respectively, there exists $\Hat{A} \subseteq \Omega$, $\Pr[(\Hat{A})^c] = O(1/t)$ and $\Hat{s}= O(\log t)$ and $\Hat{A}$ such that
\begin{equation*}
    \mathbb{E}\left[ \left\|\sum_{\tau=\Hat{s}+1}^t p^\tau x^\tau - tB \right\| \mid \Hat{A} \right]
    =
    \begin{cases}
        O\!\left(n^2 \sqrt{\iota \log n}\,\log t \cdot \sqrt{t} + \sqrt{n^3\delta}\cdot t\right),
        & \gamma \sim Q \in \mathsf{C}^{\mathrm{E}}(\delta,\iota),\\[1ex]
        O\!\left(n^2 |\mathcal{P}|_\infty \sqrt{\log n}\,\log t \cdot \sqrt{t} + \sqrt{n^3\delta}\cdot t\right),
        & \gamma \sim Q \in \mathsf{C}^{\mathrm{B}}(\mathcal{P},\delta).
    \end{cases}
\end{equation*}
\end{theorem}

\begin{theorem}[Convergence to demand utility under nonstationary input]
Under ergodic or blockwise independent inputs respectively, it holds that for any agent $i\in [n]$,
\begin{equation*}
    \mathbb{E}\left[ \left(\mathrm{\mathrm{Demand}}_i(p^1, \cdots, p^\tau)-\bar u_i^t\right)^+ \right]
    =
    \begin{cases}
        O\!\left(n^2 \sqrt{\iota \log n}\,\log t \cdot \sqrt{t} + \sqrt{n^3\delta}\cdot t\right),
        & \gamma \sim Q \in \mathsf{C}^{\mathrm{E}}(\delta,\iota),\\[1ex]
        O\!\left(n^2 |\mathcal{P}|_\infty \sqrt{\log n}\,\log t \cdot \sqrt{t} + \sqrt{n^3\delta}\cdot t\right),
        & \gamma \sim Q \in \mathsf{C}^{\mathrm{B}}(\mathcal{P},\delta).
    \end{cases}
\end{equation*}
\end{theorem}

To prove these results, we follow the same route as in \Cref{section: stationary}. We first prove the conditional convergence of the pacing multipliers $\beta^t$, which leads to the unconditional convergence of average utility vectors $\bar u^t \to u^*$. The results on fairness, efficiency and market equilibria can be then reduced to it. To avoid unnecessary repetitions of the same structure, we present two nonstationary cases in a parallel manner.

\paragraph{Convergence of pacing multipliers.} The proof is divided into three major technical components, as in the proof of \Cref{thm: convergence-multiplier-iid}. Notice that \Cref{lem: deterministic-mean-square} is a deterministic result that also holds for nonstationary inputs. Therefore, we only need to update the dual online learning regret bound (counterpart of \Cref{lem: dual-regret-iid}) and the specification of the high-probability event $\Hat{A}$ (counterpart of \Cref{lem: high-prob-bounds-iid}), see the next two lemmas. Their proofs are deferred to \Cref{proof: dual-regret-non} and \Cref{proof: high-prob-bounds-non}, repsectively.

\begin{restatable}[Dual online learning regret bound, nonstationary]{lemma}{nonlemB}
    \label{lem: dual-regret-non}
    For any $t>s$, it holds that
    \begin{align*}
        &
        \mathbb{E}\left[\textnormal{Dual-Regret}(s,\beta^*) \mid A_{h,s}\right] \geq 
        \begin{cases}
            - \left(4G(h)(t-s)\left( p_{h,s} +\delta\right) +2 G(h)\iota+\frac{\sqrt{2}\|v\|_{\infty}^2 \cdot \iota \cdot \log t}{\sigma} \right)& \textnormal{(Ergodic)} \\
            - \left( 4G(h)\left((t-s) p_{h,s} +t\delta + |\mathcal{P}|_{\infty}\right) + \frac{\sqrt{2}\|v\|_{\infty}^2 \cdot |\mathcal{P}|^2 (\log t +1)}{ \sigma}\right)& \textnormal{(Block)}
        \end{cases}
    \end{align*}
where $p_{h,s} = 1-\Pr\left[{ A_{h,s}}\right]$ is the failure probability, and $\sigma = \min_i\left\{B_i / h_i^2\right\}$ is the strong convexity modulus of $\mathcal{F}_h$.
\end{restatable}

\begin{restatable}[High-probability implicit bound, nonstationary]{lemma}{nonlemC}
    \label{lem: high-prob-bounds-non}
    For two types of nonstationary inputs respectively, there exists $\Hat{h}(n) = (\Hat{h}_i(n))_{i=1}^n$ and $\Hat{s}(t,n)$:
    \begin{enumerate}
        \item For each $i\in [n]$, $\Hat{h}_i(n) >1 $ and $\Hat{h}_i(n) = O(n)$ and is independent of $t$.
        \item $$\Hat{s}(t,n) \in\begin{cases}
             O(\iota\cdot n (\log n + \log \iota +\log t)) & \textnormal{(Ergodic)} \\
             O({|\mathcal{P}|_{\infty}}\cdot n(\log n +  \log t)) & \textnormal{(Block)}
        \end{cases}.$$
        \item Let $\Hat{A}$ denote $ A_{\Hat{h}(n), \Hat{s}(t,n)}$. Then, under both of the two cases, the failure probability satisfies 
        $1-\Pr[\Hat{A}] = O(1/t).$
    \end{enumerate}
\end{restatable}

We show how to combine \Cref{lem: deterministic-mean-square}, \Cref{lem: dual-regret-non}, and \Cref{lem: high-prob-bounds-non} to get the desired result. For each input model respectively, let $\Hat{h}(n), \Hat{s}(t,n)$ and $\Hat{A}$ be specified as in \Cref{lem: high-prob-bounds-non}.  For time step $r$ with $\Hat{s}(t,n)<r \leq t$, we have
\begin{align}
    \gamma \in \Hat{A}  &\implies \nonumber\\ 
    \|\beta^{r+1}-\beta^*\|^2
    &\leq \frac{2}{\Hat{\sigma}(n) \cdot r} \left[\frac{\max_i\|v_i\|_{\infty}^2}{2\Hat{\sigma}(n)} \log r + \left(\Bar{F}(\beta^*) - \Psi(\Hat{h}(n))\right) \cdot \Hat{s}(t,n)
    -\textnormal{Dual-Regret}(\Hat{s}(t,n),\beta^*)
    \right],\label{eq: deterministic-mean-square-specified-non}
\end{align}

Since $\Hat{h}(n) = O(n)$ still holds for each nonstationary input model (independent of the nonstationarity parameters), the following items have the same asymptotic order as in the stationary case: 
\begin{align}
    \Hat{\sigma}(n) & = \min_{i\in[n]} \left\{B_i / (\Hat{h}_i(n))^2\right\}  \in \Omega(1/n^2), \nonumber \\
    \Bar{F}(\beta^*)-\Psi(\Hat{h}(n)) &= \max_i \beta_i^* \|v_i\|_{\infty} + \sum_{i}B_i\log \frac{\Hat{h}_i}{\beta_i^*}  = O(n \log n), \label{eq: asymptotic-items-non} \\
    G(\Hat{h}(n)) &= \max_{\beta\in \mathcal{F}_{\Hat{h(n)}}} \left\{\max_i \beta_i\|v_i\|_\infty - \sum_{i}B_i \log \beta_i \right\} = O(n). \nonumber
\end{align}

Combining \eqref{eq: deterministic-mean-square-specified-non}, \eqref{eq: asymptotic-items-non} with the dual regret bound (\Cref{lem: dual-regret-non}) and the bound on $\Hat{s}(t,n)$ (\Cref{lem: high-prob-bounds-non}) of each nonstationary model, we have for any $r \in \{\Hat{s}(t,n)+1, \cdots, t\}$, 
\begin{equation}
    \label{eq: deterministic-mean-square-r-non}
    \mathbb{E}\left[ \|\beta^{r+1}-\beta^*\|_2^2 \mid \Hat{A}\right ] = \begin{cases}
    O\left({n^4  \iota \log n \cdot (\log n + \log t+\log \iota )}/{r} + n^3 \delta\right), & \text{(Ergodic)} \\
    O\left({n^4  |\mathcal{P}|_{\infty}^2 \log n (\log n + \log t + \log |\mathcal{P}|_{\infty})}/{r} + n^3 \delta\right), & \text{(Block)} 
    \end{cases}
\end{equation}
Setting $r = t$ and using the fact that $\max\left\{\{n, \iota, |\mathcal{P}|_\infty\right\} \leq t$, we prove \Cref{thm: convergence-multiplier-non}.

We can also derive another inequality by summing up \eqref{eq: deterministic-mean-square-r-non} over $r = \{\Hat{s}(t,n)+1, \cdots, t\}$,
\begin{equation}
    \label{eq: deterministic-mean-square-sum-non}
     \mathbb{E}\left[ \sum_{r=\Hat{s}(t,n)+1}^t\|\beta^{r+1}-\beta^*\|_2^2 \mid \Hat{A}\right ] = \begin{cases}
    O\left({n^4  \iota \log n (\log t)^2} + n^3 \delta t\right), & \text{(Ergodic)} \\
    O\left({n^4  |\mathcal{P}|_{\infty}^2 \log n (\log t)^2} + n^3 \delta t\right), & \text{(Block)} 
    \end{cases}.
\end{equation}
The above inequality \eqref{eq: deterministic-mean-square-sum-non} will be useful in our later analysis for envy and online market equilibrium. 

\paragraph{Convergence of time-averaged utilities.}
The unconditional mean-square convergence of time-averaged utilities can be derived using similar dual-to-primal translation steps as in the \textit{i.i.d.} case:
\begin{align*}
    \mathbb{E}\left[ \|\Bar{u}^t - u^*\|^2 \right]
    &= \mathbb{E}\left[ \|\Bar{u}^t - u^*\|^2 \mid \Hat{A}\right] \Pr\left[\Hat{A}\right] + \mathbb{E}\left[ \|\Bar{u}^t - u^*\|^2 \mid ({\Hat{A}})^c \right] \Pr\left[({\Hat{A}})^c\right]\\
    &\leq \max_i \frac{\|v_i\|_\infty^2}{B_i^2} \cdot \mathbb{E}\left[ \|\beta^{t+1} - \beta^*\|_2^2\mid \Hat{A}\right] + n\max_i\|v_i\|_{\infty}^2 \cdot (1-\Pr[\Hat{A}]) .
\end{align*}

\paragraph{Fairness properties.}
Let $\hat A$ be specified as in \Cref{lem: high-prob-bounds-non}. By \eqref{eq: envy-decomposition-0} and \eqref{eq: envy-decomposition-1}, we have
\begin{align*}
    \mathbb{E}[\eta_{\mathrm{EF}}^t]  \leq t \sqrt{\mathbb{E}\left[\max_{i, k \in [n]}E_{ik}\mid \Hat{A}\right] + \frac{\|v_i\|_{\infty}^2}{\min_{j}B_j^2} \cdot (1-\Pr[\Hat{A}])}.
\end{align*}

Applying \Cref{lem: envy} and \eqref{eq: deterministic-mean-square-sum-non} for the two input model respectively, we get the desired bound on envy. For the proportionality result, by \eqref{eq: proportionality-decomposition} we have
\begin{equation*}
    \mathbb{E}[\eta_{\mathrm{Prop}}^t] \leq \sqrt{\mathbb{E}[(\eta_{\mathrm{{Prop}}}^t)^2]} \leq \sqrt{ t^2 \mathbb{E}(\bar u_i^t - u_i^\gamma)^2}.
\end{equation*}
The result then follows from \Cref{thm: convergence-utility-non}. 

\paragraph{Efficiency properties.}
Under the two nonstationary input models respectively, let $\hat A$ be the corresponding implicit bound event, and $L_{\hat h}$ be the corresponding (constant) Lipschitzness parameter of the weighted geometric mean on the induced utility area. By \eqref{eq: efficiency-nw-decomposition}, 
\begin{equation*}
        \begin{aligned}
    \mathbb{E}[\eta_{\mathrm{NW}}^t]
    &\leq t \|v\|_\infty \cdot \mathrm{Pr}[(\hat A)^c] +  t L_{\hat h} \mathbb E\left[\|\bar u^t - u^\gamma\|^2\right].
        \end{aligned}
    \end{equation*}
Applying \Cref{lem: high-prob-bounds-non} and \Cref{thm: convergence-utility-non} gives us the bounds on expected Nash welfare suboptimality. 

For core stability, by \eqref{eq: core-stability-decomposition} and Jensen's inequality, we have  
\begin{equation*}
        \mathbb{E}\left[\eta^t_{\mathrm{Core}}(\gamma)\right]\leq t \sqrt{\mathbb{E}\|\bar u^t - u^\gamma\|_2^2}.
\end{equation*}
The result then follows from \Cref{thm: convergence-utility-non}. 

\paragraph{Online market equilibrium results.}
By \Cref{lem: convergence-spending} and \Cref{lem: regret}, we have
\begin{align*}
        \left\| \frac{1}{t}\sum_{\tau= s+1}^t b^\tau - B\right\|^2 
        \leq  3\left[ \max_i(\beta_i^*)^2 \cdot 
    \|\Bar{u}^t - u^*\| + n \cdot \|v\|_{\infty}^2 \cdot \left(\frac{s}{t}\right)^2  + \frac{\|v\|_{\infty}^2}{t} \sum_{\tau= s+1}^t \|\beta^\tau - \beta^*\|^2\right]
    ,
\end{align*}
\begin{equation*}
     \left((\mathrm{\mathrm{Demand}}_i(p^1, \cdots, p^\tau)-\bar u_i^t)^+ \right)^2 \leq \frac{2\|v_i\|_{\infty}^2}{(\beta_i^*)^2}\cdot  \left({s^2} + t \sum_{\tau = s+1}^t \|\beta^\tau - \beta^*\|^2\right).
\end{equation*}
For each type of nonstationary input, we take conditional expectation on the corresponding event $\Hat{A}$. The convergence result is then reduced to \eqref{eq: deterministic-mean-square-sum-non} and existing bound for utility convergence, $1-\Pr[\hat A]$ and $\Hat{s}(t,n)$.

\subsection{Proof of \Cref{lem: dual-regret-non}.}
\label{proof: dual-regret-non}
\nonlemB*
We prove \Cref{lem: dual-regret-non} for our two types of inputs respectively. We first introduce some notations in the following proofs. Let $\Hat{Q}$ be the distribution of $v|A_{h,s}$, $\Hat{Q}^{1:\tau}$ be the marginal distribution of $\Hat{Q}$ on the first $\tau$ rounds, and $\Hat{Q}^\tau(\cdot | v^{1:\tau-1})$ be the distribution of $v^\tau$ conditioned on $A_{h,s}$ and the information of previous rounds. We remind the readers that $\phi(\beta)$, i.e. the underlying dual of Eisenberg-Gale program, is defined \textit{w.r.t.} the average distribution across all time steps.

\subsubsection{Proof for ergodic data.} 
Decompose dual regret as follows,

\begin{align}
    \sum_{\tau = s+1}^t \left(F(\beta^\tau, v^\tau) - F(\beta^*, v^\tau)\right) = & 
    \sum_{\tau = s + 1}^{t-\iota} \left(
    \left( F(\beta^\tau, v^{\tau+\iota}) - F(\beta^*, v^{\tau+\iota})  \right)
    - \left( \phi(\beta^\tau) - \phi(\beta^*)\right)
    \right)
     \tag{A}\\
     + &\sum_{\tau = s + 1}^{t-\iota} \left(
        F(\beta^{\tau+\iota}, v^{\tau+\iota}) - F(\beta^{\tau}, v^{\tau+\iota})
     \right) \tag{B}\\
     + &\sum_{\tau = s + 1}^{t-\iota} \left(
        \phi(\beta^{\tau}) - \phi(\beta^{*})
     \right) \tag{C}\\
     + & \sum_{\tau = s + 1}^{s+\iota} \left(
        F(\beta^{\tau}, v^{\tau}) - F(\beta^*, v^{\tau})
     \right) \tag{D}.
\end{align}
\paragraph{Bounding (A).} 
For each $\tau$ we decompose the corresponding term in (A) as 
\begin{equation*}
    \underbrace{\left( F(\beta^\tau, v^{\tau+\iota}) - \phi(\beta^\tau) \right)}_{\text{(A1)}}
    -  \underbrace{\left( F(\beta^*, v^{\tau+\iota}) - \phi(\beta^*)\right)}_{\text{(A2)}}.
\end{equation*}

Note that
\begin{equation*}
    \mathbb{E}\left[\text{(A1)}|A_{h,s}\right] = \mathbb{E}_{v^{1:\tau-1} \sim \Hat{Q}^{1:\tau-1}}\left[ \mathbb{E}_{v^\tau \sim \Hat{Q}^\tau}\left[\left( F(\beta^\tau, v^{\tau+\iota}) - \phi(\beta^\tau)  \right) \mid v^{1:\tau-1}\right]  \right].
\end{equation*}
Let us investigate the inner expectation. Conditional on $v^{1:\tau-1}$,
\begin{align*}
    & \ \ \  \left|\mathbb{E}_{v^\tau \sim \Hat{Q}^\tau}\left[\left( F(\beta^\tau, v^{\tau+\iota}) - \phi(\beta^\tau)  \right) \mid v^{1:\tau-1}\right]\right|\\
    &\leq \mathbb{E}_{v^\tau \sim \Hat{Q}^\tau}\left[\left| F(\beta^\tau, v^{\tau+\iota}) - \phi(\beta^\tau)  \right| \mid v^{1:\tau-1}\right]\\
    & = \left| \int_{0}^{\|v\|_{\infty}} F(\beta^\tau, v^{\tau+\iota}) \Hat{Q}^{\tau+\iota}(\mathrm{d}v|v^{1:\tau-1}) - \int_{0}^{\|v\|_{\infty}}F(\beta^\tau, v) \Bar{Q}(\mathrm{d}v) \right|\\
    &\leq 2G(h) \cdot \|\Hat{Q}^{\tau+\iota}(\cdot|v^{1:\tau}) - \Bar{Q}\|_{\mathrm{TV}}\\
    &\leq 2G(h) \cdot \left(\|\Hat{Q}^{\tau+\iota}(\cdot|v^{1:\tau}) - {Q}^{\tau+\iota}(\cdot|v^{1:\tau})\|_{\mathrm{TV}} + \|{Q}^{\tau+\iota}(\cdot|v^{1:\tau}) - \Bar{Q}\|_{\mathrm{TV}}\right)\\
    &\leq 2G(h) \cdot \left(\|\Hat{Q}^{\tau+\iota}(\cdot|v^{1:\tau}) - {Q}^{\tau+\iota}(\cdot|v^{1:\tau})\|_{\mathrm{TV}} + {\delta}\right),
\end{align*}
where we used the boundedness of $F$ within the implicit bounds.

For the outer expectation, using the same analysis as \eqref{eq: dual-regret-2} we have
\begin{align}
     \left|\mathbb{E}\left[\text{(A1)} \mid A_{h,s}\right]\right|\leq 2G(h) \left( 1-\Pr\left[{ A_{h,s}}\right] + {\delta}\right).
    \label{eq: dual-regret-2-1}
\end{align}

The analysis of (A2) goes through similarly without the outer expectation. 

\begin{align}
    \left|\mathbb{E}\left[\text{(A2)} \mid A_{h,s}\right]\right| \leq 2 \Bar{F}(\beta^*) \left( 1-\Pr\left[{ A_{h,s}}\right] + {\delta}\right).
    \label{eq: dual-regret-2-2}
\end{align}

Combining \eqref{eq: dual-regret-2-1} and \eqref{eq: dual-regret-2-2}, and summing up over $\tau = s+1, \cdots, t$, we have
\begin{align*}
    \left|\mathbb{E}\left[\text{(A)} \mid A_{h,s}\right]\right| &\leq 2 \left(1-\Pr\left[{ A_{h,s}}\right] +\delta\right) \cdot \left(\Bar{F}(\beta^*)+ G(h)\right) \cdot (t-s)\\
    &\leq 4 G(h) \left(1-\Pr\left[{ A_{h,s}}\right] +\delta\right)(t-s).
\end{align*}

\paragraph{Bounding (B).} We use the following property of strongly-convex functions.

\begin{lemma}
    \label{lem: auxiliary-nesterov}
    Let $\Pi_{\Psi, \mathcal{S}}(g):= \arg \min_{w \in \mathcal{S}}\left\{\langle g, w \rangle + \Psi(w) \right\}$. If $\Psi$ is $\sigma$-strongly convex on $\mathcal{S}$, then 
    $$
   \| \Pi_{\Psi, \mathcal{S}}(g) - \Pi_{\Psi, \mathcal{S}}(g^\prime) \|
   \leq 
   (1/\sigma) \|g -g^\prime\|_{*}.
    $$
\end{lemma}

\paragraph{Proof.} See \citet{nesterov2013introductory}, Lemma 6.1.2.
$\hfill\square$

Noticing that $\beta^{\tau+1} = \Pi_{\Psi, \mathcal{F}_h}(\Bar{g}^\tau)$, and that $\Psi$ is $\sigma$-strongly convex on $\mathcal{F}_h$ where $\sigma = \min_{i} B_i/h_i^2$, we know from \Cref{lem: auxiliary-nesterov} that
\begin{equation}
\label{eq: dual-regret-3-2}
    \|\beta^{\tau+1}-\beta^\tau\|_2
    \leq
    \|\Bar{g}^{\tau}-\Bar{g}^{\tau-1}\|_2/\sigma
    \leq
    \|\Bar{g}^{\tau-1}-{g}^{\tau}\|_2/\tau\sigma
    \leq
    \sqrt{2}\cdot\|v\|_{\infty}/\tau\sigma.
    =
    \sqrt{2}\cdot\|v\|_{\infty}/\tau
\end{equation}

For each $\tau$, the map $\beta \mapsto F(\beta,v^{\tau+\iota})$ is Lipschitz with parameter $\|v\|_{\infty}$. Then, 

\begin{align*}
    \left| \mathbb{E}\left[F(\beta^{\tau+\iota}, v^{\tau+\iota} ) - F(\beta^{\tau}, v^{\tau+\iota}) \mid A_{h, s} \right] \right|
    &\leq \|v\|_{\infty} \mathbb{E}\left[\|\beta^{\tau+\iota}-\beta^\tau\| \mid A_{h, s}\right]
    \\
    &\leq \|v\|_{\infty} \sum_{t^\prime = \tau}^{\tau+\iota-1}  \mathbb{E}\left[  \|\beta^{t^\prime + 1}-\beta^{t^\prime}\| \right]\\
    &\leq \|v\|_{\infty}\sum_{t^\prime = \tau}^{\tau+\iota-1}\frac{\sqrt{2} \cdot \|v\|_{\infty}}{\tau \sigma}\\
    &=\frac{\sqrt{2} \|v\|_{\infty}^2 \cdot \iota }{\tau \sigma}.
\end{align*}

Summing over $\tau = s+1, \cdots, t-\iota$, 
\begin{equation*}
    \mathbb{E}\left[|B|\mid A_{h, s}\right]\leq \frac{\sqrt{2}\|v\|_{\infty} \cdot \iota }{\sigma } \cdot \log t.
\end{equation*}

\paragraph{Bounding (C) and (D).}
By optimality of $\beta^*$ we have $\textnormal{(C)} \geq 0$. By the boundedness of $F$ conditional on $A_{h, s}$, we have $|\textnormal{(D)}| \leq 2 \iota \cdot G(h)$. 

\paragraph{Putting together.} Combining the bounds of (A), (B), (C), and (D), we have
\begin{align*}
& \ \ \ \ \mathbb{E}\left[\textnormal{Dual-Regret}(h,s) \mid A_{h, s} \right] \\
&\geq \mathbb{E}\left[A+B+D \mid A_{h, s} \right]\\
&\geq - \mathbb{E}\left[ |A|+|B|+|D| \mid A_{h, s}\right]\\
&\geq - \left(4 \left(1-\Pr\left[{ A_{h,s}}\right] +\delta \right) \cdot G(h) \cdot (t-s)+  \frac{\|v\|_{\infty}^2 \cdot \iota \cdot \sqrt{2} \cdot \log t}{\sigma} + 2\iota \cdot G(h)\right).
\end{align*}

\subsubsection{Proof for block data.}
We decompose the dual regret by blocks. Define $k^\prime:= \max\{k: \tau_k \leq s+1 \}$. For $k\geq k^\prime$, define $\tau^\prime_k = \max\{\tau_k, s+1\}$. Then, 
\begin{align}
    \sum_{\tau = s+1}^t \left(F(\beta^\tau, v^\tau) - F(\beta^*, v^\tau)\right) = & 
    \sum_{k=k^\prime}^K \underbrace{\left(\sum_{\tau\in I_k, \tau>s}\left(F(\beta^{\tau_k^\prime}, v^\tau)-\phi(\beta^{\tau_k^\prime})\right) \right)}_{:=X_k}
     \tag{A}\\
     + &\sum_{k=k^\prime}^K \underbrace{\left(\sum_{\tau\in I_k, \tau>s}\left(F(\beta^\tau, v^\tau) - F(\beta^{\tau_k^\prime}, v^\tau)\right) \right)}_{:=Y_k} \tag{B}\\
     + &\sum_{k=k^\prime}^K \sum_{\tau\in I_k, \tau>s} \left(\phi(\beta^{\tau_k^\prime})-\phi(\beta^*)\right) \tag{C}\\
     + & \sum_{\tau=s+1}^t(\phi(\beta^*)-F(\beta^*, v^\tau)).
      \tag{D}
\end{align}

\paragraph{Bounding (A).}
Notice that
\begin{equation*}
    \mathbb{E}\left[X_k|A_{h,s}\right] = \mathbb{E}_{v^{1:\tau_k^\prime-1} \sim \Hat{Q}^{1:\tau_k^\prime-1}}\left[ \sum_{\tau\in I_k, \tau>s}\mathbb{E}\left[\left(F(\beta^{\tau_k^\prime}, v^\tau) - \phi(\beta^{\tau_k^\prime})\right) \mid v^{1:\tau_k^\prime-1}\right]  \right].
\end{equation*}

We first deal with the inner expectation. The key is, conditional on $v^{1:\tau_k^\prime - 1}$, the dual iterate $\beta^{\tau_k^\prime}$ is deterministic. Let $\Hat{Q}^{(k)}:= \sum_{\tau \in I_k}(\Hat{Q}^{\tau}/|I_k|)$ be the average value distribution within block $k$ conditioned on $A_{h,s}$. Then, 
\begin{align*}
   \ \ \  &\left|\sum_{\tau\in I_k, \tau>s}\mathbb{E}\left[\left(F(\beta^{\tau_k^\prime}, v^\tau) - \phi(\beta^{\tau_k^\prime})\right) \mid v^{1:\tau_k^\prime-1}\right]\right|\\
   = &\left|
    \sum_{\tau \in I_k, \tau>s} \left( \int_{0}^{\|v\|_{\infty}} F(\beta^{\tau_k^\prime}, v) \Hat{Q}^{\tau}(\mathrm{d}v|v^{1:\tau_k^\prime-1}) - \int_{0}^{\|v\|_{\infty}}F(\beta^{\tau_k^\prime}, v) \Bar{Q}(\mathrm{d}v)\right)
   \right|\\
   \leq & \sum_{\tau \in I_k, \tau>s}\left|
     \left( \int_{0}^{\|v\|_{\infty}} F(\beta^{\tau_k^\prime}, v) \Hat{Q}^{\tau}(\mathrm{d}v|v^{1:\tau_k^\prime-1}) - \int_{0}^{\|v\|_{\infty}}F(\beta^{\tau_k^\prime}, v) \Bar{Q}(\mathrm{d}v)\right)
   \right|\\
   \leq & |I_k|\cdot \left|
     \int_{0}^{\|v\|_{\infty}} F(\beta^{\tau_k^\prime}, v) \Hat{Q}^{(k)}(\mathrm{d}v|v^{1:\tau_k^\prime-1}) - \int_{0}^{\|v\|_{\infty}}F(\beta^{\tau_k^\prime}, v) \Bar{Q}(\mathrm{d}v)\right|\\
   \leq & 2 |I_k| \cdot G(h) \cdot \|\Hat{Q}^{\tau}(\cdot|v^{1:\tau_k^\prime-1}) - \Bar{Q}\|_{\mathrm{TV}}\\
   \overset{\text{(a)}}{\leq} & 2 |I_k| \cdot G(h)\cdot \left(\|\Hat{Q}^{(k)}(\cdot|v^{1:\tau_k^\prime-1}) -  {Q}^{(k)}(\cdot|v^{1:\tau_k^\prime-1})\|_{\mathrm{TV}} +\|{Q}^{(k)}- \Bar{Q}\|_{\mathrm{TV}}\right).
\end{align*}
where in the final step (a) we used block-wise independence.

For the outer expectation, using similar analysis to \eqref{eq: dual-regret-2}, we have
\begin{align*}
    \left|\mathbb{E}[X_k \mid A_{h,s}]\right| &\leq 2\cdot |I_k|\cdot G(h)\cdot \left(1-\Pr[A_{h,s}]+\|Q^{(k)}-\Bar{Q}\|_{\mathrm{TV}}\right) + \mathbb{I}\left\{s \in I_k\right\} \cdot 2G(h)|I_k|.
\end{align*}
Summing up over $k\geq k^\prime$, we get
\begin{equation*}
    \left|\mathbb{E}[\text{(A)} \mid A_{h,s}]\right| 
    \leq 2G(h) (t-s)\left(1-\Pr[A_{h,s}]\right) + 2G(h) t \delta + 2 G(h)|\mathcal{P}|_\infty.
\end{equation*}
\paragraph{Bounding (B).}
Using \eqref{eq: dual-regret-3-2}, and the fact that mapping $\beta\mapsto F(\beta, v^{\tau})$ is Lipschitz with parameter $\|v\|_{\infty}$, 

\begin{align*}
    \left|\mathbb{E}[Y_k \mid A_{h,s}]\right|
    &\leq \|v\|_{\infty} \sum_{\tau\in I_k, \tau>s} \mathbb{E}\left[ \|\beta^{\tau}-\beta^{\tau_k^\prime}\|_2 \mid A_{h,s} \right]\\
    &\leq \frac{\sqrt{2}\|v\|_{\infty}^2 \cdot \sum_{\tau\in I_k, \tau>s}(\tau-\tau_k^\prime)}{\tau_k^\prime\cdot \sigma}\\
    &\leq\frac{\sqrt{2}\|v\|_{\infty}^2 \cdot |\mathcal{P}|_{\infty}^2}{\tau_k^\prime\cdot \sigma}.
\end{align*}
Summing over $k = k^\prime, \cdots, K$, we get
\begin{align*}
    \left|\mathbb{E}[\text{(B)} \mid A_{h,s}]\right| \leq\frac{\sqrt{2}\|v\|_{\infty}^2 \cdot |\mathcal{P}|^2 (\log t +1)}{ \sigma}
\end{align*}

\paragraph{Bounding (C) and (D)}
By the optimality of $\beta^*$ we have $\text{(C)}\geq 0$. Meanwhile, (D) can be bounded in a similar way to \eqref{eq: dual-regret-2-4}.
\begin{equation*}
    \left|\mathbb{E}[\text{(D)} \mid A_{h,s}]\right| \leq 2\Bar{F}(\beta^*) (t-s)\left(1-\Pr[A_{h,s}]\right) + 2G(h) t\delta+2G(h)|\mathcal{P}|_{\infty}.
\end{equation*}
\paragraph{Putting together.} Combining the bounds on (A), (B), (C), and (D), we have
\begin{align*}
& \ \ \ \ \mathbb{E}\left[\textnormal{Dual-Regret}(h,s) \mid A_{h, s} \right] \\
&\geq \mathbb{E}\left[A+B+D \mid A_{h, s} \right]\\
&\geq - \mathbb{E}\left[ |A|+|B|+|D| \mid A_{h, s}\right]\\
& \geq -\left( 4G(h) (t-s)\left(1-\Pr[A_{h,s}] \right) + 4G(h) t\delta+ 4G(h)|\mathcal{P}|_{\infty}+\frac{\sqrt{2}\|v\|_{\infty}^2 \cdot |\mathcal{P}|^2 (\log t +1)}{ \sigma}\right).
\end{align*}

\subsection{Proof of \Cref{lem: high-prob-bounds-non}}
\nonlemC*
\label{proof: high-prob-bounds-non}
We use the following notations in the proof: For agent $j\in [n]$, let $T_j:= \left\{\tau \in [T]: x^\tau = 1\right\}$ be the set of items that are allocated to agent $j$ by the PACE algorithm. 
For $\lambda>0$, let $M_j(\lambda)$ be the index set of items which agent $j$ has valuation at least $\lambda$ and was not allocated to agent $j$ by PACE. Define $m_j(\lambda) = |M_j(\lambda)|$. In this proof, we define $i = \min \arg \max_{k\neq j} |M_j(\lambda) \cap T_k|$, and $r:= \max \left\{M_j(\lambda) \cap T_i\right\}$.

We use the proof strategy in \Cref{proof: high-prob-bounds-iid}: For each agent $j$, we are going to show that the following two events occur with high probability, and imply that the pacing multiplier $\beta_j^t$ is bounded as desired. 
Here, ${w}_j$ is a parameter denoting the deviation below the  mean (suitably defined for nonstationary inputs) for the monopolist utility for agent $j$.
\begin{equation}
    \label{eq: high-prob-events-non}
    X:= \left\{ \gamma : \sum_{t^\prime=1}^\tau v_j^{t^\prime} > \left(\mu_j - {w}_j\right) \tau\right\},  \ Y:= \left\{\gamma: \sum_{t^\prime \in M_j(\lambda)\cap T_i} v_i^{t^\prime}> (\mu_i^\prime - {w}_i) \frac{m_j(\lambda)}{n}\right\}, 
\end{equation}
where the ``mean'' constants $\mu_k, \mu_k^\prime (k\in[n])$ are defined as
\begin{align}
    \label{eq: def-mu-non}
    \mu_k &= \begin{cases}
       (1/t)\cdot \sum_{\tau=1}^t \mathbb{E}[v_k^\tau] & \textnormal{(Block)}\\
        (1/t)\cdot \left(\sum_{\tau=1}^{\iota} \mathbb{E}[v_k^\tau] + \sum_{\tau=\iota+1}^t \min_{v^{1:\tau-\iota}} \mathbb{E}[v_k^\tau \mid v^{1:\tau-\iota}]\right)& \textnormal{(Ergodic)} 
    \end{cases}, \\
    \mu_k^\prime &= \begin{cases}
       \min_{\tau \in [t]} \mathbb{E}[v_k^\tau \mid v_k^\tau >0] & \textnormal{(Block)}\\
        \min \left\{ \min_{1\leq \tau\leq \iota} \mathbb{E}[v_k^\tau \mid v_k^\tau >0], \ \ \min_{\iota+1\leq \tau \leq t, v^{1:\tau-\iota} } \mathbb{E}[v_k^\tau|v_k^\tau>0, v^{1:\tau-\iota}] \right\}& \textnormal{(Ergodic)} 
    \end{cases}.
    \label{eq: def-mu-prime-non}
\end{align}
Notice that the above definitions are different from the ones we provide in \Cref{proof: high-prob-bounds-iid}, as we relaxed them due to the presence of non-stationarity. This will give weaker (yet still constant) bounds on agent utilities and multipliers. 

Despite changes in the definition, we can still show that $X$ and $Y$ together imply a constant bound on agent $j$'s time-averaged utility, and then his pacing multiplier. Its derivation is the same as  \eqref{eq: high-prob-bounds-iid-max}:
\begin{equation}
    X \cap Y \implies \Bar{u}_j^\tau \geq \frac{B_j\left(\mu_j - {w}_j - \lambda - \frac{n\|v_i\|_{\infty}(\|v_j\|_{\infty}-\lambda)}{(\mu_i^\prime - {w}_i)\tau}\right)\lambda}{B_j\lambda + \frac{ \|v_i\|_\infty B_i}{(\mu_i^\prime - {w}_i)} \cdot (\|v_j\|_\infty -\lambda) \cdot n}.
\end{equation}

By fixing the value of parameters ${w}_i = \mu_i^\prime/2$ and ${w}_j = \lambda = \mu_j/4$, and taking minimum over the possibility of $i$, we have that for $\tau \geq \frac{8n\max_{k\in[n]}\|v_k\|_{\infty}\|v_j\|_{\infty}}{\mu_k^\prime \mu_j}$,
\begin{equation}
    \label{eq: fix-parameter-non}
    X\cap Y\implies \Bar{u}_j^\tau \geq \Bar{u}_j^\prime := \min\left\{\frac{\mu_j}{4}, \frac{B_j \mu_j^2}{4 B_j \mu_j + 32 n\cdot(\|v_j\|_\infty -\mu_j/4)\cdot \max_k\left\{B_k\|v_k\|_{\infty} / \mu_k^\prime\right\}}\right\}\in \Omega(1/n).
\end{equation}
It remains to show that both $X$ and $Y$ are indeed high-probability events for the two types of nonstationary input models respectively.

\begin{lemma}
\label{lem: high-prob-events-non}
    Let $X$ and $Y$ be defined as in \eqref{eq: high-prob-events-non}, then for the nonstationary input models, it holds that
\begin{equation}
    \label{eq: high-prob-events-x}
    \Pr[X^c] \leq \begin{cases}
        \iota \cdot \exp \left(-\frac{2{w}_j^2 \tau}{\iota \cdot\|v_j\|_\infty^2}\right) & \textnormal{(Ergodic)}\\
                \exp \left(-\frac{2{w}_j^2 \tau}{{|\mathcal{P}|_{\infty}}\cdot\|v_j\|_\infty^2}\right) & \textnormal{(Block)}
    \end{cases},
\end{equation}
and 
\begin{equation}
    \label{eq: high-prob-events-y}
     \Pr[Y^c] \leq \begin{cases}
        \iota \cdot \exp\left(- \frac{2{w}_i^2 \cdot m_j(\lambda)}{\iota\cdot n\cdot \|v_i\|_{\infty}^2}\right) & \textnormal{(Ergodic)}\\
        \exp\left(- \frac{2{w}_i^2 \cdot m_j(\lambda)}{{|\mathcal{P}|_{\infty}}\cdot n\cdot \|v_i\|_{\infty}^2}\right) & \textnormal{(Block)}
    \end{cases}.
\end{equation}
\end{lemma}
\subsubsection{Proof of \Cref{lem: high-prob-events-non}.}
We give separate proofs for the two types of nonstationary input.

\paragraph{Ergodic case.}
For simplicity we assume $\tau/\iota\in \mathbb{N}$; the proof easily generalizes to the case that $\iota$ does not divide $\tau$. We rearrange the value sequence by defining $z_{a, b} = v_j^{a\iota+b}, $ where $a\in\{0, \cdots,  \tau / \iota -1\}, b \in \{1, \cdots, \iota\}$. Consider the sum
\begin{equation*}
    Z_b = \sum_{a=0}^{\tau / \iota -1} z_{a,b}=\sum_{a=0}^{ \tau / \iota -1} v_{j}^{a\iota+b}.
\end{equation*}
By the property of ergodic inputs and the definition of $\mu_j$ in \eqref{eq: def-mu-non}, for $a>0$ we have
\begin{equation*}
    \mathbb{E}[z_{a, b}|z_{a-1,b}] = \mathbb{E}[v_j^{a\iota +b} \mid v^{1:a(\iota-1)+b}] \geq \mu_j.
\end{equation*}
Then, for each $b\in[\iota]$, we apply Azuma-Hoeffding inequality to the sum $Z_b$, 
\begin{equation*}
    \Pr[Z_b \leq (\mu_j - {w}_j) \frac{\tau}{\iota}] \leq \exp\left(\frac{-2 {w}_j^2\tau}{\iota \cdot \|v_j\|_{\infty}^2}\right).
\end{equation*}
Summing up over $b\in [\iota]$, and applying union bound, 
\begin{equation*}
    \Pr[ \sum_b Z_b \leq (\mu_j - {w}_j)\cdot \tau] \leq \iota \cdot \exp\left(\frac{-2 {w}_j^2\tau}{\iota \cdot \|v_j\|_{\infty}}\right).
\end{equation*}
This proves the bound on $\Pr[X^c]$. For the bound on $\Pr[Y^c]$, we notice that the subsequence $\{v^{t^\prime}\}_{t^\prime \in M_j(\lambda)\cap T_i}$ is also ergodic with parameter $\iota$. Therefore, we can run the similar proof on the random variables $S_i^{t^\prime}$, whose sum are stochastically dominated by the sum defined in $Y$. Remark that for ergodic case the definition of $S_i^{t^\prime}$ should be adapted according to \eqref{eq: def-mu-prime-non}.
\begin{equation*}
    S_i^{t^\prime}:= \begin{cases}
        \mathbb{E}\left[v_k^{t^{\prime}} \mid v_k^{t^{\prime}}>0 \right] & 1\leq {t^{\prime}} \leq \iota \\
        \min_{v^{1:{t^{\prime}}-\iota}}\mathbb{E}\left[v_k^{t^{\prime}} \mid v_k^{t^{\prime}}>0, v^{1:{t^{\prime}}-\iota} \right] & \iota+1 \leq t^\prime \leq t
    \end{cases}. \ \ \ \ \text{(Ergodic)}
\end{equation*}

\paragraph{Block case.} We aggregate random variables by blocks. Specifically, define
\begin{equation*}
    V_k := \sum_{t^\prime \leq \tau, t^\prime \in I_k} v_j^{t^\prime}.
\end{equation*}
Then by blockwise independence, $V_1, \cdots, V_K$ are independent random variables in $[0,{|\mathcal{P}|_{\infty}}\cdot \|v_j\|_{\infty}]$.  By Hoeffding's inequality, 

\begin{equation*}
    \Pr\left[\sum_{k=1}^K V_k \leq \sum_{k=1}^K \mathbb{E}[V_k] - {w}_j\tau\right] \leq \exp\left(- \frac{2{w}_i^2 \cdot m_j(\lambda)}{{|\mathcal{P}|_{\infty}}\cdot n\cdot \|v_i\|_{\infty}^2}\right).
\end{equation*}

This proves the bound on $\Pr[X^c]$. The bound on $\Pr[Y^c]$ is by a similar analysis by aggregating $S_i^{t^\prime}$ by blocks, noticing that the subsequence $\{v^{t^\prime}\}_{t^\prime \in M_j(\lambda)\cap T_i}$ also satisfies block-wise independence. For block input, the definition of $S_i^{t^\prime}$ is the same as in \Cref{proof: high-prob-bounds-iid}.

$\hfill \square$

Putting \Cref{lem: high-prob-events-non} and \eqref{eq: fix-parameter-non} together, we have 
\begin{equation}
    \label{eq: high-prob-1-non}
    \Pr[\Bar{u}_j^\tau < \Bar{u}_j^\prime] \in \begin{cases}
       O\left(\iota e^{-c_2\tau/(\iota n)}\right)  & \textnormal{(Ergodic)}\\
       O\left(e^{-c_3\tau/(|\mathcal{P}|_\infty n)}\right)  & \textnormal{(Block)}
    \end{cases}, \ \forall \tau > c_4n, 
\end{equation}
where $c_4 = 8\max_i\|v_i\|_{\infty}\|v_j\|_{\infty}/(\mu_i^\prime \mu_j), \Bar{u}_j^\prime \in \Omega(1/n)$, and $c_1, c_2, c_3$ are also constants independent of $n, t$, and the nonstationary parameters.

Finally, we show how \eqref{eq: high-prob-1-non} leads to \Cref{lem: high-prob-bounds-non}. Notice that for the independent case, the analysis is identical to how \eqref{eq: high-prob-1} leads to \Cref{lem: high-prob-bounds-iid}, see \Cref{proof: high-prob-bounds-iid}. The derivation for ergodic and block case are also similar, which we will specify.

\paragraph{Ergodic case.} The sum of $\Pr[\Bar{u}_j^\tau < \Bar{u}_j^\prime ]$ over $\tau\geq s$ can be bounded by the sum of geometric series, where $s>c_4n$,
\begin{equation*}
    \sum_{\tau\geq s}\Pr[\Bar{u}_j^\tau < \Bar{u}_j^\prime ] = O\left(\iota^2 \cdot n \cdot e^{-c_2 s/(\iota n)}\right).
\end{equation*}
By union bound over $j\in [n]$ and $\tau\geq s$,
\begin{equation*}
    \Pr\left[\exists \tau \geq s, j\in [n], \ \ \Bar{u}_j^\tau < \Bar{u}_j^\prime \right] = O\left(\iota^2 \cdot n^2 \cdot e^{-c_2 s/(\iota n)}\right), \ s>c_4n.
\end{equation*}
Since we have $\beta_j^\tau \geq B_j/\|v_j\|_{\infty} $ always, 
\begin{equation*}
    \forall \tau\geq s, j \in [n], \Bar{u}_j^\tau \geq \Bar{u}_j^\prime  \implies A_{h,s}, \ \left(h_j = B_j/\Bar{u}_j^\prime  = O(n)\right).
\end{equation*}
Hence for $h = (h_1, \cdots, h_n), $ where $h_j = B_j/\Bar{u}_j^\prime $,
\begin{equation*}
    1-\Pr\left[A_{h,s}\right] = O\left(\iota^2 \cdot n^2 \cdot e^{-c_2 s/(\iota n)}\right), \ s>c_4n.
\end{equation*}

Choosing $\Hat{s}(t,n) = \max\{c_4n+1, c_2^{-1}\cdot n\cdot \iota(2\log n+\log t+2\log \iota)\}$ will result in an $O(1/t)$ failure probability, which proves the ergodic case of \Cref{lem: high-prob-bounds-non}. 

\paragraph{Block case.} The sum of $\Pr[\Bar{u}_j^\tau < \Bar{u}_j^\prime ]$ over $\tau\geq s$ can be bounded by the sum of geometric series, where $s>c_4n$,
\begin{equation*}
    \sum_{\tau\geq s}\Pr[\Bar{u}_j^\tau < \Bar{u}_j^\prime ] = O\left({|\mathcal{P}|_{\infty}} \cdot n \cdot e^{-c_3 s/({|\mathcal{P}|_{\infty}} n)}\right).
\end{equation*}
By union bound over $j\in [n]$ and $\tau\geq s$,
\begin{equation*}
    \Pr\left[\exists \tau \geq s, j\in [n], \ \ \Bar{u}_j^\tau < \Bar{u}_j^\prime \right] = O\left({|\mathcal{P}|_{\infty}} \cdot n^2 \cdot e^{-c_3 s/({|\mathcal{P}|_{\infty}} n)}\right), \ s>c_4n.
\end{equation*}
Since we have $\beta_j^\tau \geq B_j/\|v_j\|_{\infty} $ always, 
\begin{equation*}
    \forall \tau\geq s, j \in [n], \Bar{u}_j^\tau \geq \Bar{u}_j^\prime \implies A_{h,s}, \ \left(h_j = B_j/\Bar{u}_j^\prime  = O(n)\right).
\end{equation*}
Hence for $h = (h_1, \cdots, h_n), $ where $h_j = B_j/\Bar{u}_j^\prime $,
\begin{equation*}
    1-\Pr\left[A_{h,s}\right] = O\left({|\mathcal{P}|_{\infty}} \cdot n^2 \cdot e^{-c_3 s/({|\mathcal{P}|_{\infty}} n)}\right), \ s>c_4n.
\end{equation*}

Choosing $\Hat{s}(t,n) = \max\{c_4n+1, c_3^{-1}\cdot n\cdot {|\mathcal{P}|_{\infty}}\cdot (2\log n+\log t+\log {|\mathcal{P}|_{\infty}})\}$ will result in an $O(1/t)$ failure probability, which proves the block case of \Cref{lem: high-prob-bounds-non}.

%% file: best-of-many-worlds/proof-adv.tex
\section{Missing Results and Proofs in \Cref{section: adversarial}}
\label{proof: adv}
\subsection{Proof of \Cref{thm: mul-envy-upper}}
\label{proof: mul-envy-upper}
\advthmA*
According to \Cref{lem: reductive}, it suffices to bound multiplicative envy for all $2$-agent instances. We first show for $2$-agent inputs where PACE gives allocation $x$, 
\begin{equation}
    \label{eq: envy-bound-with-utility}
    \frac{U_{i}(x_j)}{U_i(x_i)} \leq 1+2 \log \frac{1}{\varepsilon}+O\left(\frac{1}{U_i^t}\right), \ \forall j\neq i. 
\end{equation}

Consider the following transformation on any $2$-agent input $\gamma$ to $\gamma^\prime$ as follows:
\begin{enumerate}
    \item Set $v_1^\tau = 0$ for all $\tau \in T_2$.
    \item Reorder the items by moving the columns in $T_2$ to the beginning, and $T_1$ to the end (the order is preserved for items within $T_1$).
\end{enumerate}

We claim that the allocation result under PACE is preserved after the transformation. Clearly, all items in $T_2$ will be allocated to agent $2$, since we have set agent $1$'s valuation on them to zero. For item $\tau \in T_1$, let $\tau^\prime$ be its position in the transformed sequence. We can check that
\begin{equation*}
    \frac{U_1^{\tau^\prime}[\gamma^\prime]}{U_2^{\tau^\prime}[\gamma^\prime]} \leq \frac{U_1^\tau[\gamma]}{U_2^\tau[\gamma]} \leq \frac{B_1 v_1^\tau}{B_2 v_2^\tau},
\end{equation*}
where we use bracket to specify the input instance that results in the utilities. The first inequality is by $U_1^{\tau^\prime}[\gamma^\prime]= U_1^\tau[\gamma]$ and $U_2^{\tau^\prime}[\gamma^\prime] = U_2^{t}[\gamma]\geq U_2^\tau[\gamma]$ in the transformation, and the second inequality is due to $\tau\in T_1$.

Therefore, it suffices to consider only the transformed sequence, i.e., instances with agent $2$ getting his item only in the beginning $R$ consecutive time steps, and agent $1$ gets all items afterwards. Now, the question will be to identify the transformed sequence with worst envy for agent $2$.
A question from an adversarial point of view is: Given that agent $1$ has nothing and agent $2$ has utility $U$ (which is also his total utility) at time step $R$, how to design a forthcoming value sequence, such that 1) agent $2$'s total value over the forthcoming items is maximized, and 2) agent $1$ receives all these forthcoming items? This is characterized by an optimization program (the rounds are re-indexed for the forthcoming sequence):
\begin{equation}
\label{eq: canonical-weighted}
    \begin{aligned}
    \max_{v_1^\tau, v_2^\tau}  & \ \ \frac{B_2}{B_1U}\sum_{\tau=1}^{\infty} v_2^t \\
    \text{s.t.} 
                & \ \ v_1^\tau, v_2^\tau\in \{0\}\cup [\varepsilon,1], \ & \forall \tau \geq 1. \\
                & \ \ \frac{v_2^\tau}{v_1^\tau} \leq \frac{B_1 U}{B_2 U_1^{\tau-1}}, \ &\forall \tau \geq 1.\\
                 & \ \ U_1^{\tau} \geq \sum_{r=1}^{\tau} v_{1}^r, \  &\forall \tau \geq 1. \\
    \end{aligned}
\end{equation}
The constraint in \eqref{eq: canonical-weighted} inherently indicates that all items in the forthcoming sequence are allocated to agent $1$. Notice that infinite sum in the criterion will always converge: when $U_1^{\tau}$ becomes large enough after certain time-steps, any positive $v_2^\tau$ will violate the second constraint; the envy from agent $2$ cannot increase afterwards. It is also clear that assigning $v_1^\tau >0, v_2^\tau =0$ to an item makes no sense. Hence, we can assume without loss of generality that there are only $S$ items in the forthcoming sequence, and each has positive values from both agents; the rest has no effect to the envy bound.

Combining the $\{0\}\cup [\varepsilon,1]$ range with the second constraint, we have a bound on $v_2^\tau/v_1^\tau$ for each time step $\tau$:
\begin{equation*}
    \frac{v_2^\tau}{v_1^\tau} \leq Z(U_1^{\tau-1}) := \begin{cases}
        \min \left\{\frac{B_1U}{B_2U_1^{\tau-1}}, \frac{1}{\varepsilon}\right\}, & U_1^{\tau-1} \in \left[0,\frac{B_1 U}{B_2 \varepsilon}\right] \\
        0, & U_1^{t-1} \in \left(\frac{B_1 U}{B_2 \varepsilon}, \infty\right)
    \end{cases}.
\end{equation*}
By summing up this bound, we can then upper-bound the criterion value of \eqref{eq: canonical}:
\begin{equation}
    \label{eq: q-bound}
    \frac{B_2}{B_1U}\sum_{\tau=1}^\infty v_2^\tau = \frac{B_2}{B_1U}\sum_{\tau=1}^\infty v_1^\tau \cdot \frac{v_2^\tau}{v_1^\tau} \leq \frac{B_2}{B_1U}\sum_{t=1}^S v_1^\tau \cdot Z(U_1^{\tau-1}).
\end{equation}

Next, we show that the right-hand side of \eqref{eq: canonical} can be approximated by an integration:
\begin{align*}
    \frac{B_2}{B_1U} \sum_{\tau =1}^S v_1^\tau Z(U_1^{\tau-1}) - \frac{B_2}{B_1U}  \int_{0}^{U_1^S} Z(u) \mathrm{d}u
    &= 
    \frac{B_2}{B_1U} \sum_{\tau=1}^S \left( v_1^\tau Z(U_1^{\tau-1})- \int_{U_1^{\tau-1}}^{U_1^\tau} Z(u) \mathrm{d}u\right) \\
    &=\frac{B_2}{B_1U} \sum_{\tau=1}^S \left(\int_{U_1^{\tau-1}}^{U_1^\tau} \left( Z(U_1^{\tau-1}) - Z(u)\right) \mathrm{d}u\right)\\
    &\overset{\text{(a)}}{\leq} \frac{B_2}{B_1U} \sum_{\tau=1}^S v_1^\tau \cdot \left(Z(U_1^{\tau-1}) - Z(U_1^\tau)\right)\\
    &\overset{\text{(b)}}{\leq} \frac{B_2}{B_1U} \sum_{\tau=1}^S \frac{B_1 (v_1^\tau)^2}{B_2 \varepsilon^2 U}\\
    &=\frac{1}{\varepsilon^2 (U)^2}\sum_{\tau=1}^S (v_1^\tau)^2 \\
    &\overset{\text{(c)}}{\leq} \frac{U_1^S}{\varepsilon^2 (U)^2},
\end{align*}
where (a) is because $Z(u)$ is non-increasing, (b) is by the fact that the right-side derivative of $Z(u)$ is upper-bounded by $\frac{B_1}{B_2\varepsilon^2 V}$ for all $u>0$, and (c) is simply by $v_1^\tau \leq 1$. Combined with \eqref{eq: q-bound}, 
we have
\begin{equation}
    \label{eq: envy-to-integration-1}
    \frac{U_{2}(x_1)}{U_2(x_2)}\leq \frac{B_2}{B_1 U} \int_{0}^{U_1^{S}}Z(u) \mathrm{d}u +\frac{U_1^S}{\varepsilon^2 (U)^2},
\end{equation}
where $U$ is the total utility of agent $2$.

Now we deal with the integration term in \eqref{eq: envy-to-integration-1}. Since nonzero values are bounded by $[\varepsilon,1]$, the next item (with nonzero value from both agents) in the forthcoming sequence will surely be allocated to agent $2$ if the current utility of agent $1$ is larger than $\frac{B_1U}{B_2\varepsilon}$. Hence, we know the final utility of agent $1$ is bounded by
\begin{equation}
    \label{eq: envy-to-integration-2}
    U_1^S \leq \frac{B_1U}{B_2\varepsilon}+1.
\end{equation}
We can then bound the integration term by
\begin{equation}
    \label{eq: envy-to-integration-3}
    \frac{B_2}{B_1 U} \int_{0}^{U_1^{S}}Z(u) \mathrm{d}u \leq \frac{B_2}{B_1 U} \int_{0}^{\frac{B_1U}{B_2\varepsilon}}Z(u) \mathrm{d}u + \int_{\frac{B_1U}{B_2\varepsilon}}^{\frac{B_1U}{B_2\varepsilon}+1} Z(u) \mathrm{d}u  \leq 1+ 2\log \frac{1}{\varepsilon} + \frac{\varepsilon}{U}.
\end{equation}
Combining \eqref{eq: envy-to-integration-1}, \eqref{eq: envy-to-integration-2}, \eqref{eq: envy-to-integration-3} and the fact that PACE's allocation is invariant to the transformation from $\gamma$ to $\gamma^\prime$, we arrive at
\begin{equation*}
    \rho_{\mathrm{EF}} =\frac{U_{2}(x_1)}{U_2(x_2)} \leq 1+2 \log \frac{1}{\varepsilon}+ \frac{1}{U_2^t} \left(\frac{1}{\varepsilon^3}+\varepsilon\right) + \frac{1}{\varepsilon^2(U_2^t)^2}.
\end{equation*}
As we have discussed, any $n$-agent instance can be reduced to two agents. Since the overall multiplicative envy of agent $i$ is defined by taking maximum over all competing agents $k\neq i$, we have 
\begin{equation*}
    \rho_{\mathrm{EF}}\leq 1+2 \log \frac{1}{\varepsilon}+ \frac{1}{U_i^t} \left(\frac{1}{\varepsilon^3}+\varepsilon\right) + \frac{1}{\varepsilon^2(U_i^t)^2}.
\end{equation*}
This proves \eqref{eq: envy-bound-with-utility}. Now it remains to show how \eqref{eq: envy-bound-with-utility} leads to \Cref{thm: mul-envy-upper}. We only need to notice that
\begin{align*}
    W_i  &= \sum_{k\in[n]}\sum_{\tau \in T_k} v_i^\tau\\
    &\leq U_i^t \cdot \left(1 + \sum_{k\neq i} \frac{B_k}{B_i}\cdot \frac{U_i(x_k)}{U_i^t} \right)\\
    &\leq \frac{\|B\|_1}{B_i}\left(1+ 2(1- B_i/\|B\|_1) \log \frac{1}{\varepsilon}\right)\cdot U_i^t + O(1),
\end{align*}
where the $O(1)$ term is bounded by constants that only depend on $\varepsilon$. The above inequality directly gives the proportionality result, and that $U_i^t \sim W_i$ in the asymptotic term in the multiplicative envy-freeness bound. 

\begin{table}[!htbp]
\caption{A worst-case instance for envy}
\label{tbl: table-envy}
\centering
\begin{tabular}{  | l | l | l | l | l | l | l }
\hline
  Phase & Length & Valuation 1 & Valuation 2 & $U_1$ (after the phase) & $U_2$ (after the phase)\\ \hline
  & & & & &  \\
A1& $R$ & $0$ & $1$ & $0$ & $R$ \\ 
& &  &  & &\\\hline
  & & & & &  \\
A2& $R$ & $\varepsilon$ & $1$ & $\varepsilon \cdot R$ & $R$ \\ 
& &  &  & &\\\hline
  & & & & &  \\
B1& $\left(1 - \frac{1}{a}\right) R$ & $\varepsilon \cdot a^1$ & $1$ & $\varepsilon\cdot a^1 \cdot R$ & $R$ \\ 
& &  &  & &\\\hline
  & & & & &  \\
B2& $\left(1 - \frac{1}{a}\right) R$ & $\varepsilon \cdot a^2$ & $1$ & $\varepsilon\cdot a^2 \cdot R$ & $R$ \\ 
& &  &  & &\\\hline
...& ... & ... & ... & ... & ... \\ \hline
  & & & & &  \\
B$k$& $\left(1 - \frac{1}{a}\right) R$ & $\varepsilon \cdot a^k = 1$ & $1$ & $\varepsilon\cdot a^k \cdot R = R$  & $R$ \\ 
& &  &  & &\\\hline
  & & & & &  \\
C1& $ \frac{1}{\varepsilon}\left(1 - \frac{1}{a}\right) R$ & $\varepsilon \cdot a^1$ & $\varepsilon$ & $a^1 \cdot R$ & $R$ \\ 
& &  &  & &\\\hline
  & & & & &  \\
C2& $ \frac{1}{\varepsilon}\left(1 - \frac{1}{a}\right) R$ & $\varepsilon \cdot a^2$ & $\varepsilon$ & $a^2 \cdot R$ & $R$ \\ 
& &  &  & &\\\hline
...& ... & ... & ... & ... & ... \\ \hline
  & & & & &  \\
C$k$& $ \frac{1}{\varepsilon}\left(1 - \frac{1}{a}\right) R$ & $\varepsilon \cdot a^k = 1$ & $\varepsilon$ & $a^k \cdot R = \frac{1}{\varepsilon} R$ & $R$ \\ 
& &  &  & &\\ \hline
\end{tabular}

\end{table}

\subsection{Proof of \Cref{thm: mul-envy-lower}}
\label{proof: mul-envy-lower}
\advthmB*
    We prove \Cref{thm: mul-envy-lower} by constructing a sequence of instances with growing length, and reaches $1+2\log \frac{1}{\varepsilon}$ multiplicative envy asymptotically. The construction follows the spirit of the transformed sequence instances described in the proof of \Cref{thm: mul-envy-upper}, where agent $2$ receives items in the beginning $R$ consecutive rounds, but nothing afterwards. We construct a hard instance for $B_1 = B_2$; for unequal weights the construction is similar. 
    
    For $R\in \mathbb{N}_{+} $ and $a>1$, we construct an instance $\gamma_{a, R}$ with $2k+2$ phases. Each phase is the repetition of an item with same agent valuations, see \Cref{tbl: table-envy} for details. The table also gives both agents' utilities at the end of each phase.  

One can check that the decision rule is at boundary, i.e. $v_1^\tau/v_2^\tau = U_1^\tau /U_2^\tau$ at the end of each phase, and all items after phase A1 is allocated to agent $1$. Agent $2$ has monopolist utility of 
$$ W_2 = 2\left( 1+ \left(1-\frac{1}{a}\right)\cdot \log_{a}{\frac{1}{\varepsilon}}\right) R.$$

Out of his monopolist utility, agent $2$'s realized utility is only $R$. This gives a multiplicative envy of
\begin{equation*}
    \frac{W_2-R}{R} = 1+ 2\left(1-\frac{1}{a}\right)\cdot \log_{a} \frac{1}{\varepsilon}.
\end{equation*}

Now we let $R \rightarrow \infty$ and $a \rightarrow 1$, then the horizon length of $\gamma_{a,R}$ goes to infinity, and 
\begin{equation*}
    \frac{W_2-R}{R} = 1+ 2\left(1-\frac{1}{a}\right)\cdot \log_{a} \frac{1}{\varepsilon} \rightarrow 1+ 2\log \frac{1}{\varepsilon}.
\end{equation*}

Finally, for the tightness construction of proportionality, one only need to duplicate period A2 to C$k$ multiple times, letting agent $j \in \{3, \cdots, n\}$ take the role of agent $1$ in \Cref{tbl: table-envy}. This proves \Cref{thm: mul-envy-lower}.

\subsection{Proof of \Cref{thm: cr-upper}}
\advthmC*
By scale invarariance to budgets we assume $\sum_{i\in[n]} B_i = 1$ without loss of generality in this proof. We notice that the ratio between monopolist and realized utility under PACE can be bounded by a function of multiplicative envy $\rho_{\mathrm{EF}}$:
\begin{align}
    \frac{W_i}{U_i^t} &= \frac{1}{U_i^t} \sum_{k\in [n]} \sum_{\tau \in T_k} v_i^\tau \nonumber \\
    &= \frac{1}{U_i^t} \sum_{\tau \in T_i} v_i^\tau + \frac{1}{U_i^t} \sum_{k\neq i} \sum_{\tau \in T_k} v_i^\tau \nonumber \\
    &\leq 1 + \frac{\sum_{j\neq i} B_j }{B_i}\cdot \rho_{\mathrm{EF}}. \nonumber
\end{align}
Recall that $U_i^\gamma$ is the utility vector that maximizes the Nash welfare in hindsight. Since the realized utility from any feasible allocation is bounded by the monopolist utility, we have $U_i^\gamma \leq W_i$. Then, by AM-GM inequality, 
\begin{align*}
    \rho_{\mathrm{NW}} &= \sum_{i=1}^n B_i\left(\frac{U_i^\gamma}{U_i^t}\right) \leq \sum_{i=1}^n B_i \left(\frac{W_i}{U_i^t}\right) \\
    &\leq \sum_{i=1}^n B_i \left(1 + \frac{\sum_{j\neq i} B_j}{B_i}\cdot \rho_{\mathrm{EF}}\right)
\end{align*}
Combining the above with the bound for multiplicative envy in \Cref{thm: mul-envy-upper}, we prove the competitive ratio \textit{w.r.t.} Nash welfare. 

For the core stability result, notice that for any coalition $B(S)$ and re-allocation $y$, the utility received by $i$ under that re-allocation is no larger than $W_i$. Therefore, we have $\rho_{\mathrm{Core}}\leq \max_iW_i / U_i^t \leq n \cdot \max_i (1+2\log \frac{1}{\varepsilon} + O(1/W_i)).$

\subsection{Proof of \Cref{thm: cr-lower}}
\advthmD*
Assume agents have equal weights. We consider partitioning the horizon of length $t$ into $n$ phases by $0 = t_0 <\cdots t_{n} = t$. Each phase $S_i$ contains items $\{t_{i-1}+1, \cdots, t_{i}\}$ and has length $|S_i| = t_{i} - t_{i-1}$. As $t\rightarrow \infty$, we let the partition satisfies
\begin{equation*}
    |S_1| \rightarrow \infty, \ \ \lim_{t\rightarrow \infty}\frac{|S_{j-1}|}{|S_j|} = 0 \ \ (\forall j \in\{2, \cdots, n\}).
\end{equation*}
That is to say, each phase has unbounded length as $\gamma^t$ grows infinitely long, but each phase is only negligibly small compared to the next. 

Now we set the values for each phase. For the first phase, we simply set
\begin{equation*}
    v_i^\tau = 1, \ \ \forall i \in [n], \tau \in S_1.
\end{equation*}
For any given online algorithm, we select one agent in each phase $k$ recursively for $k=1, \cdots, n-1$:
$$i_k:=\arg\min_{j} \{U_j^{t_k} : j \not \in \{i_1, \cdots, i_{k-1}\}\},$$
and we set the agent values for phase $k+1$ as
\begin{equation*}
    v_i^{\tau} = \begin{cases}
        0 & i\in\{i_1, \cdots, i_{k}\} \\
        1 & \text{otherwise}
    \end{cases}.
\end{equation*}
That is to say, we erase an agent's value in a phase (by setting it to zero for all items) if the agent has been selected, and we select an agent who has the smallest utility among those whose value are not killed, at the end of each phase. Then, we can recursively show that
\begin{equation*}
    U_{i_k}^t = U_{i_k}^{t_k} \leq \frac{t_k}{n-k+1}.
\end{equation*}
Then, for any given online algorithm the product of agent realized utilities can be upper-bounded by
\begin{equation*}
    \prod_{i=1}^n U_i^{t} = \prod_{k=1}^n U_{i_k}^t \leq \prod_{k=1}^n \frac{t_k}{n-k+1} = \frac{\prod_{k=1}^n t_k}{n!}
\end{equation*}

However, consider the hindsight allocation $x^\prime$ that gives all items in $S_k$ to agent $i_k$. Its realized utilities will be $U_{i_k}^\prime = t_k - t_{k-1}$. We can then bound
\begin{align*}
    \left(\rho_{\mathrm{NW}}\right)^n = \prod_{i=1}^n \frac{U_i^{\gamma_t}}{U_i^t} 
    \geq \prod_{i=1}^n \frac{U_i^\prime}{U_i^t} 
    = \prod_{k=1}^n \frac{U_{i_k}^\prime}{U_{i_k}^t} 
    \geq n! \prod_{k=1}^n \frac{t_k - t_{k-1}}{t_k} \rightarrow n! \ (t\rightarrow \infty)
\end{align*} 
This proves our desired result. 
\subsection{Pseudocode of PACE with seed utilities}
\begin{algorithm}
\caption{PACE with seeds (Unconstrained)}
\label{alg: seed}

    \SetKwInput{KwInit}{Initialization}
    \KwIn{number of agents $n$, time horizon $t$, agent weights $B$, seed $\xi$.}
    \KwInit{$\beta_i^1 = +\infty$.}
    \For{$\tau = 1, \cdots, t$,}{
        Observe agent valuation $v^\tau$. Simulate an auction where agent $i$ bids $\beta_i^\tau v_i^\tau$. 
        The whole item $t$ is allocated to the highest bidder, with lexicographical tie-breaking:
        $$i^\tau := \min \left\{\arg \max_{i\in[n]} \beta_i^\tau v_i^\tau\right\}, \ \  x_i^\tau = \bm{1}(i =i^\tau).$$
        Agent $i$ updates his estimated \textit{seeded} utility
        $ \Bar{u}_i^\tau= \frac{1}{\tau} \cdot x_i^\tau v_i^\tau + \frac{\tau-1}{\tau}\Bar{u}_i^{\tau-1} + \frac{\xi}{\tau}.$\\
        Agent $i$ updates the pacing multiplier 
        $ \beta_i^{\tau+1} = {B_i}/{\Bar{u}_i^\tau}.$
    }
\end{algorithm}

The unconstrained PACE algorithm with seeds is described in \Cref{alg: seed}. Notice that the seeded algorithm is almost the same as \Cref{alg: PACE}, the only difference being that it computes the time-averaged utility as if each agent has an \textit{initial seed utility} $\xi>0$, instead of zero utility, before the arrival of items. 

\subsection{Proof of \Cref{thm: seed-envy}}
\advthmEnvyB*
\begin{proof}
    As shown in the proof of \Cref{thm: mul-envy-upper}, it suffices to consider the class of $2$-agent transformed sequences, where agent $2$ potentially ``envies'' agent $1$'s bundle; the seed utility part is unchanged under the transformation. 
    In this proof we let $Y(U_1)$ be the maximum possible value of $U_2(x_1)$, i.e., agent $2$'s valuation on agent $1$'s bundle under PACE, subject to the constraint that agent 1 receives total utility $U_1$. For simplicity we assume equal weights. For general weights the proof is similar with multiplying $Y(U_1)$ by the weight ratio $B_2 / B_1$. 
    
    Since the seed utility part does not contribute to $U_2(x_1)$, we have $Y(\xi) = 0$. To derive $Y(U_1)$ recursively, we consider the last item in the transformed sequence before the cumulative utility of agent $1$ reaches $U_1$. Suppose agent $1$'s value on that item is $u$. Then this item contributes to $Y(\cdot)$ at most by $U_2^t  u/ (U_1 - u)$. By taking the maximum over $u \in [0, \|v\|_\infty]$, and noticing that $U_1 - u \geq \xi$ in the seed utility setting, we have 
    \begin{equation} 
\label{eq:Y-recursion}
Y(U_1)
\leq 
\max_{\substack{u\in[0,\|v\|_\infty]\\ U_1-u\ge \xi}}
\left\{
Y(U_1-u)+\frac{U_2^t}{U_1-u}\,u
\right\}.
\end{equation}

We show that for any $U_1 \geq \xi$, 
\begin{equation}
    \label{eq: proof-seed-envy-target-0}
    Y(U_1)\le\Psi(U_1):=
\frac{U_2^t\|v\|_\infty}{\xi\,\log\!\bigl(1+\|v\|_\infty/\xi\bigr)}
\log\frac{U_1}{\xi}.
\end{equation}

Fix any admissible $u\in[0,\|v\|_\infty]$ such that $U_1-u\ge \xi$. Then
\[
0\le \frac{u}{U_1-u}\le \frac{\|v\|_\infty}{\xi}.
\]
Also,
\[
\Psi(U_1)-\Psi(U_1-u)
=
\frac{U_2^t\|v\|_\infty}{\xi\,\log\!\bigl(1+\|v\|_\infty/\xi\bigr)}
\log\frac{U_1}{U_1-u}
=
\frac{U_2^t\|v\|_\infty}{\xi\,\log\!\bigl(1+\|v\|_\infty/\xi\bigr)}
\log\!\left(1+\frac{u}{U_1-u}\right).
\]
Notice that $U_2^t \geq \xi >0$. Therefore it suffices to prove that
\begin{equation}
\label{eq: proof-seed-envy-target-1}
    \frac{u}{U_1-u}
\le
\frac{\|v\|_\infty}{\xi\,\log\!\bigl(1+\|v\|_\infty/\xi\bigr)}
\log\!\left(1+\frac{u}{U_1-u}\right).
\end{equation}

Now note that the function $z/\log (1+z)$ is increasing on $(0,+\infty)$. By $\frac{u}{U_1-u}\le \frac{\|v\|_\infty}{\xi}$ we have
\[
\frac{\frac{u}{U_1-u}}{\log\!\left(1+\frac{u}{U_1-u}\right)}
\le
\frac{\|v\|_\infty/\xi}{\log\!\bigl(1+\|v\|_\infty/\xi\bigr)}.
\]
Rearranging gives the desired inequality \eqref{eq: proof-seed-envy-target-1}. 
Taking the maximum over all admissible $u$ in \eqref{eq:Y-recursion}, we prove \eqref{eq: proof-seed-envy-target-0}. Finally, notice that for a length-$t$ online input, the eventual utility of agent $1$ satisfies $U_1^t \leq \xi + t\|v\|_\infty$. Therefore, 
\begin{equation*}
      \frac{U_2(x_1)}{U_2(x_2)} = \frac{Y(U_1^t)}{U_2^t} \leq \frac{Y(\xi + \|v\|_\infty \cdot t)}{U_2^t} \leq \frac{\|v\|_\infty\log \left(1+ \|v\|_\infty t/\xi\right)}{\xi \log (1+\|v\|_\infty/ \xi)} . 
\end{equation*}
We then have
\begin{equation*}
    \rho_{\mathrm{EF}} \leq \max\{1,U_2(x_1)/U_2(x_2)\} = \max\left\{ 1,\frac{\|v\|_\infty \log (1+\|v\|_\infty t/\xi)}{\xi \log (1+\|v\|_\infty/\xi)} \right\}.
\end{equation*}

{For the proportionality result, notice that $\rho_{\mathrm{Prop}} \leq (B_i + \rho_{\mathrm{EF}} \sum_{j\neq i} B_j )/\|B\|_1$ for any algorithm that fully allocates all items. }

\end{proof}

\subsection{Proof of \Cref{thm: seed}}
\label{proof: seed}
\advthmE*
\begin{proof}
    For any input $\gamma$ and given $\xi>0$, define
\begin{equation*}
    R(\gamma):=
    \sup_{\widetilde{U}}\left\{
    \frac{1}{\sum_{i=1}^n B_i}
    \sum_{i=1}^n B_i\cdot\frac{\widetilde{U}_i}{U_i^t}
    \right\},
\end{equation*}
where $\widetilde{U}$ is taken over the space of feasible seeded utilities. We first show that
\begin{equation}
    \label{eq: seed}
    R(\gamma) \leq 3+ \frac{4}{\xi}\|v\|_\infty +2\log \left(1+\frac{\|v\|_\infty\cdot t}{\xi}\right).
\end{equation}

For simplicity we denote the total utility of agent $i$ in seeded PACE by $U_i:=U_i^t$, where $U_i^t$ here includes the initial seed utility $U_i^0\geq \xi$. For the competing allocation, we let $\tilde{x}_i^\tau$ be agent $i$'s received fraction for item $\tau$, and $\widetilde{U}_i$ be the resulting total seeded utility of that allocation, i.e.,
\[
    \widetilde{U}_i=U_i^0+\sum_{\tau=1}^t \tilde{x}_i^\tau v_i^\tau .
\]
Also, as in the Fisher market interpretation of PACE, we denote $p^\tau = \max_{i\in [n]} \beta_i^\tau v_i^\tau$ as the \textit{price} at time step $\tau$, and define $q^\tau$ as
\begin{equation}
    \label{eq: price-bound-0}
    q^\tau :=( \tau-1) \cdot p^\tau = \max_{i\in[n]} \frac{B_i v_i^\tau}{U_i^{\tau-1}} .
\end{equation}

The proof is divided into three major steps.

\paragraph{Step 1. Bounding the ratio with prices.} 
For any feasible hindsight allocation $\tilde{x}_i^\tau$ and its corresponding total seeded utility $(\widetilde{U}_i)$, we show that our target ratio can be bounded by the sum over $q^\tau$ as follows:
\begin{align}
    \sum_{i=1}^n B_i\frac{\widetilde{U}_i}{U_i}
  &=
    \sum_{i=1}^n B_i\frac{U_i^0+\sum_{\tau=1}^t \tilde{x}_i^\tau v_i^\tau}{U_i} 
    \nonumber\\
    &\leq
    \sum_{i=1}^n B_i +\sum_{i=1}^n B_i\frac{\sum_{\tau=1}^t \tilde{x}_i^\tau v_i^\tau}{U_i} 
    \nonumber\\
    &=
    \sum_{i=1}^n B_i +\sum_{i=1}^n  \sum_{\tau=1}^t \left(B_i \cdot \frac{v_i^\tau}{U_i}\cdot \tilde{x}_i^\tau  \right)
    \nonumber \\
    &\leq\sum_{i=1}^n B_i + \sum_{i=1}^n  \sum_{\tau=1}^t \left(B_i \cdot \frac{v_i^\tau}{U_i^{\tau-1}}\cdot \tilde{x}_i^\tau  \right)
    \nonumber \\
    &\overset{\text{(a)}}{\leq}\sum_{i=1}^n B_i + \sum_{i=1}^n \sum_{\tau=1}^t q^\tau \tilde{x}_i^\tau 
    \nonumber \\
    &\overset{\text{(b)}}{\leq} \sum_{i=1}^n B_i + \sum_{\tau=1}^t q^\tau,
    \label{eq: price-bound-1}
\end{align}
where (a) is by the definition of $q^\tau$ in \eqref{eq: price-bound-0}, and (b) is due to the supply feasibility constraint. 

\paragraph{Step 2. Rewrite the price as ratios.} 
Next, we use the integrality of our algorithm to rewrite the prices as ratios. Notice that at time step $\tau$, the entire item is allocated to agent $i^\tau$, which means $x^\tau_{i^\tau} = 1$, and $x^\tau_{i} = 0$ for $i\neq i^\tau$. Hence, we have
\begin{align}
     \sum_{\tau=1}^t q^\tau 
     &= \sum_{\tau=1}^t \frac{B_{i^\tau} v_{i^\tau}^\tau}{U_{i^\tau}^{\tau-1}}
     \nonumber\\
     &= \sum_{\tau=1}^t \frac{B_{i^\tau} v_{i^\tau}^\tau}{U_{i^\tau}^{\tau-1}}\cdot x_{i^\tau}^\tau
     \nonumber\\
     &= \sum_{\tau=1}^t \left(\frac{B_{i^\tau} v_{i^\tau}^\tau}{U_{i^\tau}^{\tau-1}}\cdot x_{i^\tau}^\tau
     + \sum_{i\neq i^\tau }\frac{B_i v_{i}^\tau}{U_{i}^{\tau-1}}\cdot x_{i}^\tau\right)
     \nonumber\\
     &= \sum_{\tau=1}^t \sum_{i=1}^n \frac{B_i v_{i}^\tau}{U_{i}^{\tau-1}}\cdot x_{i}^\tau
     \nonumber\\
     &= \sum_{i=1}^n B_i \left( \sum_{\tau=1}^t \frac{u_i^\tau}{U_{i}^{\tau-1}}\right).
     \label{eq: price-bound-2}
\end{align}

\paragraph{Step 3. Introducing an auxiliary inequality.}
\begin{lemma}[\citet{bach2019universal}]
\label{lem: bach2019}
    For non-negative real numbers $a_1, \cdots, a_n \in [0,a]$ and any $a_0 \geq 0$, it holds that 
    \begin{equation*}
        \sum_{i=1}^n \frac{a_i}{a_0+\sum_{j=1}^{i-1}a_j} \leq 2+\frac{4a}{a_0} + 2\log \left(1+\sum_{i=1}^{n-1} \frac{a_i}{a_0} \right).
    \end{equation*}
\end{lemma}
We can use the above inequality to bound the right hand side of \eqref{eq: price-bound-2}. Notice that $u_i^\tau \in [0,\|v\|_{\infty}]$ and $U_i^0\geq \xi$, we have
\begin{align*}
     \sum_{\tau=1}^t \frac{u_i^\tau}{U_{i}^{\tau-1}} 
     &=
     \sum_{\tau=1}^t \frac{u_i^\tau}{U_i^0+\sum_{s=1}^{\tau-1} u_i^s} \\
     &\leq 
     2 + 4 \frac{\|v\|_{\infty}}{U_i^0} + 2\log \left(1+\sum_{\tau=1}^{t-1}\frac{u_i^\tau}{U_i^0}\right)
     \\
     &\leq  2 + 4 \frac{\|v\|_{\infty}}{\xi} + 2\log \left(1+\frac{\|v\|_{\infty}\cdot t}{\xi}\right).
\end{align*}
Summing up over $i\in[n]$, we have
\begin{equation}
    \label{eq: price-bound-3}
    \sum_{i=1}^n B_i \left( \sum_{\tau=1}^t \frac{u_i^\tau}{U_{i}^{\tau-1}}\right) \leq \left(\sum_{i=1}^n B_i\right) \cdot \left( 2 + 4 \frac{\|v\|_{\infty}}{\xi} +  2\log \left(1+\frac{\|v\|_{\infty}\cdot t}{\xi}\right) \right).
\end{equation}
Putting \eqref{eq: price-bound-1}, \eqref{eq: price-bound-2}, and \eqref{eq: price-bound-3} together, we have
\[
    R(\gamma)
    \leq
    3 + 4 \frac{\|v\|_{\infty}}{\xi}
    +  2\log \left(1+\frac{\|v\|_{\infty}\cdot t}{\xi}\right).
\]

Finally, we relate this ratio bound to Nash welfare. For any feasible competing seeded utility vector $\widetilde{U}$, by the weighted AM-GM inequality,
\begin{align*}
    \prod_{i=1}^n
    \left(\frac{\widetilde{U}_i}{U_i^t}\right)^{B_i/\sum_{j=1}^n B_j}\leq
    \frac{1}{\sum_{j=1}^n B_j}
    \sum_{i=1}^n B_i\frac{\widetilde{U}_i}{U_i^t}
    \leq R(\gamma).
\end{align*}
Taking the supremum over all feasible competing allocations gives the bound on $\rho_{\mathrm{NW}}$. For a bound on multiplicative core stability, notice that for any coalition $B(S)$ and re-allocation $y$, the utility received by $i$ under that re-allocation is no larger than $W_i$. Therefore, we have $\rho_{\mathrm{Core}}\leq \max_iW_i / U_i^t \leq n \rho_{\mathrm{EF}}$. 
\end{proof}

\subsection{Set-aside PACE algorithm for the known monopolist utility case.}
\label{proof: adv-set-aside}

We show how the seed algorithm can extend to interesting variants when the monopolist utility of each agent $W_i$ is known a priori. 
The Set-aside PACE algorithm in \Cref{alg: set-aside} uses this prior knowledge to give stronger adversarial performance guarantees by allocating half of each item uniformly among the agents.

\begin{algorithm}
\caption{Set-aside PACE (Unconstrained)}
\label{alg: set-aside}
    \SetKwInput{KwInit}{Initialization}
    \KwIn{number of agents $n$, time horizon $t$, agent weights $B$, monopolist utilities $W_1, \cdots, W_n$.}
    \KwInit{$\beta_i^1 = +\infty$, $\gamma_i = W_i/2n$}
    \For{$\tau = 1, \cdots, t$,}{
        Allocate half the item using the decision rule of PACE. The other half is proportionally allocated to each agent according to their budgets.
        $$i^\tau := \min \left\{\arg \max_{i\in[n]} \beta_i^\tau v_i^\tau\right\}, \ \  x_i^\tau = \frac{1}{2n}+ \frac{1}{2} \cdot \bm{1}(i =i^\tau).$$
        Agent $i$ updates his estimated seeded utility
        $$ \Bar{u}_i^\tau= \frac{1}{\tau} \cdot x_i^\tau v_i^\tau + \frac{\tau-1}{\tau}\Bar{u}_i^{\tau-1} + \frac{\xi_i}{\tau}.$$
        Agent $i$ updates the pacing multiplier, normalized by his monopolist utility: 
        $$ \beta_i^{\tau+1} = \frac{B_i W_i}{\Bar{u}_i^\tau}$$
    }
\end{algorithm}

\Cref{alg: set-aside} can be interpreted as the seeded PACE algorithm with seed utility $\xi_i = W_i/2n$ for each agent. Interestingly, each agent does not receive this seed as an initial utility. Instead, the algorithm sets aside half of each item for pure proportional allocation, and it knows that this will become a guaranteed share reserved for each agent in the beginning. Hence, it treats this reserved half as the seed, and runs the decision rule of \Cref{alg: seed} to allocate the remaining half of each item. By normalizing each agent's value by $W_i$, we can invoke \Cref{thm: seed} with $\xi=1/2n$. 

\begin{lemma}
\label{lem: set-aside}
For arbitrary input $\gamma$ with known monopolist utility, \Cref{alg: set-aside} satisfies
    \begin{equation}
        \label{eq: set-aside}
        \mathrm{CR}(\gamma) \leq 6+ \frac{16n}{\min_{i} W_i} \|v\|_\infty+4\log \left(1+\frac{2n\|v\|_\infty}{\min_{i} W_i}\right)+4\log t.
    \end{equation}
\end{lemma} 

We remark that \Cref{alg: set-aside} gives a bound on the competitive ratio \textit{w.r.t.} NW of the set-aside algorithm. \Cref{lem: set-aside} yields a result similar to a previous set-aside type algorithm proposed by \citet{banerjee2022online}. Their result also has a $O(\log t)$ dependence on the horizon length, though their result can also fit the more general case where the predictions on $W_i$ are inaccurate. On the other hand, their algorithm requires solving a convex program at each time step, and is thus more complicated than \Cref{alg: set-aside}. Both set-aside algorithms require additional knowledge about monopolist utilities compared to (vanilla) PACE, and neither is desirable for stochastic input since they allocate half of each item uniformly.

%% file: best-of-many-worlds/proof-applications.tex
\section{Missing Results and Proofs in \Cref{section: app}}
\label{section: blood-donation-proof}
\subsection{Proof of \Cref{thm:blood-fairness-efficiency-metrics}}

\appthmA*

\begin{proof}
Notice that \Cref{alg: blood} has unknown number of iterations, since the number of active donors is stochastic, and thus does not belong to any existing input models in this paper. However, for the purpose of analysis, we notice that \Cref{alg: blood} is equivalent to the following padded algorithm \Cref{alg: blood-with-padding}, in the sense that they give identical allocation for the same input. When an inactive donor appears, \Cref{alg: blood-with-padding} still explicitly calls a zero-valued iteration of PACE and increases the iteration counter $\tau$. This padding trick forces the input to PACE to have fixed length $|\mathcal{J}|D$, and length-$|\mathcal{J}|$ block independent structure. One can easily check that padding does not change the decision of the algorithm. It then makes sense to restrict our attention to the performance of \Cref{alg: blood-with-padding}. We remark that, while the blockwise case in \Cref{thm: convergence-utility-non} gives a direct bound for \Cref{alg: blood-with-padding} due to block-structure, here we are going to derive a better bound that involves the graph structure parameter $\Delta_{\mathcal{G}}$, and the fact that some donors are inactive within a day in this setting. 

\begin{algorithm}[!htbp]
\caption{{PACE for Fair Matching in Blood Donation Platforms (with padding)}}
\label{alg: blood-with-padding}
    \SetKwInput{KwInit}{Initialization}
    \KwIn{recipient set $\mathcal{I}$, donor set $\mathcal{J}$, number of days $D$, recipient weights $B=(B_i)_{i\in\mathcal{I}}$.}
    \KwInit{$n\leftarrow |\mathcal{I}|$, horizon length $t\leftarrow |\mathcal{J}|D$, iteration count $\tau\leftarrow 1$, cumulative utilities $U_i^0\leftarrow 0$ for all $i\in\mathcal{I}$.}
    
    \For{$d = 1, \cdots, D$}{
        Observe the daily donation graph $\mathcal{G}^d=(\mathcal{I},\mathcal{J},\mathcal{E}^d,w^d)$.\\
        Initialize $x^d_{ij}\leftarrow 0$ for all $i\in\mathcal{I}$ and $j\in\mathcal{J}$.\\
        Find the active and inactive donor sets
        \[
            \mathcal{J}_1^d:=\{j\in \mathcal{J}: \exists i\in \mathcal{I},\, w^d_{ij}>0\},
            \qquad 
            \mathcal{J}_0^d:=\mathcal{J}\backslash\mathcal{J}_1^d .
        \]
        
        \For{each donor $j \in \mathcal{J}_1^d$, in arbitrary order}{
            Set the current value vector by $v_i^\tau\leftarrow w_{ij}^d$ for all $i\in\mathcal{I}$.\\
            Choose the recipient according to one iteration of PACE:
            \[
                i^\tau \in \min \arg\max_{i\in\mathcal{I}} B_i\frac{v_i^\tau}{U_i^{\tau-1}} .
            \]
            Recommend donor $j$ to recipient $i^\tau$ by setting $x^d_{i^\tau j}\leftarrow 1$.\\
            Update cumulative utilities: for all $i\in\mathcal{I}$,
            \[
                U_i^\tau \leftarrow U_i^{\tau-1}+\mathbb{I}\{i=i^\tau\}v_i^\tau .
            \]
            Set $\tau \leftarrow \tau+1$.
        }
        \For{each donor $j \in \mathcal{J}_0^d$, in arbitrary order}{
            Run one padded zero-value iteration with $v_i^\tau\leftarrow 0$ for all $i\in\mathcal{I}$.\\
            Set $U_i^\tau\leftarrow U_i^{\tau-1}$ for all $i\in\mathcal{I}$.\\
            Set $\tau \leftarrow \tau+1$.
        }
    }
\end{algorithm}

Now we focus on the performance of \Cref{alg: blood-with-padding} as a block input case with special structure; all notations are consistent with \Cref{sec: blockwise}, where the number of agents is $n = |\mathcal{I}|$, the number of blocks is $D$, and the total number of padded PACE iterations is $t=|\mathcal{J}|D$. For compactness, define
\[
    \Lambda_{\mathrm{BD}}
    :=
    \Delta_{\mathcal{G}}\log n + \mathbb{E}[m_d] + \frac{\mathbb{E}[m_d^2]}{|\mathcal{J}|}.
\]
We show that there exists high probability event $\Hat{A}$ such that
\begin{equation}
\label{eq: proof-blood-1}
    1-\Pr[\Hat{A}] = O\!\left(\frac{1}{t|\mathcal{J}|^2}\right), \ \ \
    \mathbb{E}\left[\|\beta^{t+1}-\beta^*\|_2^2 \ | \ \Hat{A}\right]
    =
    O\left(\frac{n^4\log t}{t}\Lambda_{\mathrm{BD}}\right).
\end{equation}

The above \eqref{eq: proof-blood-1} follows from the combination of \Cref{lem: deterministic-mean-square}, the dual regret bound \Cref{lem: dual-regret-blood}, and the implicit high-probability bound \Cref{lem: high-prob-bounds-blood}, in the same pattern as in the proof of the stationary and non-stationary cases. The proof of \Cref{lem: dual-regret-blood} and \Cref{lem: high-prob-bounds-blood} are deferred to the next subsections. 
\begin{restatable}[Dual regret bound, blood donation]{lemma}{applemA}
    \label{lem: dual-regret-blood}
    It holds that
    \begin{align*}
        &
        \mathbb{E}\left[\textnormal{Dual-Regret}(s,\beta^*) \mid A_{h,s}\right] \geq 
            - \left(\left(4 G(h)\cdot (t-s)) \cdot p_{h,s} \right) + \frac{4G(h)\cdot\mathbb{E}[m_d]}{1-p_{h,s}} + \frac{\sqrt{2}\|v\|_\infty^2\cdot \mathbb{E}[m_d^2]}{\sigma (1-p_{h,s})|\mathcal{J}|} (\log D+1)\right),
    \end{align*}
where $p_{h,s} = 1-\Pr\left[{ A_{h,s}}\right]$ is the failure probability, and $\sigma = \min_i\left\{B_i / h_i^2\right\}$ is the strong convexity modulus of $\mathcal{F}_h$.
\end{restatable}
\begin{restatable}[High-probability implicit bound, blood donation]{lemma}{applemB}
    \label{lem: high-prob-bounds-blood}
    There exists $\Hat{h}(n) = (\Hat{h}_i(n))_{i=1}^n$ and $\Hat{s}(t,n)$:
    \begin{enumerate}
        \item For each $i\in [n]$, $\Hat{h}_i(n) >1 $ and $\Hat{h}_i(n) = O(n)$ and is independent of $t$.
        \item $$\Hat{s}(t,n) =
             O({\Delta_{\mathcal{G}}} \cdot n \cdot (\log n +  \log t + \log |\mathcal{J}|)) 
            $$
        \item Let $\Hat{A}$ denote $ A_{\Hat{h}(n), \Hat{s}(t,n)}$. Then, the failure probability satisfies 
        $1-\Pr[\Hat{A}] = O(1/(t\cdot |\mathcal{J}|^2)).$
    \end{enumerate}
\end{restatable}

Finally, we show how to get mean-square convergence in average weights from \eqref{eq: proof-blood-1}. Notice that $|\Bar{w}_i-w_i^*|$ is upper-bounded by $|\mathcal{J}|\cdot \|v\|_\infty$ even when $\Hat{A}$ does not hold. 

\begin{align*}
    \mathbb{E}\left[ \|\Bar{w} - w^*\|^2 \right]
    &= \mathbb{E}\left[ \|\Bar{w} - w^*\|^2  \mid \Hat{A}\right] \Pr\left[\Hat{A}\right] + \mathbb{E}\left[ \|\Bar{w} - w^*\|^2 \mid ({\Hat{A}})^c \right] \Pr\left[({\Hat{A}})^c\right]\\
    &\leq |\mathcal{J}|^2 \cdot\max_i \frac{\|v_i\|_\infty^2}{B_i^2} \cdot \mathbb{E}\left[\sum_{i=1}^n \left(\frac{B_i}{\Bar{w}_i} - \frac{B_i}{w_i^*}\right)^2 \ \big{|} \Hat{A}\right] + n|\mathcal{J}|^2\max_i\|v_i\|_{\infty}^2 \cdot (1-\Pr[\Hat{A}])\\
    &= \max_i \frac{\|v_i\|_\infty^2}{B_i^2} \cdot \mathbb{E}\left[ \|\beta^{t+1}-\beta^*\|^2 \mid \Hat{A} \right] + n|\mathcal{J}|^2\max_i\|v_i\|_{\infty}^2 \cdot (1-\Pr[\Hat{A}]),
\end{align*}
where in the last step we used the fact that $\Bar{w}_i = |\mathcal{J}|B_i/\beta_i^{t+1}$ since a day has $|\mathcal{J}|$ padded iterations. Combined with \eqref{eq: proof-blood-1}, this gives us the desired bound for utility convergence in \Cref{thm:blood-fairness-efficiency-metrics}. The same argument gives convergence to the realized hindsight benchmark $w^\gamma$, and the stated fairness and efficiency guarantees follow from the same utility-convergence-to-fairness-efficiency conversion as in \Cref{section: stationary} and \Cref{section: nonstationary}.

\end{proof}
\subsection{Proof of \Cref{lem: dual-regret-blood}.} 
\applemA*
\begin{proof}
For consistency of notation, we use $\tau_d:=|\mathcal{J}|(d-1)+1$ to denote the index of the first PACE iteration on day $d$. Define $d^\prime:= \max\{d: \tau_d \leq s+1 \}$. For $d\geq d^\prime$, define $\tau^\prime_d = \max\{\tau_d, s+1\}$. 

As in the block case, we decompose the dual regret by days, which are independent, length-$|\mathcal{J}|$ blocks with identical distribution, and we will use the term block or day interchangeably in the proof. Further, we notice that zero-valued iterations, i.e., steps with $v^{\tau} = 0$, do not add to the dual regret. We make this explicit using an indicator $\mathbb{I}\{v^\tau \neq 0\}$.
\begin{align}
    \sum_{\tau = s+1}^t \left(F(\beta^\tau, v^\tau) - F(\beta^*, v^\tau)\right) = & 
    \sum_{\tau = s+1}^t \mathbb{I}\left\{\tau>s, v^\tau \neq 0\right\}\left(F(\beta^\tau, v^\tau) - F(\beta^*, v^\tau)\right)\nonumber\\
    = &
    \sum_{d=d^\prime}^D \underbrace{\left(\sum_{\tau=\tau_d}^{\tau_{d}+|\mathcal{J}|-1}\mathbb{I}\left\{\tau>s, v^\tau \neq 0\right\}\left(F(\beta^{\tau_d^\prime}, v^\tau)-\phi(\beta^{\tau_d^\prime})\right) \right)}_{:=X_d}
     \tag{A}\\
     + &\sum_{d=d^\prime}^D \underbrace{\left(\sum_{\tau=\tau_d}^{\tau_{d}+|\mathcal{J}|-1}\mathbb{I}\left\{\tau>s, v^\tau \neq 0\right\}\left(F(\beta^\tau, v^\tau) - F(\beta^{\tau_d^\prime}, v^\tau)\right) \right)}_{:=Y_d} \tag{B}\\
     + &\sum_{d=d^\prime}^D \sum_{\tau=\tau_d}^{\tau_{d}+|\mathcal{J}|-1}\mathbb{I}\left\{\tau>s, v^\tau \neq 0\right\}\left(\phi(\beta^{\tau_d^\prime})-\phi(\beta^*)\right) \tag{C}\\
     + & \sum_{\tau=s+1}^t\mathbb{I}\left\{\tau>s, v^\tau \neq 0\right\}(\phi(\beta^*)-F(\beta^*, v^\tau)).
      \tag{D}
\end{align}

\paragraph{Bounding (A).}
To bound (A), we treat each day as an independent block, and re-use the corresponding part in the proof of \Cref{lem: dual-regret-non}, with an additional indicator $\mathbb{I}\{v^\tau \neq 0\}$. Since we assumed that days have identical distribution, we can drop the perturbation term in the analysis. 

\begin{align*}
    \left|\mathbb{E}[X_d \mid A_{h,s}]\right| &\leq 2\cdot |\mathcal{J}|\cdot G(h)\left(1-\Pr[A_{h,s}]\right) + \mathbb{I}\left\{s \in \{\tau_d, \cdots, \tau_{d+1}-1\}\right\} \cdot 2G(h)\mathbb{E}[m_d | A_{h,s} ].
\end{align*}
Notice that the second term is nonzero for the block $d$ where $s$ is in. Summing up over $d\geq d^\prime$, we get
\begin{equation*}
    \left|\mathbb{E}[\text{(A)} \mid A_{h,s}]\right| 
    \leq 2G(h) (t-s)\left(1-\Pr[A_{h,s}]\right) + \frac{2G(h)\mathbb{E}[m_d]}{\Pr[A_{h,s}]}.
\end{equation*}
\paragraph{Bounding (B).}
Using \eqref{eq: dual-regret-3-2}, and the fact that mapping $\beta\mapsto F(\beta, v^{\tau})$ is Lipschitz with parameter $\|v\|_{\infty}$, we have

\begin{align*}
    \left|\mathbb{E}[Y_d \mid A_{h,s}]\right|
    &\leq \|v\|_{\infty}  \cdot \mathbb{E}\left[ \sum_{\tau = \tau_d^\prime}^{\tau_{d+1}^\prime-1} \mathbb{I}\left\{v^\tau \neq 0\right\}\|\beta^{\tau}-\beta^{\tau_d^\prime}\|_2 \mid A_{h,s} \right]\\
    &\leq \frac{\sqrt{2}\|v\|_{\infty}^2 \cdot \mathbb{E}\left[\sum_{\tau = \tau_d^\prime}^{\tau_{d+1}^\prime-1} \mathbb{I}\left\{v^\tau \neq 0\right\}(\tau - \tau_d^\prime)\right] \ | \ A_{h,s}}{\tau_d^\prime\cdot \sigma}\\
    &\leq\frac{\sqrt{2}\|v\|_{\infty}^2 \cdot \mathbb{E}[m_d^2]}{\tau_d^\prime\cdot \sigma \Pr[A_{h,s}]} .
\end{align*}
Notice that in the last inequality we applied the structure of the block padding that all non-zero valued donors appear in the beginning rounds within a block. 
Summing over $d = d^\prime, \cdots, D$, and recall that $\tau_d = (d-1)|\mathcal{J}|+1$, we get
\begin{align*}
    \left|\mathbb{E}[\text{(B)} \mid A_{h,s}]\right| \leq
     \frac{\sqrt{2}\|v\|_\infty^2\cdot \mathbb{E}[m_d^2]}{\sigma (1-p_{h,s})|\mathcal{J}|} (\log D+1).
\end{align*}

\paragraph{Bounding (C) and (D)}
By the optimality of $\beta^*$ we have $\text{(C)}\geq 0$. Meanwhile, (D) can be bounded in a similar way to \eqref{eq: dual-regret-2-4}, which results in the same upper bound with (A).
\paragraph{Putting together.} Combining the bounds on (A), (B), (C), and (D), we have
\begin{align*}
& \ \ \ \ \mathbb{E}\left[\textnormal{Dual-Regret}(s,\beta^*) \mid A_{h, s} \right] \\
&\geq \mathbb{E}\left[\text{(A)}+\text{(B)}+\text{(D)} \mid A_{h, s} \right]\\
&\geq - \mathbb{E}\left[ |\text{(A)}|+|\text{(B)}|+|\text{(D)}| \mid A_{h, s}\right]\\
& \geq - \left(\left(4 G(h)\cdot (t-s)) \cdot p_{h,s} \right) + \frac{4G(h)\cdot\mathbb{E}[m_d]}{1-p_{h,s}} + \frac{\sqrt{2}\|v\|_\infty^2\cdot \mathbb{E}[m_d^2]}{\sigma (1-p_{h,s})|\mathcal{J}|} (\log D+1)\right).
\end{align*}

\end{proof}

\subsection{Proof of \Cref{lem: high-prob-bounds-blood}.} 
\applemB*
\begin{proof}
We show how to bound the failure probability of the events $X$ and $Y$ defined in \eqref{eq: high-prob-events-non}.
This will give us a bound that depends on the maximum degree $\Delta_{\mathcal{G}}$, which is more refined than directly applying the bounds from block case. The rest of the proof follows exactly the same pattern as previous cases. 

\begin{lemma}
\label{lem: high-prob-events-app}
    Let $X$ and $Y$ be defined as in \eqref{eq: high-prob-events-non}, with $\mu_i, \mu_i^\prime$ given by
\begin{equation*}
     \mu_i = \sum_{\tau=1}^t \mathbb{E}[v_i^\tau], \ \ \mu_i^\prime=\min_{\tau \in [t]} \mathbb{E}[v_i^\tau \mid v_i^\tau >0].
\end{equation*}
    Then for the stochastic blood donation setting, 
\begin{equation}
    \label{eq: high-prob-events-x-y-app}
    \Pr[X^c] \leq 
                \exp \left(-\frac{2{w}_i^2 \tau}{\Delta_{\mathcal{G}}\cdot\|v_i\|_\infty^2}\right), \ 
    \Pr[Y^c] \leq 
                \exp\left(- \frac{2{w}_i^2 \cdot m_i(\lambda)}{\Delta_{\mathcal{G}}\cdot n\cdot \|v_i\|_{\infty}^2}\right).
\end{equation}
\end{lemma}
\subsubsection{Proof of \Cref{lem: high-prob-events-app}.}

We aggregate random variables by days, and let $I_d:=\{(d-1)|\mathcal{J}|+1, \cdots, d|\mathcal{J}|\}$ denote the block generated from day $d$. Specifically, define
\begin{equation*}
    V_d := \sum_{t^\prime \leq \tau, t^\prime \in I_d} v_i^{t^\prime}.
\end{equation*}
Then by independence of input from different days, $V_1, \cdots, V_D$ are independent random variables in $[0,{\Delta_{\mathcal{G}}}\cdot \|v_i\|_{\infty}]$. Note that the upper bound ${\Delta_{\mathcal{G}}}\cdot \|v_i\|_{\infty}$ is by the fact that recipient $i$ has non-zero value for at most $\Delta_{\mathcal{G}}$ active donors in day $d$. By Hoeffding's inequality, 
\begin{equation*}
    \Pr\left[\sum_{k=1}^K V_k \leq \sum_{k=1}^K \mathbb{E}[V_k] - {w}_i\tau\right] \leq \exp\left(- \frac{2{w}_i^2 \tau}{\Delta_{\mathcal{G}}\cdot \|v_i\|_{\infty}^2}\right).
\end{equation*}

This proves the bound on $\Pr[X^c]$. The bound on $\Pr[Y^c]$ is by a similar analysis by aggregating $S_i^{t^\prime}$, with the same definition as in \Cref{proof: high-prob-bounds-iid} and \Cref{proof: high-prob-bounds-non}. 
\end{proof}

\subsection{Pseudocode of Quasilinear PACE}
\label{section: alg-ql-appendix}

We give the detailed description of how to cast PACE as a multi-bidder dynamics in \Cref{alg: QL-1}. 
\begin{algorithm}[!htbp] \caption{PACE for Repeated First-Price Auctions with Budgets} \label{alg: QL-1} \SetKwInput{KwInit}{Initialization} \KwIn{number of bidders $n$, per-round budgets $B$.} \KwInit{$\beta^1 = 1^n$, and $\Bar{u}_i^0=0$ for all $i\in[n]$.} \For{$\tau = 1, \cdots, t$}{ Observe bidder values $v^\tau$. Each bidder $i$ submits the paced bid $\beta_i^\tau v_i^\tau$.\\ Allocate the impression to the highest paced bidder, with lexicographical tie-breaking: \[ i^\tau := \min \left\{\arg \max_{i\in[n]} \beta_i^\tau v_i^\tau\right\}, \qquad x_i^\tau = \bm{1}(i=i^\tau). \] Charge the first-price payment and record bidder spending: \[ b_i^\tau = x_i^\tau \cdot \max_{j\in[n]} \beta_j^\tau v_j^\tau . \] Each bidder updates her time-averaged gross utility: \[ \Bar{u}_i^\tau = \frac{1}{\tau}\,x_i^\tau v_i^\tau + \frac{\tau-1}{\tau}\Bar{u}_i^{\tau-1}. \] Each bidder updates her pacing multiplier: \[ \beta_i^{\tau+1} = \min \left\{\frac{B_i}{\Bar{u}_i^\tau},1\right\}. \] } \end{algorithm}

\subsection{Missing Results for Quasi-linear PACE, Stochastic Input}

\begin{theorem}[Nonstationary input, FPPE] \label{thm: ergodic-QL} \label{thm: block-QL} For \Cref{alg: QL-1} under ergodic or blockwise independent inputs, it holds that \begin{equation*} \mathbb{E} \left[\left\| \Bar{u}^{\ql,t} - u^{\mathrm{\ql, *}}\right\|^2\right] = \begin{cases} O\left(\dfrac{n^4 \iota \log n (\log t)^2}{t}+n^3 \delta\right), & \gamma \sim Q \in \mathsf{C}^{\mathrm{E}}(\delta,\iota),\\[0.5em] O\left(\dfrac{n^4 |\mathcal{P}|_{\infty}^2 \log n (\log t)^2}{t}+n^3 \delta\right), & \gamma \sim Q \in \mathsf{C}^{\mathrm{B}}(\mathcal{P},\delta). \end{cases} \end{equation*} \end{theorem}

\paragraph{Proof of \Cref{thm: convergence-ql-iid} and \Cref{thm: ergodic-QL}}
We show the proof for the stationary case; for the ergodic and block case, the proof is similar to the \textit{i.i.d.} case, with same treatment of nonstationary as in previous proofs. 

Now that \Cref{alg: QL-1} has an explicit upper bound $1$ on dual variable, we know that the event $A_{-,1} = \{\forall \tau \geq 0,  \forall i\in[n], \beta_i^\tau \leq 1\}$ holds deterministically. We can then drop the preliminary in \Cref{lem: deterministic-mean-square}, and set $s = 0$. This will give us 
\begin{equation}
\label{eq: ql-iid-proof}
    \|\beta^{t+1} - \beta\|^2 \leq \frac{2}{\sigma t}\left[\frac{\max_i\|v_i\|_\infty^2}{2\sigma}\log t - \text{Dual-Regret}(1,\beta)\right].
\end{equation}
Similarly, the dual regret (with respect to $\beta^*$) can be independently bounded without the implicit restricting technique (or equivalently, setting $h=  1, s=1$ unconditionally in the previously defined boundedness event $A_{h,s}$). Combined with \eqref{eq: ql-iid-proof}, we have 
\begin{equation}
\label{eq: ql-iid-beta}
    \mathbb{E}\left[\|\beta^{t+1}- \beta^*\|^2\right] = O\left(\frac{n^4\log n \log t}{t}\right). 
\end{equation}

To further get convergence in FPPE utilities, we introduce the following KKT conditions~\cite{cole2017convex} for the quasi-linear convex program.

\begin{proposition}[KKT conditions for FPPE] The optimal primal-dual solutions to \eqref{eq: inf-QL-primal} and \eqref{eq: inf-QL-dual} satisfies
\begin{itemize}
    \item $\delta_i^*(1-\beta_i^*) = 0$  for all $i\in[n]$. 
    \item $u_i^* = B_i/\beta_i^* - \delta_i^*$ for all $i\in[n]$. 
    \item $\langle p_i^* - \beta_i^*v_i^*, x_i^*\rangle = 0$  for all $i\in[n]$. 
\end{itemize}
\end{proposition}
We then consider two cases: 1) $\beta_i^* = 1$, and 2) $\beta_i^*<1$. 

\paragraph{Case $\beta_i^* = 1$: } By the KKT conditions, the underlying FPPE has $u_i^{\ql, *} = B_i$. We have
\begin{align*}
    (\Bar{u}_i^{\ql, t} -u_i^{\ql, *})^2 &= (\Bar{u}_i^t - \Bar{b}_i^t)^2 \\
    &= \frac{1}{t^2} \left( \sum_{\tau=1}^t (u_i^\tau - b_i^\tau)\right)^2 \\
    &\leq  \frac{1}{t} \sum_{\tau=1}^t (u_i^\tau - b_i^\tau)^2 \\
    &\leq \frac{1}{t}\|v_i\|_\infty^2 \cdot \sum_{\tau=1}^t (1-\beta_i^\tau)^2 \\
    &= \frac{1}{t}\|v_i\|_\infty^2 \cdot \sum_{\tau=1}^t (\beta_i^\tau - \beta_i^*)^2.
\end{align*}
Taking expectations and applying \eqref{eq: ql-iid-beta} gives us the desired result. 

\paragraph{Case $\beta_i^* < 1$:} By the KKT conditions, in the underlying FPPE agent $i$ completely depletes the budget, i.e., $b_i^* = B_i, \delta_i^* = 0$, and therefore $u_i^* = B_i/\beta_i^*$. Hence, 
\begin{align}
    \nonumber \mathbb{E}(\Bar{u}_i^t - u_i^*)^2 &= \mathbb{E}\left[\mathbb{I}\{\beta_i^{t+1}<1\} \left(\frac{B_i}{\beta_i^{t+1}} - \frac{B_i}{\beta_i^*} \right)^2\right] + \mathbb{E}\left[\mathbb{I}\left\{\beta_i^{t+1} = 1\right\}\left(u_i^t - u_i^*\right)^2 \right] \\
    \nonumber & \leq \frac{\|v_i\|_\infty^2}{B_i^2} \cdot \mathbb{E}(\beta^{t+1} - \beta^*)^2 + \|v_i\|_\infty^2\Pr[\beta_i^{t+1} = 1] \\
    \nonumber & \leq \frac{\|v_i\|_\infty^2}{B_i^2} \cdot \mathbb{E}(\beta^{t+1} - \beta^*)^2 + \|v_i\|_\infty^2\Pr[(\beta_i^{t+1} - \beta_i^*)^2 \geq (1-\beta_i^*)^2] \\
    & \leq \|v_i\|_\infty^2\left( \frac{1}{B_i^2} +\frac{1}{(1-\beta_i^*)^2}\right)\cdot \mathbb{E}(\beta^{t+1} - \beta^*)^2. \label{eq: ql-iid-gross}
\end{align}
The last step is by Markov's inequality. Next, we show that the time-averaged expenditure $\Bar{b}_i^t$ of agents is close to $B_i$. Using similar decomposition as in the proof of \Cref{lem: convergence-spending}, we have
\begin{equation}
    \label{eq: ql-iid-spending}
    \mathbb{E}(\Bar{b}_i^t - B_i)^2 \leq 2\cdot \mathbb{E}(\Bar{u}_i^t - u_i^*)^2 + \frac{\|v_i\|_\infty^2}{t} \sum_{\tau=1}^t\mathbb{E}(\beta_i^\tau - \beta_i^*)^2 .
\end{equation}
Combining \eqref{eq: ql-iid-gross} and \eqref{eq: ql-iid-spending} by the inequality $(x-y)^2 \le 2(x^2 + y^2)$, we get
\begin{equation*}
    (\Bar{u}_i^{\ql, t} - u_i^{\ql, *})^2 \leq 3\|v_i\|_\infty^2\left( \frac{1}{B_i^2} +\frac{1}{(1-\beta_i^*)^2}\right)\cdot \mathbb{E}(\beta^{t+1} - \beta^*)^2 + \frac{\|v_i\|_\infty^2}{t} \sum_{\tau=1}^t\mathbb{E}(\beta_i^\tau - \beta_i^*)^2. 
\end{equation*}
Combining with \eqref{eq: ql-iid-beta} gives us the desired bound. 

$\hfill \square$

\subsection{Proof of \Cref{thm: ql-adv}}

\qlthmmain*
\begin{proof}
Similar to the proof of \Cref{thm: seed}, we define a utility ratio metric which is an upper bound on $\rho_{\mathrm{QLNW}}$ by the AM-GM inequality:
\begin{equation*}
    R^{\mathrm{QL}}({\gamma}):=\sup_{\widetilde{U}, \widetilde{\Delta}}\left\{\sum_{i=1}^n \frac{B_i}{\|B\|_1}\cdot\frac{\widetilde{U}_i + \widetilde{\Delta}_i}{U_i^{\ql, t}}\right\},
\end{equation*}
Since in \Cref{alg: QL-1} an agent never bids more than his or her true valuation, the net profit is always non-negative in each round, i.e., the quasi-linear utility is at least the budget. Thereby we have $U_i^t+\Delta_i^t \geq t B_i \geq \widetilde{\Delta}_i$. Then,
\begin{equation}
    \sum_{i=1}^n B_i\frac{\widetilde{U}_i +\widetilde{\Delta}_i}{U_i^t + \Delta_i^t} =  \sum_{i=1}^n B_i\frac{\widetilde{\Delta}_i}{U_i+\Delta_i} 
    +\sum_{i=1}^n B_i\frac{\widetilde{U}_i}{U_i^t + \Delta_i^t} \leq \sum_{i=1}^n B_i + \sum_{i=1}^n B_i\frac{\widetilde{U}_i}{U_i^t + \Delta_i^t} . 
\label{eq: adv-ql-decomposition}
\end{equation}

We further bound the second term on the right-hand side. 
\begin{align}
   \sum_{i=1}^n B_i\frac{\widetilde{U}_i}{U_i^t + \Delta_i^t} 
    &\leq
    \sum_{i=1}^n B_i\frac{\sum_{\tau=1}^t \tilde{x}_i^\tau v_i^\tau}{U_i^t + \Delta_i^t} 
    \nonumber\\
    &= \sum_{i=1}^n  \sum_{\tau=1}^t \left(B_i \cdot \frac{v_i^\tau}{U_i^t + \Delta_i^t} \cdot \tilde{x}_i^\tau \right)
    \nonumber \\
    &\leq \sum_{i=1}^n  \sum_{\tau=1}^t \left( \frac{B_i}{U_i^t + \Delta_i^t}  v_i^\tau \cdot \tilde{x}_i^\tau \right)
    \nonumber \\
    &\overset{\text{(a)}}{\leq} \sum_{i=1}^n  \sum_{\tau=1}^t \left( \min \left\{\frac{1}{t}, \frac{B_i}{U_i^{\tau-1}} \right\} \cdot v_i^\tau  \tilde{x}_i^\tau \right)
    \nonumber \\
    &\overset{\text{(b)}}{\leq} \frac{\|v\|_\infty}{t} + \sum_{\tau=2}^t \underbrace{\max_{i\in[n]} \min\left\{\frac{v_i^\tau}{t}, \frac{B_i v_i^\tau}{U_i^{\tau-1}}\right\}}_{:= q^\tau}. 
    \nonumber 
\end{align}
where (a) is by the non-negative profit property i.e., $U_i^t + \Delta_i^t \geq B_i$, and (b) is by the unit supply constraint on each item. For simplicity, we define the summand on the right-hand side as $q^\tau$ for each $\tau \geq 2$. definition of $q^\tau$. Combined with \eqref{eq: adv-ql-decomposition}, we have
\begin{equation}
    R^{\mathrm{QL}}(\gamma) \leq \sum_{i=1}^n B_i +  \frac{\|v\|_\infty}{t} + \sum_{\tau=2}^t q^\tau 
        \label{eq: ql-adv-price-bound-1}
\end{equation}

Next, we show how to bound the sum of $q^\tau$. By the contraction property of the min operator, we have for $\tau \geq 2$,
\begin{equation*}
    \arg \max_{i\in[n]} \beta_i^\tau v_i^\tau  = \arg \max_{i\in[n]} \min\left\{\frac{v_i^\tau}{\tau - 1}, \frac{B_i v_i^\tau}{U_i^{\tau-1}}\right\} \subseteq\arg \max_{i\in[n]} \min\left\{\frac{v_i^\tau}{t}, \frac{B_i v_i^\tau}{U_i^{\tau-1}}\right\}.
\end{equation*}
Therefore for the $i^\tau$ selected by \Cref{alg: QL-1} we have $i^\tau \in \arg \max_{i\in[n]}\min\left\{\frac{v_i^\tau}{t}, \frac{B_i v_i^\tau}{U_i^{\tau-1}}\right\}$. By the integrality of the algorithm, we have
\begin{align}
     \sum_{\tau=2}^t q^\tau 
     &= \sum_{\tau=2}^t \min\left\{\frac{1}{t}, \frac{B_i }{U_{i^\tau}^{\tau-1}} \right\}v_{i^\tau}^\tau
     \nonumber\\
     &= \sum_{\tau=2}^t \min\left\{\frac{1}{t}, \frac{B_i }{U_{i^\tau}^{\tau-1}} \right\}\cdot v_{i^\tau}^\tau x_{i^\tau}^\tau
     \nonumber\\
     &= \sum_{\tau=1}^t \left(\min\left\{\frac{1}{t}, \frac{B_i }{U_{i^\tau}^{\tau-1}} \right\}\cdot v_{i^\tau}^\tau x_{i^\tau}^\tau
     + \sum_{i\neq i^\tau }\min\left\{\frac{1}{t}, \frac{B_i }{U_{i}^{\tau-1}} \right\}
     \cdot v_i^\tau x_{i}^\tau\right)
     \nonumber\\
     &= \sum_{\tau=1}^t \sum_{i=1}^n \min\left\{\frac{1}{t}, \frac{B_i }{U_{i}^{\tau-1}} \right\}\cdot u_{i}^\tau.
     \label{eq: ql-adv-price-to-ratio}
\end{align}

For any $\eta >0$ and $i\in [n]$, consider a partition of $[t]$ by $T_i^-(\eta) = \{\tau: U_i^\tau \leq \eta\}$ and $T_i^+(\eta) = [t] \backslash T_i^-(\eta)$. Then for $\tau$ such that $\tau - 1 \in T_i^+(\eta)$, we have
\begin{equation}
    \label{eq: ql-adv-parameter-auxiliary}
    \min\left\{ \frac{u_i^\tau}{tB_i}, \frac{u_i^\tau}{U_i^{\tau-1}}\right\} \leq \frac{2u_i^\tau}{U_i^{\tau-1}+\eta}.
\end{equation}
Using the partition to decompose the sum on the right-hand side of \eqref{eq: ql-adv-price-to-ratio}, we get
\begin{align*}
    \sum_{\tau=2}^t q^\tau &\leq \sum_{i=1}^n \sum_{ \tau: \tau-1\in T_{i}^-(\eta)} \min\left\{\frac{1}{t}, \frac{B_i }{U_{i}^{\tau-1}} \right\}\cdot u_{i}^\tau +\sum_{i=1}^n \sum_{ \tau: \tau-1\in T_{i}^+(\eta)} \min\left\{\frac{1}{t}, \frac{B_i }{U_{i}^{\tau-1}} \right\}\cdot u_{i}^\tau\\
    &\leq\frac{1}{t}\sum_{i=1}^n \sum_{ \tau: \tau-1\in T_{i}^-(\eta)}u_i^\tau + \sum_{i=1}^n B_i \cdot \sum_{\tau:\tau-1\in T_i^+(\eta)} \min \left\{\frac{u_i^\tau}{tB_i},\frac{u_i^\tau}{U_i^{\tau-1}}\right\}  \\
    &\overset{\text{(c)}}{\leq} \frac{n}{t}(\eta+\|v\|_\infty) + \sum_{i=1}^n B_i \cdot \sum_{\tau:\tau-1\in T_i^+(\eta)} \frac{2 u_i^\tau}{U_i^{\tau-1} + \eta}\\
    &\overset{\text{(d)}}{\leq} \frac{n}{t}(\eta+\|v\|_\infty) + \left( \sum_{i=1}^n B_i\right) \left(4 + 8\frac{\|v\|_\infty}{\eta} + 4 \log \left(1+\frac{t\|v\|_\infty}{\eta} \right)\right).
\end{align*}

In step (c), we used the definition of $T_i^-(\eta)$ and the inequality \eqref{eq: ql-adv-parameter-auxiliary}. In step (d) we used the auxiliary inequality from \Cref{lem: bach2019}. By choosing $\eta = t\cdot \sqrt{(\sum_{i=1}^n B_i)\|v\|_\infty/n}$ and combining with \eqref{eq: ql-adv-price-bound-1}, we get
\begin{equation*}
    \sum_{i=1}^n B_i \frac{\widetilde{U}_i + \widetilde{\Delta}_i}{U_i^t + \Delta_i^t} \leq \left(1+\frac{8}{t}\right) \sqrt{n\|v\|_\infty \sum_{i=1}^n B_i} +  \frac{(n+1)\|v\|_\infty}{t} + \left( \sum_{i=1}^n B_i\right)\left(5 +   4 \log \left(1+ 
    \sqrt{\frac{n\|v\|_\infty}{\sum_{i=1}^n B_i}}\right)\right).
\end{equation*}

By the AM-GM inequality, this proves the desired upper bound on ${\rho}_{\mathrm{QLNW}}$
\end{proof}

\subsection{Proof of \Cref{thm: justified-envy}}
\appthmB*
\begin{proof}
Without loss of generality assume all $B_i$ are equal; the general budget case is follows similarly after adapting by the budget ratio. We first prove the justified envy result. For any given $i, j \in [n], i\neq j$, we investigate $i$'s value on $j$'s allocation, restricted to the subset that $v_i^\tau > p^\tau$. Recall that $T_j = \{\tau : x_j^\tau = 1\}$, and by the first-price rule we have $p^\tau = \beta_j^\tau v_j^\tau$ for items $\tau \in T_j$. We partition $T_j = T_j^{(1)}\cup T_j^{(2)}$ according to the pacing multiplier of $i$ as follows. 
\begin{equation*}
    T_j^{(1)} := T_j \cap \left\{\tau: \beta_i^\tau = 1\right\}, \ T_j^{(1)} := T_j - T_j^{(1)} = T_j \cap \left\{\tau: \beta_i^\tau < 1\right\}.
\end{equation*}
For items in $T_j^{(1)}$, we have $v_i^\tau = \beta_i^\tau v_i^\tau \leq \beta_j^\tau v_j^\tau = p_j $, i.e., the price is no less than agent $i$'s valuation. We then know that $T_j^{(1)}$ does not contribute to agent $i$'s justified envy. 
For item $ \tau \in T_j^{(2)}$, we have
\begin{equation*}
    v_i^\tau \leq \frac{p^\tau}{\beta_i^\tau} = \frac{\beta_j^\tau v_j^\tau }{\beta_i^\tau} \leq \frac{B_j v_j^\tau }{\Bar{u}_j^\tau \beta_i^\tau }  =\frac{B_j \Bar{u}_i^\tau v_j^\tau }{B_i\Bar{u}_j^\tau } = \frac{U_i^\tau v_j^\tau}{ U_j^\tau} \leq\frac{U_i^t v_j^\tau}{U_j^\tau} \leq \frac{U_i^t v_j^\tau}{ \sum_{s \in T_j^{(2)}, s \leq \tau} v_j^s}.
\end{equation*}
Summing up over $\tau \in T_j^{(2)}$, this gives
\begin{equation*}
    \sum_{\tau \in T_j^{(2)}}v_i^\tau \leq   U_i^t\cdot \sum_{\tau \in T_j^{(2)}} \frac{v_j^\tau}{\sum_{s \in T_j^{(2)},s \leq \tau} v_j^s}.
\end{equation*}
Similar to the proof of multiplicative envy bound in the standard setting, now we reduce the problem of finding the maximum value from agent $i$ on the subset $T_j^{(2)}$ to the sum of a series of utility increments as in the above right-hand side. Notice that on subset $T_j^{(2)}$ the projection of \Cref{alg: QL-1} is not active, and thus has the same decision as PACE. We can then reuse the analysis from the proof of \Cref{thm: mul-envy-upper}.

Next we show the quasi-linear envy bound. Let $\tau^*:= \max\{ \tau \in [t]: \sum_{s\leq\tau} p^{s} \cdot x_i^s  \leq tB_i\}$ be the last timestep where the total spending of agent $i$ does not exceed the budget $tB_i$; $\tau^* = t$ if the initial budget is not depleted. Let $k\neq i$ be the agent that gives the maximum envy value for $i$. Notice that the total quasi-linear utility is always no less than $tB_i$, we can then decompose

\begin{align*}
    \mathrm{\rho}_{\mathrm{QLEF}} &= \frac{\sum_{\tau\leq \tau^*} (v_i^\tau-p^\tau)\cdot x_k^s + \sum_{\tau \geq \tau^*} (v_i^\tau - p^\tau)\cdot x_k^\tau +tB_i}{U_i^t+ \Delta_i^t}\\
    &\leq \underbrace{\frac{\sum_{\tau\leq \tau^*} v_i^\tau \cdot x_k^\tau\cdot \mathbb{I}\{p^\tau \leq v_i^\tau\}} {\max\{U_i^{\tau^*}, tB_i\}}}_{\text{(I)}} +  \underbrace{\frac{(t-\tau^*) \|v\|_\infty + tB_i}{tB_i}}_{\text{(II)}}.
\end{align*}

For term (I), we notice that it can be upper-bounded by the justified (multiplicative) envy on the instance that consists only by the first $\tau^*$ rounds. We can thus reuse the analysis in the first half of the proof. Notice that denominator is at least the order of $t$, we then have $\text{(I)} \leq 1+ 2\log (1/\varepsilon) + O(1/t).$

For term (II), we first lower-bound $\tau^*$. Notice that the budget expenditure of agent $i$ per round is at most $\|v\|_\infty.$ Then $\tau^* \geq tB_i /\|v\|_\infty$ and $\text{(II)} \leq \|v\|_\infty /B_i$.

\end{proof}

%% file: best-of-many-worlds/appendix-experiments.tex
\section{More Discussions on the Experiments}
\label{appendix: experiments}
\subsection{Dataset, Pre-processing, and Benchmark}
We introduce the two datasets, and how they have been preprocessed to generate online fair allocation instances.

\paragraph{The Instagram notification dataset.} The Instagram notifications dataset~\citep{kroer2023fair} contains information of generated notifications for $T_0=41072$ users, each being a potential recipient of $n=4$ notification \emph{types}: Comment Subscribed, Feed Suite Organic Campaign, Like, and Story Daily Digest. Each row of the dataset represents a notification event (i.e. the opportunity to send a user one notification), which includes (1) available notification types; (2) the user for whom the notification is generated; (3) the utility that each notification type would receive if their notification gets shown; (4) timestamp.  
\citet{kroer2023fair} describes the motivation for this model in detail. Briefly, the idea is that different engineering teams at Instagram own each notification type, and they are each optimizing their own metrics with respect to their notifications. However, if each team is allowed to send as many notifications as they like, it can lead to a tragedy-of-the-commons situation where users receive too many notifications even though each individual team is rationally maximizing their own metric. A CEEI-type auction system is thus used to control how many notifications are sent to users, and this is where the data is derived from.
We assume that each user has unit capacity, i.e., only one notification can be sent for a given notification event. The task for online fair allocation is to distribute recommendations (items to be allocated) to the notification types (agents in the market) in a fair manner, given the arriving order of the users. 

\paragraph{The MovieLens dataset.} The MovieLens dataset we use in our experiment describes rating activities by $m=610$ users on the MovieLens recommendation platform across 9742 movies, including the timestamps of their $T_0=100836$ rating activities~\citep{harper2015movielens}. 
We use this dataset to construct a fair online allocation problem where the buyers that wish to be recommended are different genres of movies, and the items are impression opportunities that arise every time a user visits the platform. We note that in a real-world system, it would be more typical if each movie is a buyer. We use genres because it gives us a better complete matrix of valuations for conducting simulations. 
Using the labels from the platform, we select the $n=10$ most frequent movie genres, and assume that a user's average score for a genre is correlated linearly with their probability of accepting a recommended movie from this genre (0 means always reject, and 5 means always accept). Using matrix completion methods, we get a user-genre matrix with acceptance probabilities. By further assuming that each genre's utility is the expected number of acceptances it gets, the matrix induces the problem instance where recommendation opportunities to a user upon an access (items) are allocated to movie genres (agents), given certain arrival sequence across $m$ user types. 

Given the valuation matrix generated from the above two datasets, we simulate PACE on three different types of item arrival sequences: (1) stationary (\textit{i.i.d.}), (2)  periodic, and (3) true temporal, respectively. For both datasets, we set agent weights to be equal, i.e. $B_i = 1/n$, and normalize the valuations such that $(1/t)\cdot\sum_{\tau=1}^t v_i^\tau = 1$ for each agent $i$.

\begin{itemize}
    \item \textit{Stationary (\textit{i.i.d.}) input.} We sample from the $t_0$ events in the given dataset repeatedly to generate an instance of length $t_1 = 200,000$. At each timestep, one event is sampled uniformly and independently at random among all events in the dataset (with replacement). 
    \item \textit{Periodic input.} We construct periodic input instances that fall under the model in \Cref{sec: blockwise}, with a long period length $q=t_0/8$. At the $j$'th timestep ($j\in[q]$) within a period, the event is sampled uniformly at random from $S_j$, where $S_j =\{8(j-1)+1,\cdots, 8j\}$ is the set of contiguous events, each with size $8$. The length of the generated instance is expanded to $t_1 = 200,000$.
    \item \textit{True-temporal input.} The input follows the real-world time order in which the events occurred, according to the timestamps in the datasets. 
\end{itemize}

\paragraph{Additional details on benchmark algorithms.}
We now clarify the details of how the benchmark algorithms are implemented.

\begin{itemize}
    \item The SAG algorithm. The SAG algorithm requires a prediction of $W_i=\sum_{\tau=1}^t v_i^\tau$ for each agent $i$. We run SAG with the exact values of these quantities as predictions.
    \item Online subgradient descent and mirror descent. We implement these algorithms with a diminishing learning rate of the form $c/\sqrt{\tau}$. This is the standard theoretically motivated stepsize order for general convex but non-strongly-convex objectives such as the dual of EG. We optimize over the constant $c$ separately for each instance, up to a grid discretization of $0.2$, and report the results obtained with the best choice of $c$.
\end{itemize}

\subsection{Experiments on the Failure of Constrained PACE}
\begin{figure*}[t] 
    \centering
    \begin{subfigure}
	\centering
\includegraphics[width=0.49\linewidth]{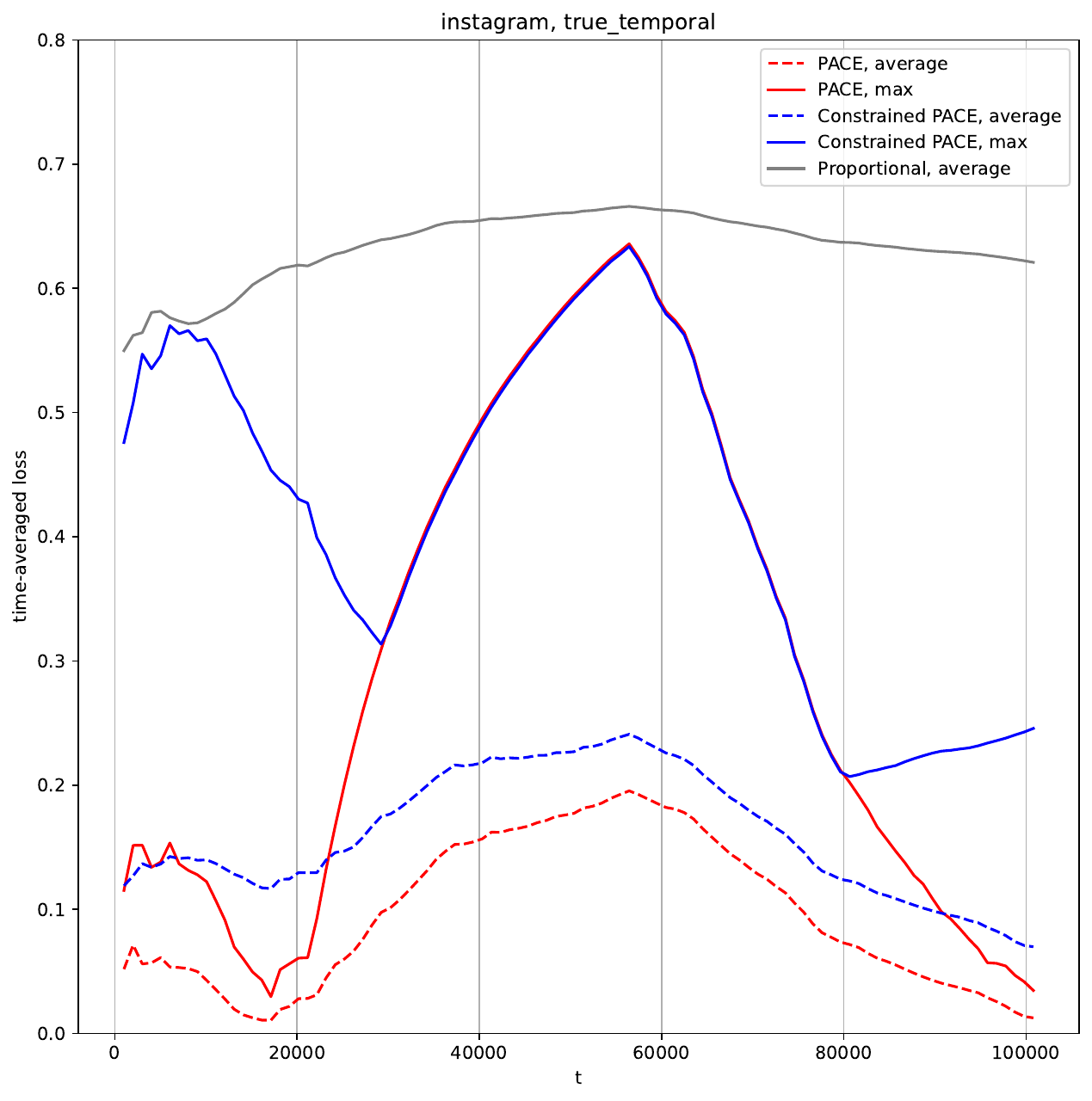}
	\end{subfigure}
     \begin{subfigure}
	\centering
\includegraphics[width=0.49\linewidth]{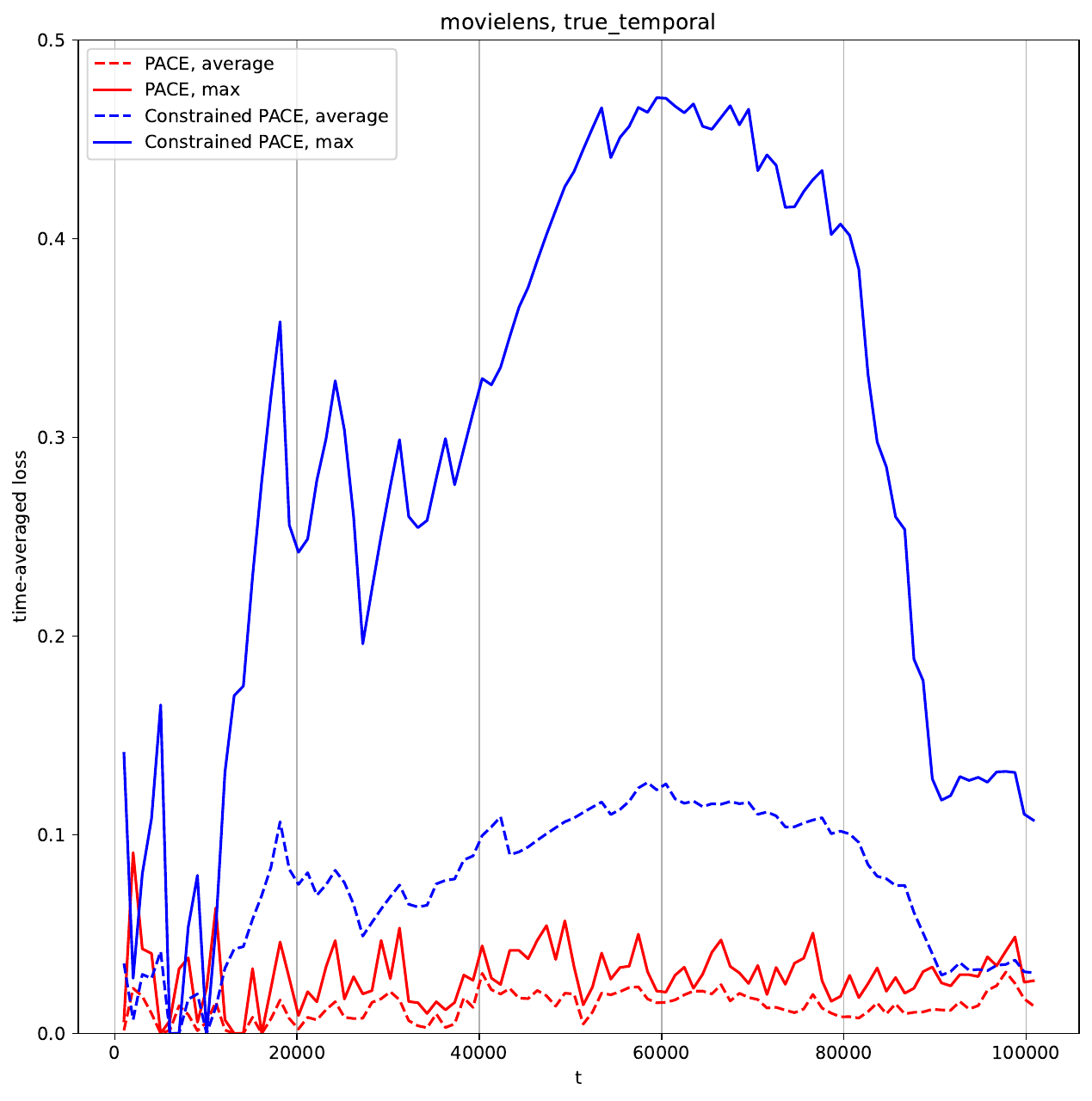}
	\end{subfigure}
    \caption{\small Simulations of constrained and unconstrained PACE on the Instagram notification dataset \textit{(left)} and the MovieLens dataset \textit{(right)} with true temporal arrival.}
    \label{fig: experiments-constrained}
\end{figure*}

We show the failure of the constrained version of PACE~\citep{gao2021online} in our simulations on true temporal arrivals; see \Cref{fig: experiments-constrained}. Notably, the projection range in the constrained algorithm is set according to \citet{gao2021online}, based on the knowledge of the underlying optimal $u_i^*$. However, the nonstationarity causes the time-average utilities to deviate, and it is difficult for the constrained algorithm to restore the bound after the highly nonstationary period.  
\subsection{More Discussions on True Temporal Arrivals} 
\label{section: true-temporal-extra}
\begin{figure}[!htbp]
    \centering
    \includegraphics[width=0.8\linewidth]{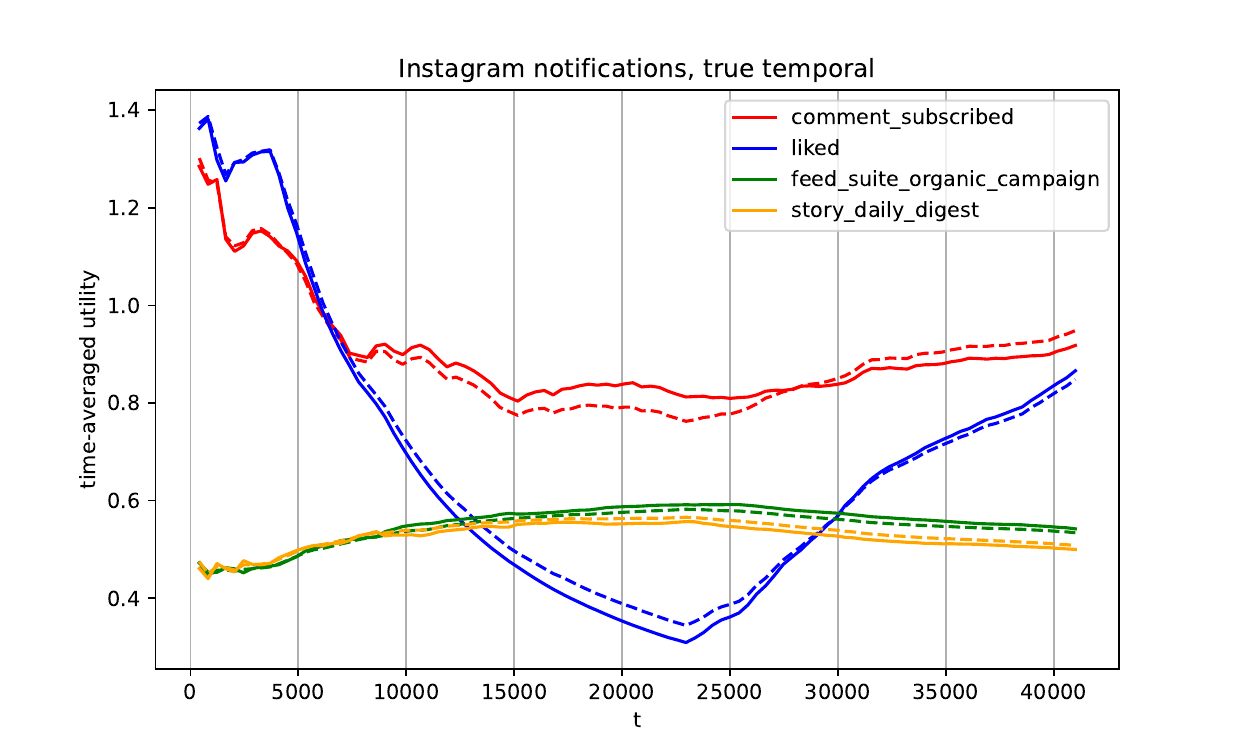}
    \caption{Time-averaged utility of each agent under PACE \textit{(solid line)} and under the optimal allocation given only the arrivals up to the current time step \textit{(dotted line)} in the true temporal experiments with the Instagram notification dataset.}
    \label{fig: experiments-2}
\end{figure}

Here we discuss more closely the structure of the true temporal arrival in the Instagram notification dataset. \Cref{fig: experiments-2} shows the utilities achieved by PACE and by the prefix optimal solution for each agent (i.e., notification type) as a function of $\tau$. We can see that even the prefix optimal solution fluctuates substantially over time. Despite the fluctuation, PACE tracks the hindsight optimal utility closely, although not as closely as in the other experiments. 

\begin{figure}[!htbp]
    \centering
    \includegraphics[width=0.77\linewidth]{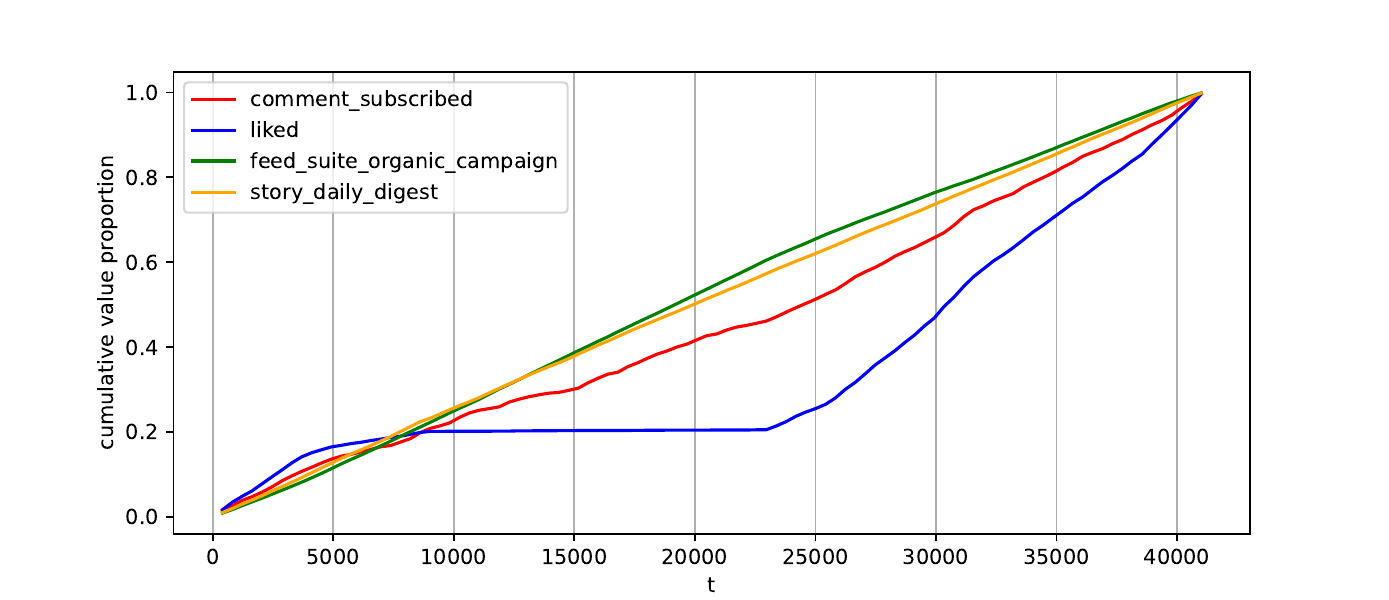}
    \caption{The cumulative value proportion (the fraction of the monopolist utility of an agent revealed up to the current step) of each agent in the true temporal Instagram notification arrival.}
    \label{fig: experiments-3}
\end{figure}

We now investigate the nonstationarity of the dataset itself and argue how it resembles certain adversarial constructions for online fair allocation. We plot the cumulative valuation (i.e., prefix monopolist utility) observed by each agent up to time step $\tau$, as shown in \Cref{fig: experiments-3}. Remarkably, agents' values are distributed across the horizon in a highly non-uniform pattern; for the ``liked'' notification type, there is \emph{no} positive-valued item from step $8,000$ until step $23,000$. This type of nonstationarity is highly undesirable for general online fair allocation algorithms: an agent is unavailable to receive utility for a large fraction of the time, and it would be optimal in hindsight to reserve more items for such an agent in advance. However, an online algorithm cannot know this in advance and cannot prioritize any agent preemptively. This matches the structure of the construction in our \Cref{thm: cr-lower} if the horizon terminates at $\tau = 23,000$. Nevertheless, we remark that after step $23,000$, the ``liked'' notification type becomes available, and PACE recovers quickly and tracks the optimal solution closely. 

